\def\commentswitch#1{\iffalse#1\fi} 

\def\commentswitchsj#1{\iffalse#1\fi} 

\newcommand\notesj[1]{\commentswitchsj{ \todo[color=green!20, inline, size=\small]{Su: #1}}}
\newcommand\notejep[1]{\commentswitch{\todo[color=blue!20, inline, size=\small]{Jack: #1} }}
\newcommand\notekh[1]{\commentswitch{\todo[color=orange!20, inline, size=\small]{Klaus: #1}}}
\newcommand\noteac[1]{\commentswitch{\todo[color=teal!20, inline, size=\small]{Ami: #1}}}
\newcommand\noteemh[1]{\commentswitch{\todo[color=Emerald!20, inline, size=\small]{Eric: #1}}}
\documentclass[a4paper,fleqn,usenatbib]{mnras}
\usepackage[T1]{fontenc}
\usepackage{ae,aecompl}

\usepackage{graphicx}	% Including figure files
\usepackage{amsmath}	% Advanced maths commands
\usepackage{amssymb}	% Extra maths symbols

\usepackage{dcolumn}     % Align table columns on decimal point
\usepackage{bm}          % bold math 
\usepackage{color}
\usepackage{lineno} 
\usepackage[dvipsnames]{xcolor}
\usepackage{todonotes,multirow,enumitem}
\usepackage{soul}
\usepackage[normalem]{ulem}
\usepackage{siunitx}

\usepackage{cprotect}

\usepackage{booktabs}

\newcommand{\ud}{\mathrm{d}}

\newcommand{\bea}{\begin{eqnarray}}
\newcommand{\eea}{\end{eqnarray}}
\newcommand{\beas}{\begin{eqnarray*}}
\newcommand{\eeas}{\end{eqnarray*}}  
\newcommand{\hMpc}{h^{-1}{\rm\;Mpc}}

\newcommand{\kpc}{{\rm\;kpc}}

\newcommand{\degsqr}{{\rm\;deg^2}}
\newcommand{\dperp}{d_{\perp}}
\newcommand{\stripe}{$123 \degsqr~$}  
\newcommand{\stripeGold}{$\sim150 \degsqr~$}
\newcommand{\spt}{$1,244 \degsqr~$}  % 1243.53 
\newcommand{\trainCMASS}{$12,639~$}
\newcommand{\Ntrain}{$6,325~$} % exact
\newcommand{\Nno}{$340,202~$}
\newcommand{\Atrain}{$62.5~$}
\newcommand{\dbsgc}{0.010^{+0.045}_{-0.052} } %0.01151 + 0.04265 - 0.05266
\newcommand{\dbfull}{0.044^{+0.044}_{-0.043}  } % 0.04303 + 0.04398 - 0.04163
\newcommand{\dzsgc}{\left( 3.46^{+5.48}_{-5.55} \right) \times 10^{-3}} %0.00358 + 0.00527 - 0.00561
\newcommand{\dzfull}{( 3.51^{+4.93}_{-5.91}) \times 10^{-3}} % 0.00383 + 0.00453 - 0.00621

\newcommand{\redmagic}{redMaGiC~}

\newcommand\cred{\color{red}}
\newcommand\cgreen{\color{green}}
\newcommand\cgray{\color{gray}}

\usepackage{eso-pic}% http://ctan.org/pkg/eso-pic

\AddToShipoutPictureBG*{%
  \AtPageUpperLeft{%
    \hspace{0.75\paperwidth}%
    \raisebox{-3.5\baselineskip}{%
      \makebox[0pt][l]{\textnormal{DES-2019-0449}}
}}}%

\AddToShipoutPictureBG*{%
  \AtPageUpperLeft{%
    \hspace{0.75\paperwidth}%
    \raisebox{-4.5\baselineskip}{%
      \makebox[0pt][l]{\textnormal{FERMILAB-PUB-19-238-AE}}
}}}%

\def\cmt#1{{\cred#1}}
\def\cmtanswer#1{{\cred#1} }

\def\sjcmt#1{[{\cgreen#1}] }
\def\reph#1{{\cgray#1}}

\title[DMASS]{Producing a BOSS-CMASS sample with DES imaging}

\author[DES Collaboration]{
\parbox{\textwidth}{
\Large
S.~Lee,$^{1,2}$
E.~M.~Huff,$^{3}$
A.~J.~Ross,$^{1}$
A.~Choi,$^{1}$
C.~ Hirata,$^{1,2}$
K.~Honscheid,$^{1,2}$
N.~MacCrann,$^{1,2}$
M.~A.~Troxel,$^{4}$
C.~Davis,$^{5}$
T.~F.~Eifler,$^{6,3}$
R.~Cawthon,$^{7}$
J.~Elvin-Poole,$^{1,2}$
J.~Annis,$^{8}$
S.~Avila,$^{9}$
E.~Bertin,$^{10,11}$
D.~Brooks,$^{12}$
A.~Carnero~Rosell,$^{13,14}$
M.~Carrasco~Kind,$^{15,16}$
J.~Carretero,$^{17}$
L.~N.~da Costa,$^{14,18}$
J.~De~Vicente,$^{13}$
S.~Desai,$^{19}$
B.~Flaugher,$^{8}$
P.~Fosalba,$^{20,21}$
J.~Garc\'ia-Bellido,$^{9}$
E.~Gaztanaga,$^{20,21}$
D.~W.~Gerdes,$^{22,23}$
D.~Gruen,$^{24,5,25}$
R.~A.~Gruendl,$^{15,16}$
J.~Gschwend,$^{14,18}$
G.~Gutierrez,$^{8}$
D.~L.~Hollowood,$^{26}$
D.~J.~James,$^{27}$
T.~Jeltema,$^{26}$
K.~Kuehn,$^{28}$
M.~Lima,$^{29,14}$
M.~A.~G.~Maia,$^{14,18}$
J.~L.~Marshall,$^{30}$
P.~Martini,$^{1,31}$
P.~Melchior,$^{32}$
F.~Menanteau,$^{15,16}$
C.~J.~Miller,$^{22,23}$
R.~Miquel,$^{33,17}$
R.~L.~C.~Ogando,$^{14,18}$
A.~Palmese,$^{8}$
A.~A.~Plazas,$^{32}$
E.~Sanchez,$^{13}$
V.~Scarpine,$^{8}$
M.~Schubnell,$^{23}$
S.~Serrano,$^{20,21}$
I.~Sevilla-Noarbe,$^{13}$
M.~Smith,$^{34}$
M.~Soares-Santos,$^{35}$
F.~Sobreira,$^{36,14}$
E.~Suchyta,$^{37}$
M.~E.~C.~Swanson,$^{16}$
G.~Tarle,$^{23}$
D.~Thomas,$^{38}$
J.~Weller,$^{39,40,41}$
and J.~Zuntz$^{42}$
\begin{center} (DES Collaboration) \end{center}
}
\vspace{0.4cm}
\\
\parbox{\textwidth}{
$^{1}$ Center for Cosmology and Astro-Particle Physics, The Ohio State University, Columbus, OH 43210, USA\\
$^{2}$ Department of Physics, The Ohio State University, Columbus, OH 43210, USA\\
$^{3}$ Jet Propulsion Laboratory, California Institute of Technology, 4800 Oak Grove Dr., Pasadena, CA 91109, USA\\
$^{4}$ Department of Physics, Duke University Durham, NC 27708, USA\\
$^{5}$ Kavli Institute for Particle Astrophysics \& Cosmology, P. O. Box 2450, Stanford University, Stanford, CA 94305, USA\\
$^{6}$ Department of Astronomy/Steward Observatory, University of Arizona, 933 North Cherry Avenue, Tucson, AZ 85721-0065, USA\\
$^{7}$ Physics Department, 2320 Chamberlin Hall, University of Wisconsin-Madison, 1150 University Avenue Madison, WI  53706-1390\\
$^{8}$ Fermi National Accelerator Laboratory, P. O. Box 500, Batavia, IL 60510, USA\\
$^{9}$ Instituto de Fisica Teorica UAM/CSIC, Universidad Autonoma de Madrid, 28049 Madrid, Spain\\
$^{10}$ CNRS, UMR 7095, Institut d'Astrophysique de Paris, F-75014, Paris, France\\
$^{11}$ Sorbonne Universit\'es, UPMC Univ Paris 06, UMR 7095, Institut d'Astrophysique de Paris, F-75014, Paris, France\\
$^{12}$ Department of Physics \& Astronomy, University College London, Gower Street, London, WC1E 6BT, UK\\
$^{13}$ Centro de Investigaciones Energ\'eticas, Medioambientales y Tecnol\'ogicas (CIEMAT), Madrid, Spain\\
$^{14}$ Laborat\'orio Interinstitucional de e-Astronomia - LIneA, Rua Gal. Jos\'e Cristino 77, Rio de Janeiro, RJ - 20921-400, Brazil\\
$^{15}$ Department of Astronomy, University of Illinois at Urbana-Champaign, 1002 W. Green Street, Urbana, IL 61801, USA\\
$^{16}$ National Center for Supercomputing Applications, 1205 West Clark St., Urbana, IL 61801, USA\\
$^{17}$ Institut de F\'{\i}sica d'Altes Energies (IFAE), The Barcelona Institute of Science and Technology, Campus UAB, 08193 Bellaterra (Barcelona) Spain\\
$^{18}$ Observat\'orio Nacional, Rua Gal. Jos\'e Cristino 77, Rio de Janeiro, RJ - 20921-400, Brazil\\
$^{19}$ Department of Physics, IIT Hyderabad, Kandi, Telangana 502285, India\\
$^{20}$ Institut d'Estudis Espacials de Catalunya (IEEC), 08034 Barcelona, Spain\\
$^{21}$ Institute of Space Sciences (ICE, CSIC),  Campus UAB, Carrer de Can Magrans, s/n,  08193 Barcelona, Spain\\
$^{22}$ Department of Astronomy, University of Michigan, Ann Arbor, MI 48109, USA\\
$^{23}$ Department of Physics, University of Michigan, Ann Arbor, MI 48109, USA\\
$^{24}$ Department of Physics, Stanford University, 382 Via Pueblo Mall, Stanford, CA 94305, USA\\
$^{25}$ SLAC National Accelerator Laboratory, Menlo Park, CA 94025, USA\\
$^{26}$ Santa Cruz Institute for Particle Physics, Santa Cruz, CA 95064, USA\\
$^{27}$ Center for Astrophysics $\vert$ Harvard \& Smithsonian, 60 Garden Street, Cambridge, MA 02138, USA\\
$^{28}$ Australian Astronomical Optics, Macquarie University, North Ryde, NSW 2113, Australia\\
$^{29}$ Departamento de F\'isica Matem\'atica, Instituto de F\'isica, Universidade de S\~ao Paulo, CP 66318, S\~ao Paulo, SP, 05314-970, Brazil\\
$^{30}$ George P. and Cynthia Woods Mitchell Institute for Fundamental Physics and Astronomy, and Department of Physics and Astronomy, Texas A\&M University, College Station, TX 77843,  USA\\
$^{31}$ Department of Astronomy, The Ohio State University, Columbus, OH 43210, USA\\
$^{32}$ Department of Astrophysical Sciences, Princeton University, Peyton Hall, Princeton, NJ 08544, USA\\
$^{33}$ Instituci\'o Catalana de Recerca i Estudis Avan\c{c}ats, E-08010 Barcelona, Spain\\
$^{34}$ School of Physics and Astronomy, University of Southampton,  Southampton, SO17 1BJ, UK\\
$^{35}$ Brandeis University, Physics Department, 415 South Street, Waltham MA 02453\\
$^{36}$ Instituto de F\'isica Gleb Wataghin, Universidade Estadual de Campinas, 13083-859, Campinas, SP, Brazil\\
$^{37}$ Computer Science and Mathematics Division, Oak Ridge National Laboratory, Oak Ridge, TN 37831\\
$^{38}$ Institute of Cosmology and Gravitation, University of Portsmouth, Portsmouth, PO1 3FX, UK\\
$^{39}$ Excellence Cluster Origins, Boltzmannstr.\ 2, 85748 Garching, Germany\\
$^{40}$ Max Planck Institute for Extraterrestrial Physics, Giessenbachstrasse, 85748 Garching, Germany\\
$^{41}$ Universit\"ats-Sternwarte, Fakult\"at f\"ur Physik, Ludwig-Maximilians Universit\"at M\"unchen, Scheinerstr. 1, 81679 M\"unchen, Germany\\
$^{42}$ Institute for Astronomy, University of Edinburgh, Edinburgh EH9 3HJ, UK\\
}
}

\date{Accepted XXX. Received YYY; in original form ZZZ}
\pubyear{2019}

\begin{document}
\label{firstpage}
\pagerange{\pageref{firstpage}--\pageref{lastpage}}
\maketitle

\begin{abstract}
We present a sample of galaxies with the Dark Energy Survey (DES) photometry that replicates the properties of the BOSS CMASS sample. The CMASS galaxy sample has been well characterized by the Sloan Digital Sky Survey (SDSS) collaboration and was used to obtain the most powerful redshift-space galaxy clustering measurements to date. A joint analysis of redshift-space distortions (such as those probed by CMASS from SDSS) and a galaxy-galaxy lensing measurement for an equivalent sample from DES can provide powerful cosmological constraints. Unfortunately, the DES and SDSS-BOSS footprints have only minimal overlap, primarily on the celestial equator near the SDSS Stripe 82 region. Using this overlap, we build a robust Bayesian model to select CMASS-like galaxies in the remainder of the DES footprint. 
The newly defined DES-CMASS (DMASS) sample consists of 117,293 effective galaxies covering $1,244\deg^2$. Through various validation tests, we show that the DMASS sample 
%Through various validation tests, we show that the newly defined DES-CMASS (DMASS) sample 
selected by this model matches well with the BOSS CMASS sample, specifically in the South Galactic cap (SGC) region that includes Stripe 82. Combining measurements of the angular correlation function and the clustering-z distribution of DMASS, we constrain the difference in mean galaxy bias and mean redshift between the BOSS CMASS and DMASS samples to be $\Delta b = \dbsgc$ and $\Delta z = \dzsgc$ for the SGC portion of CMASS, and $\Delta b = \dbfull$ and $\Delta z= \dzfull$ for the full CMASS sample. These values indicate that the mean bias of galaxies and mean redshift in the DMASS sample is consistent with both CMASS samples within $1\sigma$.
\end{abstract}


%\tableofcontents
%\appendix

\section{Introduction}
\label{sec:intro}

%Since the discovery of the accelerating expansion of the Universe two decades ago \REF{Reiss, Purmutter}, observational and theoretical work led us to a concordance cosmological model dominated by 70\% of dark energy, 25\% of dark matter, and 5\% of baryons \REF{LambdCDM ref  Lahav, Mortonson}.

Since the discovery of the accelerating expansion of the Universe two decades ago \citep{Riess98, Perlmutter99}, observational and theoretical work has led to a concordance cosmological model dominated by 70\%  dark energy, 25\% dark matter, and 5\% baryons.
Despite the fact that dark energy occupies the majority of the energy density in the universe, 
%Despite the massive fraction of dark energy inDMASS the Universe, 
little is understood about its physical nature due to the apparent lack of visible properties. 
%we barely understand it because of its invisible and undetectable properties.
%While there is overwhelming observational evidence for dark energy, there is no compelling theoretical understanding yet. 
%Aside from the observational evidence, dark energy doesn't have compelling models, 
%Even the most persuasive model $-$the cosmological constant$-$ introduces new physics beyond the standard model of particle physics \cite{Frieman2008}. 
%
%The accelerated expansion of the Universe is generally explained due to the dark energy that occupies 70\% of the Universe. 
Compelling evidence for the presence of dark energy comes from observations of the underlying matter distribution in the Universe using supernovae, Baryon Acoustic Oscillations (BAO), and measurements of large-scale structure growth \citep{FriemanReview08,  Weinberg13, HutererReview181}.
\noteemh{Try for one or two bridge paragraphs, connecting dark energy to the need for joint lensing and clustering constraints. \cmtanswer{ -- Shown in the previous paragraph and in the first sentence of the next pragraph. - Compelling evidence for the presence of dark energy comes from ovservations of the underlying matter distribution ~~. / To trace out the underlying structrue of matter~.  \\
To show the connection clearly, the first sentence was reworded}}
%\sout{In this sense, cosmologists traditionally use galaxies to trace out the underlying structure in matter, by measuring galaxy clustering as a function of scales. }

To trace out the underlying structure in matter, cosmologists traditionally use galaxies by measuring galaxy clustering as a function of spatial separation. 
%In this sense, galaxies have been traditionally used to trace out the underlying structure in matter. 
However, using galaxies as tracers results in a biased view of the matter distribution because
%It is likely that 
%galaxies will form only in regions where gas can reach high enough density to cool and form stars.
galaxies form at the peaks of the matter density field where gas reaches high enough density to cool and form stars \citep{Kaiser1984}.
The relation between the spatial distributions of galaxies and the underlying dark matter density field is known as galaxy bias.
Galaxy bias varies for different scales and galaxy properties such as luminosity or type, and those quantities are degenerate with each other. In the absence of additional information, galaxy bias is indistinguishable from the overall amplitude of matter fluctuations, which makes galaxy bias a major systematic uncertainty in cosmological analyses \citep{Seljak2003SDSSImplications}. 
%
% Focus on why it is hard to predict. Delete all the details about large and small scales. 

%The relation between the spatial distributions of galaxies and the underlying dark matter density field, known as galaxy bias $b_g$, depends on scales and galaxy properties such as luminosity, color and redshift \citep{Kaiser1984}. On small scales, the dependencies become complicated and can be predicted only by numerical simulations. On large scales, galaxy bias becomes less scale-dependent and tends towards a constant value,  however, the latest analysis of the Baryon Oscillation Spectroscopic Survey (BOSS) reported that the galaxy bias derived from the combination of redshift-space distortion parameters still yields a $\sim 10\%$ uncertainty, making this as one of the major sources of systematic error \REF{allBOSSDR12analysis}. 

Fortunately, weak gravitational lensing provides a direct way to measure the matter distribution, avoiding the issue of galaxy bias. Cosmic shear is the subtle shape distortions of background  (source) galaxies by the foreground (lens) matter distribution. It is thus directly connected to the matter distribution and thereby lets us measure the matter distribution without any galaxy bias
(see the review in \cite{Weinberg13} and references therein).
%\citep{Weinberg13}. 
However, cosmic shear is technically challenging to measure due to many sources of systematic errors. Because of the small size of the effect compared to the 
%intrinsic, randomly-oriented galaxy ellipticities, 
intrinsic random variation in galaxy orientations and ellipticities, weak lensing measurements require a substantial number of source galaxies to achieve small statistical errors.
This results in including small and faint galaxies whose systematic errors are challenging to control 
%\sout{This results in including source galaxies as faint and small as possible down to the limit where systematic errors need to be controlled} 
\citep{Mandelbaum2017WeakCosmology}.

Galaxy-galaxy lensing has been shown to be a powerful tool 
%Galaxy-galaxy lensing has been shown to be a powerful tool to overcome the aforementioned limitations 
\citep{Baldauf2010, Yoo2012, Choi2012, Vandenbosch2013, Mandelbaum2013, Park2016CombiningClustering, Miyatake2015,More2015,Alam2017TestingCMASS, Singh2018, Singh2019, Amon2018, Jullo2019}
 that is insensitive to some of the systematic errors that affect cosmic shear \citep{Hirata2004}.
%\notesj{references in Singh paper}
%Galaxy-galaxy lensing was first introduced by \REF{Baldauf et al (2010)}. 
It is the cross-correlation function between foreground galaxies and background shear, which represents a direct measurement of the galaxy-matter correlation function. 
%\cmt{\sout{This method has a lower systematic error budget and costs less computational effort. However, lensing observables fully exert their full constraining power only in combination with accurate galaxy clustering information. }}
%the genuine beauty of galaxy-galaxy lensing comes when it is combined with galaxy clustering. 
In combination with accurate galaxy clustering information, lensing observables can fully exert their  constraining power. 
In galaxy-galaxy lensing, the galaxy bias is tied to the matter clustering in a different way from galaxy clustering. Combining the two probes breaks the degeneracy between the two constraints.
%Therefore, on linear scales, it is possible to combine two clusterings and cancel out the galaxy bias 
%\citep{Baldauf2010}.
%\noteac{I would change the above "cosmic shear" to simply "shear" because cosmic shear refers specifically to shear caused by LSS.  Also, I wouldn't say GGL "has been proposed" because it's been applied successfully for decades, so maybe "has been shown to be" would be better? \cmtanswer{-- Done. now 'weak lensing shear' or 'shear'} }

Some  of the sets of galaxies most frequently used as gravitational lenses in cosmological analyses are the BOSS spectroscopic galaxy samples \citep{Reid2016} from the Baryon Oscillation Spectroscopic Survey 
\citep[BOSS;][]{Dawson2013TheSDSS-III}, which is part of the Sloan Digital Sky Survey-III \citep[SDSS-III;][]{Eisenstein2011BOSS}. 
%The BOSS CMASS galaxy sample is one of the target galaxy samples from the Baryon Oscillation Spectroscopic Survey (BOSS). 
The large sample size and availability of spectroscopic redshifts for all BOSS galaxies allowed the BOSS collaboration to measure the BAO signature with an uncertainty of only one per cent for the case of the BOSS CMASS sample, which is the most constraining BAO measurement to date  \citep{Reid2016}. 
%\notejep{"best" is maybe subjective? most constraining or most precise is better? \cmtanswer{-- Done} }
This led to several follow-up studies that combined the BOSS galaxy clustering results with galaxy-galaxy lensing measurements using the BOSS galaxies as lenses.  

%?Earlier, Mandelbaum?..? and the next few sentences are kind of a mess?..could flow better. Should say what the concept of Baldauf is briefly. Next sentence uses ?samples? twice. A few sentences later, not clear whose work is referred to with ?In their work?. Just write this section more clearly on who did what and what is being referred to.

\cite{Mandelbaum2013} constrained the amplitude of the matter fluctuations at $z<0.4$ using data from the Sloan Digital Sky Survey (SDSS) Data release 7. 
They utilized two spectroscopic samples $-$ BOSS Main and Luminous Red samples as lenses and combined galaxy-galaxy lensing between those samples and SDSS source galaxies with galaxy clustering from the same samples.
\cite{Singh2018} adopted a similar approach. They combined galaxy clustering from BOSS with galaxy-galaxy lensing and galaxy-CMB lensing signals, by utilizing the BOSS LOWZ ($0.15<z<0.43$) and CMASS ($0.43<z<0.7$) samples as lenses. However, due to the shallow depth of SDSS imaging, their measurement of galaxy-galaxy lensing was obtained only with BOSS LOWZ. 

%\cmt{ \cite{Singh2016Cross-correlatingCross-correlations} adopted a similar approach. They utilized the BOSS LOWZ sample ($0.15<z<0.43$) and CMASS ($0.43<z<0.7$) as lenses and SDSS galaxies as sources to measure weak lensing signals. However, due to the shallow depth of SDSS imaging, they instead measured galaxy-Planck CMB lensing  for CMASS. These two weak lensing measurements were combined with galaxy clustering from BOSS LOWZ and CMASS. }

\cite{Miyatake2015}, \cite{More2015} and \cite{Alam2017TestingCMASS} extended this kind of joint analysis to galaxies at a higher redshift $z\sim 0.5$ by using BOSS CMASS as lenses with the deeper and better quality imaging data from CFHTLenS \citep{CFHTLenS}.
%independent data sets - BOSS CMASS as lenses and the CFHTLenS catalog \citep{CFHTLenS} as sources. 
%the BOSS CMASS galaxies as lenses. Due to the shallow depth of SDSS imaging, they instead utilized the CFHTLenS catalog \citep{CFHTLenS}.  
%
%The CFHTLenS catalog contains galaxies with a median redshift of $z\sim0.7$ and much deeper than SDSS, therefore is adequate as sources for  CMASS lenses.
%
\cite{Jullo2019} performed a similar analysis with BOSS CMASS galaxies and two weak lensing data sets - CFHTLenS and CFHT-Stripe 82 \citep{CFHT-S82}. 
\cite{Amon2018} utilized 
three spectroscopic galaxy samples including BOSS LOWZ \& CMASS with
%BOSS LOWZ and CMASS with other two spectroscopic galaxy samples (2dFLenS \citep{2dFLenS}, GAMA \citep{GAMA}) 
 deep imaging data from KiDS \citep{KiDS} to do a joint analysis of galaxy clustering and galaxy-galaxy lensing. 
However, the lensing measurements of these analyses are limited to the small overlapping area - a few hundreds of $\degsqr$. For instance, the overlapping region between BOSS and CFHTLenS  is only $\sim 105 \degsqr$ which is about one hundredth of the BOSS area.  

The Dark Energy Survey (DES) is a large photometric survey that images over $5,000 \degsqr$ of the southern sky to 
 a $5\sigma$ limiting magnitude of $\sim24$ in the $i$-band. It observes in the $grizY$ filter bands. 
%24th $i$-band limiting magnitude in the $grizY$ filter bands. 
Precise photometry and the largest survey area among the current generation of 
%Stage-III 
experiments 
%and precise photometry 
%Its high signal-to-nosie weak lensing signal 
makes DES data an excellent source of
%strong candidate as 
imaging data for a joint analysis of galaxy clustering and galaxy-galaxy lensing. %combining weak lensing and the BOSS measurements.  
%\sout{However, as the other lensing surveys aforementioned,} 
However, as with previous measurements combining lensing and clustering,
%\noteemh{However, as with previous measurements combining lensing and clustering,} 
the overlapping region between the DES Year 1 footprint ($\sim1,800\degsqr)$ and the BOSS footprint is fairly small, consisting of only \stripeGold near the celestial equator called Stripe 82 \citep{ DESOverview, Y1GOLD}.
%
%\sout{ For instance, one of the most promising weak lensing surveys, the Dark Energy Survey (DES), covers $\sim 5000 \degsqr$ but the overlapping region between the DES Year 1 footprint ($1800 \degsqr$) and BOSS is only $\sim 124 \degsqr$ near the celestial equator called Stripe 82 \citep{ DESOverview, Y1GOLD}. }
%\notejep{should maybe put DES Y1 area here too? so you would have DES Y5, Y1, Y1-BOSS overlap \cmtanswer{-- Done} }
Simply combining BOSS galaxy clustering with galaxy-galaxy lensing from DES would be limited to the small overlapping area and fail to utilize the full statistical power of DES.
%If one attempts to combine the BOSS galaxy clustering with galaxy-galaxy lensing from DES, one would end up with a result limited to the minimal overlapping area failing to utilize the entire power of DES.

%From Troxel et al : In 2017 the DES collaboration published the analyses of its first year of data (Y1). It presented results which, for the first time, put constraints on certain cosmological parameters derived from galaxy surveys at the same level as the constraints obtained from the CMB data which is based on physical processes billions of years before galaxies were formed. These results, described in [34] (hereafter Y1KP) are based on the two-point statistics of galaxy clustering and weak gravitational lensing. The combined analysis of the three different two-point correlation functions (galaxy cluster- ing, cosmic shear, and the galaxy-shear cross-correlation, typ- ically referred to as galaxy-galaxy lensing) is the end-product of a complex set of procedures which

%\notekh{Instead we present in this paper an attempt to define a catalog of DES galaxies from the full survey footprint with properties matching the SDSS CMASS sample.  or something like this \cmtanswer{-- Done. See the next paragraph.} }
Inspired by the 
%\sout{challenging points addressed} 
potential power of combining all the available SDSS and DES measurements, we present in this paper a way of defining a catalog of DES galaxies from the full footprint of DES, whose properties match with the BOSS CMASS galaxy sample. The resulting DES-CMASS (hereafter DMASS) sample will be the best available for a cosmological analysis combining galaxy-galaxy lensing and galaxy clustering measurements. 
Thus, we will produce a sample that effectively increases the area available for such studies by a factor of 10 (\stripe~to \spt).
%
%Inspired by the challenging points addressed, this work describes our efforts to construct a DES-CMASS galaxy sample (hereafter DMASS) in the footprint of the Dark Energy Survey, most of which does not overlap SDSS, with the ultimate goal to perform a cosmological analysis combining the best available galaxy-galaxy lensing and galaxy clustering measurements.
%
%The CMASS spectroscopic sample from the BOSS survey \citep{Reid2016} is the largest spectroscopic sample that provides the most accurate %BAO and RSD measurements to date.  
%Combining the most accurate galaxy clustering measurement with the lensing measurement from the most compelling lensing survey would %achieve by far better measurements than the previous studies.  
%We suggest to select a pure CMASS galaxy sample from the the Dark Energy Survey, most of which does not overlap SDSS, for the purpose of the cosmological analysis combining galaxy galaxy lensing and galaxy clustering. %BOSS CMASS galaxy sample provides the most accurate constraints on [] and the latest shear measurements of DES 
%\cmt{extra advantages like using spectroscopic information to train photoz or something....  }

%\cmt{data, challenges}

We start by using the subset of BOSS CMASS galaxies in Stripe 82 where the BOSS footprint overlaps with DES. 
%and DES Y1 GOLD photometric sample \citep{Y1GOLD} for color matching. 
Using galaxies measured by both DES and BOSS we train a galaxy selection model using the DES photometric information.
%The model for selecting galaxies are trained with the photometric information of DES of this subset. 
%However, the intrinsic way of defining CMASS sample introduces large noise to galaxies at low redshift end, which makes defining the same cut in the DES system difficult. Not only that, a slight offset between two filter systems where $g-r$ transition happens is somewhat correlated to the $g-r$ colors of galaxies at low redshift end. This hardware difference exacerbates the situation. 
%\cmt{how we did}
Rather than classifying individual galaxies, the model assigns a membership probability to each galaxy and down-weights galaxies that are less likely to be CMASS. 
%To account for these, rather than classifiying individual galaxies we assign a membership probability to each galaxy.
%Galaxies near the noisy selection cuts are down-weighted by the low probabilities they have and reproduce the same noisy level that the original CMASS selection cuts have. 
%A galaxy way above than $\dperp$ cut will be assigned high probability but low galaxies near $\dperp$ cut are down-weighted by the value of their probability. This is same with $\dperp$ cut accepting CMASS galaxies randomly within SDSS photometric error.
To account for spatial dependence of photometric errors, we use the Extreme Deconvolution algorithm \citep{Bovy2011a} and obtain underlying color distributions of galaxies from the training sample. The underlying color distributions are convolved with photometric errors of the target region, and thereby the model correctly accounts for 
%is optimized for 
the photometric errors in the different regions.

%\cmt{structure of this paper}
This paper is organized as follows. In the following section, we will introduce the BOSS CMASS sample and the DES Y1 GOLD catalog we use for this work and present the selection criteria that were used for the BOSS CMASS sample in detail. We will  address the difference between the SDSS and DES photometric systems and explain how it will be accounted for in our probabilistic model.
Our model construction can be found in Section \ref{sec:model}. 
%We will discuss the difference between the SDSS and DES photometric systems and explain how it is accounted for in our probabilistic model. 
The systematic uncertainties of the DMASS sample will be presented in Section \ref{sec:systematics} 
%and the resulting DES galaxy sample (DMASS) will be covered in Section \ref{sec:dmass}.
%and systematic uncertainties are addressed in Section \ref{sec:systematics}. 
and the basic properties of the resulting DMASS catalog and validation tests will be discussed in Section \ref{sec:result}. We will summarize and conclude in Section \ref{sec:conclusion}.

%\cmt{cosmology used}
%Throughout this work, we use 
%the cosmology used in the SDSS DR12 BOSS publication \citep{Chuang2017, Pellejero-Ibanez2017, Alam2016, Sanchez2016, Ross}. 
The fiducial cosmological model used throughout this paper is a flat $\Lambda$CDM model with the following parameters: matter density $\Omega_{m} = 0.307$, baryon density $\Omega_b = 0.048$, 
amplitude of matter clustering $\sigma_8 = 0.8288$, 
%primordial amplitude $A_s = 2.26 \times 10^{-9}$, 
spectral index $n_s = 0.96$ and Hubble constant $h \equiv H_0/100~{\rm{km~ s^{-1} Mpc^{-1}}} = 0.677$. 

\begin{figure*}
% <-- (CMH) commented out so this will compile
%\includegraphics[width=0.8\textwidth]{./figures/colormapping_2.png}
\includegraphics[width=0.45\textwidth]{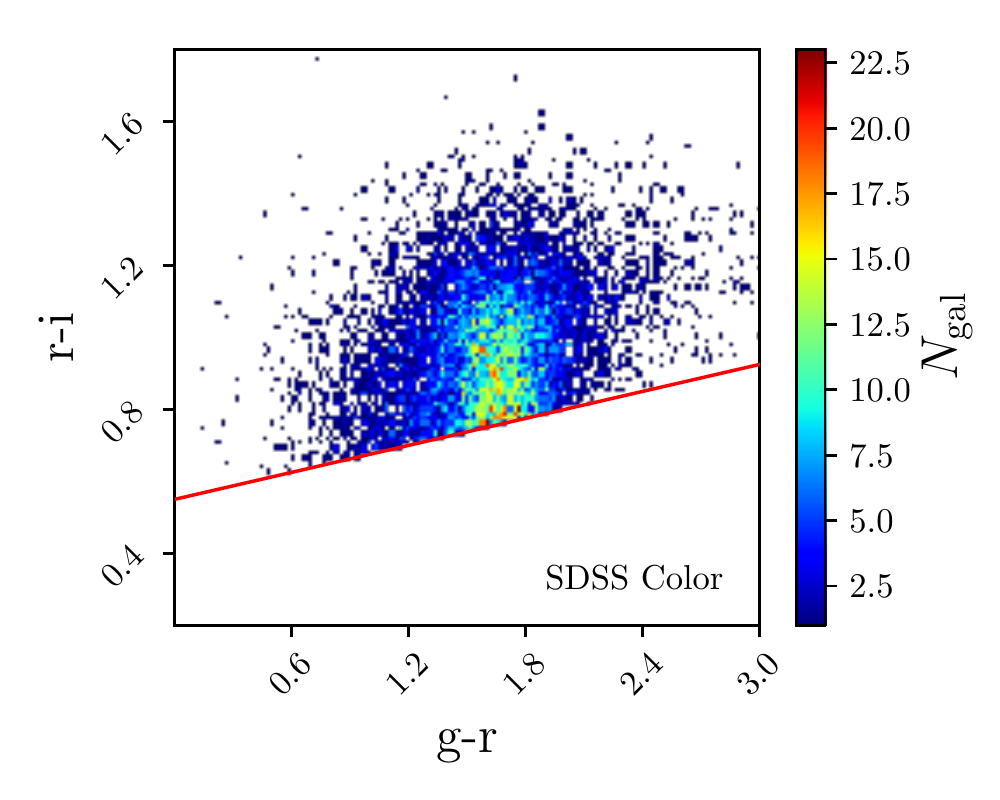}
\includegraphics[width=0.45\textwidth]{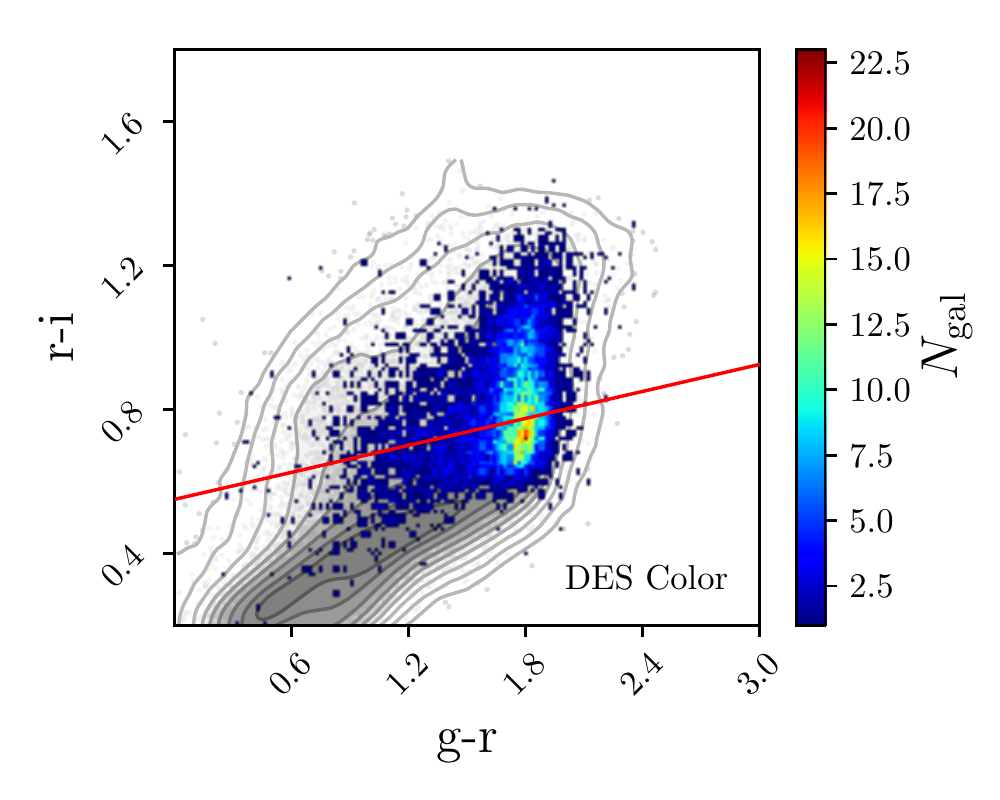}
\caption{
The two-dimensional histograms of CMASS galaxies from Stripe 82 in the $g-r$ vs. $r-i$ color plane. The left panel shows SDSS, the right panel shows DES colors of the same galaxies. 
 %in the SDSS photometry (left) and DES photometry (right) in the $g-r$ vs. $r-i$ color plane. 
The colorbar shows the number of galaxies binned in each histogram bin. 
The red line is the $d_{\perp}$ cut. CMASS galaxies look bluer in the DES photometry and the $d_{\perp}$ cut discards almost half of the CMASS galaxies by crossing the most dense region. The grey contours in the right panel show the full distribution of DES Y1 GOLD galaxies in the color plane. The grey contours show that blindly lowering the $\dperp$ cut results in accepting more non-CMASS galaxies.}
\label{fig:dperp}
\end{figure*}

\begin{figure}
\centering
\includegraphics[width=0.48\textwidth]{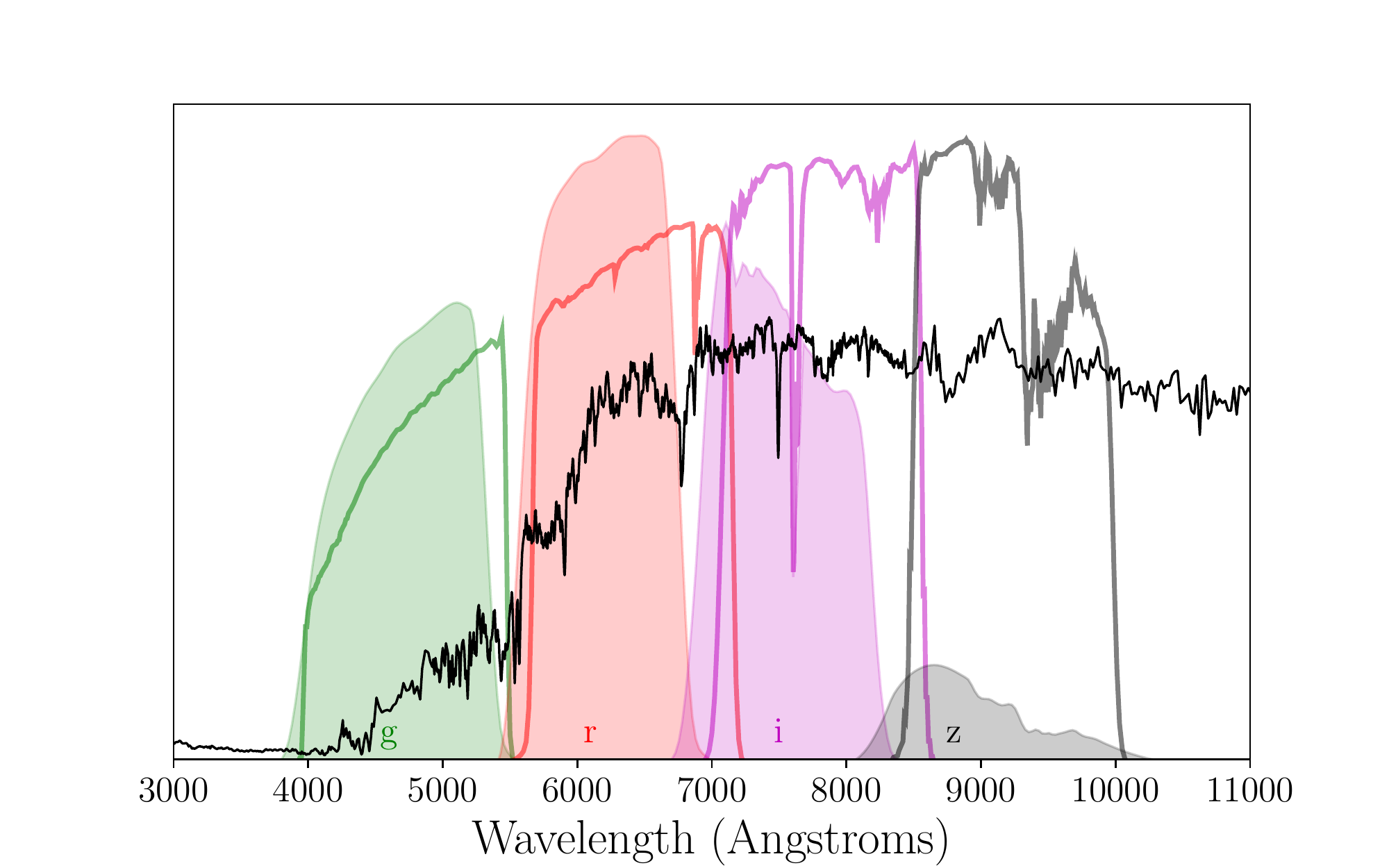}
\caption{
The response functions for the $griz$ SDSS (shaded) and DES (solid lines) filters as a function of wavelength ($\si{\angstrom}$) with the spectral energy density (SED) distribution of an elliptical galaxy at $z=0.4$ (black solid line). 
Near the $4000 \si{\angstrom}$ break where the $g-r$ transition happens, the SDSS $r$ filter (shaded) covers slightly lower wavelength than the DES $r$ filter (solid lines) does. This implies the same galaxy near $z=0.4$ looks redder in the SDSS photometry than in the DES photometry.
%\cmt{ \sout{ As the $4000\AA$ break of the SED moves towards the higher wavelength, the transition between $g$ filter to $r$ filter happens first in SDSS, which implies the same galaxy looks redder in the SDSS photometry.}}
}
\label{fig:filter}
\end{figure}

\section{Data}
\label{sec:data}

%In this section, we describe two data - BOSS CMASS galaxy sample and DES Y2 GOLD galaxy catalog - that we use for this work. We dedicated on 

%We determine selection criteria to apply to DES data in order to produce a galaxy sample matched to BOSS\REF{BOSS SURVEY}. In order to do so, we must understand how BOSS data was selected. Specifically, we match to the BOSS data published as part of data release 12 of the Sloan Digital Sky Survey III  \REF{SDSS-III ref, Eisenstein 2011}. 

\subsection{BOSS DR12 CMASS Sample}

The Baryon Oscillation Spectroscopic Survey 
\citep[BOSS;][]{Eisenstein2011BOSS, Bolton2012SpectralSurvey, Dawson2013TheSDSS-III}
was designed to measure the scale of baryon acoustic oscillations (BAO) in the clustering of matter over a larger volume than the combined efforts of all previous spectroscopic surveys of large-scale structure. 
BOSS uses the same wide field, dedicated telescope as was employed by SDSS I and II \citep{York2000SDSS}, the 2.5 m aperture Sloan Foundation Telescope \citep{Gunn2006TheSurvey} at Apache Point Observatory in New Mexico. Those surveys imaged over $10,000 \degsqr$ of high galactic latitude sky in the $ugriz$ bands, using a mosaic CCD camera  \citep{Gunn1998TheCamera} with a field of view spanning $\ang{3}$.
BOSS consists primarily of two interleaved spectroscopic surveys observed simultaneously: a redshift survey of 1.5 million luminous galaxies extending to $z = 0.7$ and a survey of the Lyman alpha forest toward 150,000 quasars in the redshift range $2.15 < z < 3.5$. 
Description of survey design, target selection, and their implications for cosmological analysis are available in \cite{Dawson2013TheSDSS-III} and \cite{Reid2016}.

The BOSS DR12 galaxy survey targeted two distinct samples known as LOWZ and CMASS \citep{Reid2016}. %The spectroscopic sample used in this work is "CMASS" sample described in Reid et al(2016) from the BOSS survey published as part of the SDSS DR12. 
The higher redshift sample CMASS covers redshifts $0.43 < z < 0.75$ and is designed to select a stellar mass-limited sample of objects of all intrinsic colors, with a color cut that selects almost exclusively on redshift.
%
%BOSS galaxy target selection mirrors the split at $z~0.4$ and selects two principal samples : "LOWZ" and "CMASS." 
%The CMASS sample is designed to select galaxies at $0.43 < z < 0.7$.
%The BOSS galaxies are selected to have approximately uniform comoving number density of $\bar{n} = 3 \times 10^{-4}$ out to a redshift $z = 0.6$, before monotonically decreasing to zero density at $z \sim 0.8$. Galaxy shot noise and sample variance make roughly equal contributions to BAO errors when $\bar{n}P_{BAO} = 1$, where $P_{BAO}$ is the gaalxy power spectrum at the BAO scale, approximately $k = 0.2$. For the strongly clustered galaxies observed by BOSS,  $\bar{n} = 3 \times 10^{-4}$ yields roughly $\bar{n}P_{BAO} = 2$, making shot noise clearly sub-dominant. 
%
% << Need to be paraphrased ---------------------------------
% General overview of CMASS : Cawthon et al. (2018)
% Need to be paraphrased ------------------------------>> 
%
% The Baryon Oscillations Spectroscopic Survey is a part of the SDSS program
% We focus on CMASS DR12 sample. Galaxies are selected from SDSS DR8 imaging (Aihara et al. 2011) according to a series of colour cuts designed to obtain a sample with ap- proximately ?constant stellar mass.
% color cuts
% weight
%
%
\noteemh{(putting the definitions of quantities immediately after the equations where they appear) \cmtanswer{-- Paragraphs were reorganized for this starting from the next paragraph:}} 
The CMASS galaxy sample is selected by the combination of the 7 different  color and magnitude cuts. %\citep{Reid2016}.
Every source satisfying the selection cuts was targeted by the BOSS spectrograph to obtain their redshifts, except for $5.8\%$ of targets in a fiber collision group and $1.8\%$ of targets for which the spectroscopic pipeline fails to obtain a robust redshift \citep{Reid2016}.
%More details about the BOSS target selection can be found in \cite{Reid2016}. 

The following three cuts simply limit colors or magnitudes to exclude redshift failures or outliers with problematic photometry:
\bea
17.5 < &i_{\rm cmod}& < 19.9 \label{eq:imag_limit}\\
i_{\rm fib2} &<& 21.5 \label{eq:ifib_limit} \\
r_{\rm mod} - i_{\rm mod} &<& 2~,  \label{eq:outlier}
\eea
where the subscript `mod' denotes model magnitudes, `cmod' denotes cmodel magnitudes, and `fib2' stands for fiber magnitude estimated in a $2''$ aperture diameter assuming $2''$ seeing. For further details of SDSS magnitudes, we refer readers to the SDSS survey website\footnote{\url{https://www.sdss.org/dr12/algorithms/magnitudes}}.
The following two cuts are applied to reject stars:
\bea
i_{\rm psf} - i_{\rm mod} &>& 0.2 + 0.2(20-i_{\rm mod}) \label{eq:sg} \\
z_{\rm psf} - z_{\rm mod} &>& 9.125 - 0.46 z_{\rm mod}~, \label{eq:sg2}
\eea
where `psf' stands for magnitudes computed from the point spread function model. 

To exclusively select galaxies on redshift, the BOSS target selection utilizes the quantity $\dperp$ defined as 
\bea
\dperp \equiv (r_{\rm mod}-i_{\rm mod}) - (g_{\rm mod}-r_{\rm mod})/8.0~.  
\eea
This quantity is designed to approximately follow the color locus of the passively evolving LRG model in \cite{Maraston2009} at $z>0.4$. Since redshift gradually increases along the color locus, $\dperp$ is a good indication of redshift for CMASS type galaxies. 

The following two cuts use $\dperp$ to select objects with respect to redshift:
\bea
i_{\rm cmod} &<& 19.86 + 1.6 (d_{\perp} - 0.8)
\label{eq:mag_color} \\
d_{\perp} &>& 0.55 \label{eq:dperp}~.
\eea
Equation \eqref{eq:mag_color} selects the brightest objects at each redshift to keep an approximately constant stellar mass limit over the redshift range of CMASS. 
Equation \eqref{eq:dperp}, the so called `$\dperp$' cut, is the most restrictive cut among all selections described above. 
This cut isolates intrinsically red galaxies at high redshift.
Considering the color/magnitude space occupied by all SDSS objects, this cut slices the densest region of the sample in the $gri$ color plane and determines the sample's redshift distribution. 
This is in contrast to the other cuts, which apply mainly to the edges of the color/magnitude distributions. Therefore, our work is mainly focused on characterizing the same cut in the DES photometry. More details about the $\dperp$ cut can be found in \cite{Eisenstein2001LRG} and \cite{Padmanabhan2007}, and our derived $\dperp$ cut in the DES system will be discussed in Section \ref{sec:model}.

The colors and magnitudes used in the selection criteria are corrected for Milky Way extinction by the galactic extinction map \citep{SFD98}.

\subsection{DES Y1 Gold Catalog}
 
The Dark Energy Survey \citep[DES;][]{DESCollaboration2006, DESCollaboration2017} 
is an imaging survey covering $5,000 \degsqr$ of the southern sky.
This photometric data set has been obtained  
in five broadband filters, $grizY$, ranging from $\sim 400 \rm nm$ to $\sim 1,060 \rm nm$
\citep{Li2016ASSESSMENTSURVEYS, Burke2017ForwardSurvey}, using 
%is a photometric survey utilizing 
the Dark Energy Camera \citep[DECam;][]{Flaugher2015THECAMERA} mounted on the Blanco 4m telescope at Cerro Tololo Inter-American Observatory (CTIO) in Chile. 
%to observe $\sim 5000 \degsqr$ of the southern sky 
%in five broadband filters , $grizY$, ranging from $\sim 400 \rm nm$ to $\sim 1060 \rm nm$
%\REF{(Li et al. 2016; Burke et al. 2018)}.
%\cmt{gold catalog}
The main goal of DES is to improve our understanding of cosmic acceleration and the nature of dark energy using four key probes: weak lensing, large-scale structure, galaxy clusters, and Type Ia supernovae. 

The Y1A1 GOLD wide-area object catalog\footnote{\url{https://des.ncsa.illinois.edu/releases/y1a1}} \citep{Y1GOLD} we use in this work 
%was published as part of the DES Year 1 public data release \citep{DESDR1}. The catalog 
consists of $\sim 137$ million objects detected in coadd images %covering $\sim 1800 \degsqr$  
covering two disjoint areas; one overlapping with the South Pole Telescope \citep[SPT;][]{SPT}, and a much smaller area near the celestial equator called Stripe 82 \citep{Annis2011The82}.
%Y1 footprint consisted of two areas: one near the celestial equator including Stripe 82 (S82; Annis, James and Soares-Santos, M. and Strauss, M. A. and others 2014), and a much larger area that was also observed by the South Pole Telescope (SPT; Carlstrom et al. 2011).
%in the DES grizY filters and the largest photometric data set at the achieved depth to date, enabling precise measurement of %cosmic acceleration at $z<1$ 

%\cmt{mask, flags, number of objects and coverage after masking} 
For this work, we refine the DES Y1 Gold catalog selection by removing imaging artifacts and areas around bright foreground objects such as bright stars and globular clusters. We only keep clean sources with 
%\sjcmt{Same in Table 4? }
%\verb|FLAGS_GOLD == 0| 
flag $bit < 1$ in Table 4 and flag  $bit < 2$ in Table 5 in \cite{Y1GOLD}. We also select sources classified as galaxies by the flag  \verb+MODEST==1+. 
Furthermore, we remove regions tagged by the DES Y1 BAO study \citep{Y1BAO} using veto masks.
%, except for the regions where multi object fitting (MOF; \cite{Y1GOLD}) pipeline didn't perform properly. 
%applied the same veto mask used for the DES Y1 BAO study \citep{Y1BAO}. 
These additional masks select only the wide area
parts of the surveys, namely those overlapping SPT, and remove a patch of $18\degsqr$ where the airmass computation is highly corrupted. 
%These additional masks exclude small disjoint fields from the DES supernova survey and two auxiliary fields used for photo-z calibration and star-galaxy separation tests, as well as regions where the airmass computation is highly corrupted. 
%The mask also removes HEALPix pixels (at resolution $N_{\rm side} = 4096$) where $80\%$ of the pixel is not occupied by the observed area in all of the $griz$ bands. 
The DES Y1 BAO study additionally removes a few$\degsqr$ sized regions where multi-object fitting \citep[MOF;][]{Y1GOLD} photometry is unreliable. However, we do not exclude these regions since we do not use MOF measurements. Further details about the Y1 BAO masks can be found in \cite{Y1BAO}. The resulting footprint after applying all masks aforementioned occupies \spt in SPT and \stripe in Stripe 82.
%In Section \ref{sec:systematics} we describe additional corrections applied to the sample to correct for observational effects. 

%\cmt{KH: fill in numbers - maybe move to next chapter}
%\sjcmt{SJ: moved color descriptions to the section for the training set}

%Throughout this work, We use DES magnitudes designed for extended sources \verb|MAG_DETMODEL| for colors and \verb|MAG_MODEL| for apparent magnitudes to substitute for SDSS magnitudes \verb|modelMag| and \verb|cModelMag| respectively. Those magnitudes are chosen because their definitions are similar to their SDSS counterparts. %\verb|cModelMag| and \verb|MAG_MODEL|.
%This choice is motivated by the similarities in definitions of \verb|modelMag| and \verb|MAG_DETMODEL|, and \verb|cModelMag| and \verb|MAG_MODEL|. 
%Full descriptions of SDSS and DES magnitudes can be found in \REF{SDSS} and \REF{DES}. 
All magnitudes in the DES Y1 GOLD catalog are shifted by stellar locus regression (SLR) which corrects for Galactic dust reddening \citep{Y1GOLD}. 
%For consistency with the original CMASS selection, we have removed this SLR correction and instead applied reddening corrections based on the SFD map \citep{SFD98} with a reddening law of $\rm A_{\rm griz} = R_{\rm griz} \times E(B - V)_{\rm SFD}$ as done in SDSS. 
%\cmt{\sout{with slightly different coefficients $R_{\rm griz} = [3.186, 2.140,1.569,1.196]$.}
For consistency with the original CMASS selection, we have removed this SLR correction and instead applied reddening corrections based on the SFD map \citep{SFD98} as done in SDSS. The correction to the DES magnitude for a band $b$ is $\rm A_b = R_{b} \times E(B - V)_{\rm SFD}$ with interstellar extinction coefficients for $griz$ bands, $R_{b} = [3.186, 2.140,1.569,1.196]$, computed in \cite{DESDR1}. 
 
We applied additional magnitude cuts to the DES Y1 Gold catalog to exclude outliers in color space as follows:
\bea
17 < &G_{\rm DET}& < 24~ \\
17 < &R_{\rm DET}& < 24~ \\
17 < &I_{\rm DET}& < 24~ \\
0 < &G_{\rm MOD} - R_{\rm MOD}& < 2.5~ \\
0 < &R_{\rm MOD} - I_{\rm MOD}& < 1.5~ \\ %~ {\rm{and}} \\
&I_{\rm AUTO}& < 21~.
\eea
Sources satisfying the magnitude cuts are kept.
Subscripts $\rm DET$ and $\rm MOD$ stand for DES \verb|MAG_DETMODEL|\footnote{
%Magnitude is measured from a fitted shape to the object in a reference detection image taken in one band or a combination of two bands. 
This magnitude is computed by fitting a galaxy model 
to the object in a reference detection image taken in one band or a combination of two bands.
%convolved by the detection PSF in the detection image, 
Then this fitted model is applied to all measurement images, by fitting only the amplitude.
%(which can be taken in one band or be a linear/non-linear combination of images taken in 2 or more filters). 
%After establishing a reference detection image, the fitting is carried out on this image using an exponential galaxy profile. The magnitude estimation in each band is then carried out with the fitted shape.
}  magnitude and \verb|MAG_MODEL|\footnote{This magnitude is measured by fitting a galaxy model to the object in each band.
% from a fitted galaxy model to the object in each band. 
} magnitude respectively, and $\rm AUTO$ stands for DES \verb|MAG_AUTO|\footnote{Magnitude is measured in an elliptical aperture, shaped by the second moments of the object and scaled using the Kron radius.} magnitudes. These three magnitudes are computed by an image-processing software called  \verb|SExtractor|\footnote{\url{https://sextractor.readthedocs.io}}. We refer interested readers to the documentation of \verb|SExtractor| \citep{SEXTRACTOR} for further details. 
Note that all DES quantities are written in upper case to avoid confusion with corresponding SDSS quantities.
%\begin{verbatim}
%17 < MAG_DETMODEL[_GRI] < 24
%0 < MAG_MODEL_R - MAG_MODEL_I < 1.5
%0 < MAG_MODEL_G - MAG_MODEL_R < 2.5
%MAG_AUTO_I < 21 * 
%\end{verbatim}
%The choice of magnitudes and colors is simply following the input color combinations for the probabilistic model that will described in the later section. 
These cuts effectively remove galaxies that are not likely to be CMASS galaxies.
%, and would clearly have corrupted photometry if they were. 
Further, these cuts reduce compute time by decreasing the sample size to 10\% of the full Y1 GOLD sample, while keeping 99.5\% of CMASS galaxies in the overlapping region, Stripe 82. 

%Since our color selection requires observations in all
%four griz bands we use the coverage maps to enforce that all pixels considered (at resolution 4096) show at least 80% coverage in each band.
%Other areas that are severely affected by imaging artifacts or other-wise have a high density of image artifacts are masked out as well.
%Objects are selected so that we avoid imaging artifacts and pernicious regions with foreground objects using the cuts on flags badregion and flags gold described therein.

%A series of veto masks, including among others masks for bright stars and the Large Magellanic Cloud, reduce the area by ∼ 300 deg2, leaving ∼1500 deg2 suitable for galaxy clustering study. 

\subsection{Differences between the SDSS and DES photometry}
\label{sec:data-difference}

%\sout{\cite{Y1GOLD} presented a set of photometric equations mapping magnitudes between the DES system and SDSS systems in $griz$ bands by matching photometry in the equatorial fields of SDSS Data Release 9 \citep{SDSSDR9}. However, these mapping equations are only valid for point sources. Unlike point sources whose magnitudes are calculated from a simple Gaussian luminosity profile, magnitudes for extended sources are derived from different models of luminosity profiles and bands optimized for each source. }
In the DES imaging pipeline, magnitudes for extended sources are derived from different models of luminosity profiles and bands optimized for each source \citep{Y1GOLD}. 
%The best-fit model from one or combination of two galaxy profile models fit in a reference band that can be different for every source. {\bf AJR didn't figure out a way to re-word this}
This complicated procedure makes magnitudes in one band highly correlated with other bands, as well as the shape or size of galaxies and instruments for each system and results in magnitudes for the same object being very different in one system from another in a way that is  challenging to predict. 
%\cgrey{Magnitude for extended sources is complicated interplay among the type and shape of galaxies }
%Therefore, magnitude for extended source is strongly correlated with galaxy profiles and image pipeline, the structure of instrument for each system. \REF{}

Figure \ref{fig:dperp} shows the difference in the $r-i$ vs. $g-r$ color space of the two different imaging systems, using only tagged CMASS galaxies in the overlap region. 
%\noteac{Have you considered having a colorbar for this plot?  Is the color scale exactly the same for both panels? \cmtanswer{--done}}
The DES colors of CMASS galaxies are obtained by cross-matching the DES Y1 GOLD catalog with the CMASS photometric sample in Stripe 82 by position with a $2''$ tolerance. For the DES data, \verb|MAG_DETMODEL|  magnitudes are used. 
%The right panel shows the SDSS CMASS galaxies from Stripe 82 in the $gri$ color plane of SDSS. The same galaxies are plotted in DES colors in the right panel. Note that DES magnitude \verb|MAG_DETMODEL| is selected for substituting SDSS colors since its definition is similar to its counterpart in SDSS. 
The grey contours in the right panel show all sources from the DES Y1 GOLD catalog. The red solid line in both panels is the  $\dperp$ cut given by Equation \eqref{eq:dperp}. 
%We select DES detmodel-mag as a counterpart of SDSS model magnitude because the DES detmodel-mag was defined in a very similar way, and then plotted the same CMASS galaxies in the DES g-r-i color plane in the right panel. 
%The $d_{\perp}$ cut in DES consists of DES det-model magnitude as $d_{\rm perp} = $ by simply replacing SDSS magnitudes with DES magnitudes.
By noting the large fraction of DES objects below this line, one can clearly see how different the two systems are. In the DES data, the $\dperp$ cut crosses the most dense part of galaxy sample. Notably, this is a dense region for the full gold sample as well. If we were to blindly apply the $\dperp$ to the DES data, we would remove almost half of the true CMASS sample. Applying a simple transformation that moves the $\dperp$ cut to lower $r-i$ values recovers most of the CMASS galaxies, but at the cost of introducing many non-CMASS galaxies into the sample. Also noticeable in Figure \ref{fig:dperp} is the larger scatter in the SDSS distribution, especially in $g-r$.
%show there should be a complicated relation between SDSS and DES colors more than a simple one-to-one relationship. 
%Scatters in the DES side are clustered more than ones in the SDSS side {\bf not sure what that means}, also $\dperp$ cut located at the same position in the right panel is cutting out almost half of galaxies from the CMASS sample.  Simply moving $\dperp$ cut downward can catch more CMASS galaxies but ends up introducing much more non-CMASS galaxies at the same time because the density of gold galaxies is higher in bluer color shown as the grey contours. 

%\cmt{reason - 1. filter difference} 

There are several reasons for the discrepancy in the color space shown in Figure \ref{fig:dperp}. One is that despite both surveys using $griz$ filters, these filters are not identical. Figure \ref{fig:filter} illustrates this fact.  The response functions for the five SDSS (shaded) and DES (solid line) filters with the spectral energy density distribution of an elliptical galaxy are shown. The break in the model spectrum at $4,000 \si{\angstrom}$, a primary feature of galaxy spectra, 
%5due to ionized metals in old late type stars’ atmospheres
migrates through the $g$, $r$ and $i$ filters as the redshift increases \citep{Eisenstein2001LRG,Padmanabhan2007}. For elliptical galaxies near $z \sim 0.4$, the $4,000 \si{\angstrom}$ break is located at wavelengths where the $g-r$ transition happens. 
%\cmt{ \sout{As the $4000 \AA$ break moves toward higher redshift, this $g-r$ transition occurs first in SDSS. }}
Near the $4,000 \si{\angstrom}$ break, the SDSS $r$ filter (shaded) covers slightly lower wavelengths than the DES $r$ filter (solid lines) does.
That implies galaxies near $z\sim 0.4$ look redder in SDSS than they do in the DES photometric system. Since the redshift $z=0.4$ is also where the $\dperp$ cut is defined, this discrepancy of the filter transition exacerbates the color mismatch.

%The discrepancies already started at the instrument level.  
%SDSS and DES use 5 filters to detect galaxies in different color bands. SDSS uses $ugriz$ filter system and DES uses $grizY$ system but we will look at only $griz$ because we use only those 4 filters for our model. 
%
%Peaks at each band are [matched..? compared..?] with templates of spectroscopic galaxy spectra to analyze the properties of galaxies and derive photometric redshift \cmt{breif explanation about how to derive photoz from the shift of the peaks at galaxy spectra}. The most significant peak at the galaxy spectra is the
%
%http://astronomy.nmsu.edu/nicole/teaching/ASTR505/lectures/lecture26/slide01.html
%

%the optimal measure of the flux of a galaxy uses a matched galaxy model. With this in mind, the code fits two models to the two-dimensional image of each object in each band:

%\cmt{2. galaxy profile}
A second cause for the discrepancy in color space arises from differences in the SDSS and DES imaging pipelines. Magnitudes for extended sources are derived from the flux of a galaxy fitted with a best matched galaxy profile. Widely used galaxy profiles are exponential and de Vaucouleurs profiles \citep{DeVaucouleurs}, which perform better for disc and bulge galaxies, respectively. The SDSS imaging pipeline uses either one of these profiles to model magnitudes depending on the shape of a galaxy and uses a linear combination of two profiles for SDSS cmodel magnitudes. The DES imaging pipeline uses only the exponential profile consistently for all magnitudes. The fitting procedure is different as well. For instance, the SDSS pipeline fits galaxies only in the $r$ band to obtain model  magnitudes\footnote{
The term `model' magnitudes here indicates \\ `modelMag' magnitudes used in the BOSS selection criteria. }, while the closest analogue produced by the DES pipeline, \verb|MAG_DETMODEL|, 
the DES pipeline fits galaxies in a reference image that can be taken from one band or a combination of more than two bands \citep{DESDR1}.
%\cmt{KH: check DES statement}
%fits galaxies only in one band as SDSS but the band where galaxy will be fitted is chosen differently for every galaxy.\REF{DES COADD PEPER?}.  

%The DES imaging pipeline fits the shape of galaxies with only one galaxy profile - exponential profile compared to the SDSS imaging pipeline that uses one more galaxy profile - de Vaucouleurs profile for [...] type galaxies. The galaxy profile is directly related to the magnitudes. The SDSS model magnitude in the $d_{\rm perp}$ cut is determined from the best fit of either de Vaucouleurs profile or exponential profile, and the SDSS cmodel magnitude is defined from a bestfit of a linear combination of two models. On the other hand, the DES detmodel and model magnitudes are derived from only one galaxy profile that contributes to a discrepancy between two photometric systems. 
%2) Reference band that the bestfit model is determined : SDSS model magnitude and DES model magnitude are calculated from the bestfit model fitted at each band. SDSS cmodel magnitude is from the bestfit model fitted in the r band. DES magdetmodel is fitted in the deepest band, or the combination of more than two bands.  
%There is no exact counterpart between DES and SDSS colors satisfying two conditions. 

%\cmt{3. SDSS photometric error and dperp cut}
The last and most significant reason for the mismatch in color distributions is the fact that SDSS has significantly larger photometric errors compared to DES. The typical photometric error of the CMASS galaxies is $\sim0.2$ along the $g-r$ axis and $\sim0.07$ for the $r-i$ axis 
%\citep{SDSSEDR} 
 which is $\sim5$ times larger than the typical error of DES\footnote{based on the information available at catalogs in Section \ref{sec:data}.}. 
 %\citep{Y1GOLD}. 
%\notesj{Reference check. The errors are for CMASS. Can't find this in literature }
%The scope of SDSS photometric error is shown in Figure 1 in black.
The CMASS selection cuts in Equations \eqref{eq:imag_limit}-\eqref{eq:dperp} are simple cuts that do not take into account photometric errors. Ignoring photometric errors does not cause a notable problem for the cuts designed to limit faint magnitudes or to exclude outliers but must be considered thoroughly when it comes to the $\dperp$ cut. This is due to the location of the $\dperp$ cut in the densest region of the color space. 
Many galaxies with true colors outside of the $\dperp$ cut have scattered into the sample, while a similar amount of galaxies with true colors within the $\dperp$ cut could have scattered out of the SDSS selection.
%\sout{The size of the photometric errors implies that many galaxies with true colors outside of the $\dperp$ cut have scattered into the sample, while many galaxies with true colors within the $\dperp$ cut could have scattered out of the SDSS selection.}
%
%Due to the large SDSS photometric error, colors of galaxies near $\dperp$ cut is pretty much deviated from its true colors. Since $\dperp$ cut accepts any galaxies within its range regardless of true color, some accepted galaxies may not be a CMASS galaxy. The  - exacerbate the situation. As a result, $\dperp$ cut ends up excluding a big portion of true CMASS galaxies that were supposed to be included while failing to remove the similar amount of low redshift galaxies whose true color should be below $\dperp$ cut. 
%As explained above and in Figure \ref{fig:cross_cmass_dmass_st82}, $\dperp$ cut crosses dense region of CMASS sample in DES system.
%However, when the $d_{\rm perp}$ cut removes unnecessary low redshift galaxies, the cut only used observed colors without a careful consideration of errors of galaxies near the cuts that resulted in excluding a big portion of true CMASS galaxies that were supposed to be included while failing to remove the similar amount of low redshift galaxies whose true color should be below the $d_{\rm perp}$ cut. 
%
%infer that the d_ ?cut? used to obtain the BOSS CMASS sample, in terms of true properties? 
From this discussion, we infer that the $\dperp$ cut used to obtain the BOSS CMASS sample, in terms of true properties, is not a sharp cut shown in Figure \ref{fig:dperp}, but should instead be a form of likelihood function that accepts or rejects galaxies in a probabilistic way based on galaxy colors and photometric errors.
\notekh{should we reword this to express that to model the dperp cut in DES we cannot use simple cuts/selection criteria but that we have develop a method/likelihood that assigns probabilities to each (DES) galaxy ?}
%4) SDSS error : SDSS photometric error is very big, and dperp cut doesn't consider this error.
%dperp cut is determined based on the observed color of galaxies. It is simply crossing very high dense region without any consideration of photometric errors.
%Thus, there must be a lot of scatters around dperp cut. If SDSS pipeline observes the same galaxies, the same cut would never be able to be restored. 
%When colors and magnitudes of two systems are simply mapped one to one, galaxies tend to look bluer in the DES system that results in the lost of 20\% of CMASS sample by the $d_{\perp}$ cut. 

\notekh{isn't this a repeat of what was said in the previous paragraph? Maybe combine \cmtanswer{--Deleted duplicated sentence}}
Based on the three reasons we listed above, 
%\sout{especially motivated by the true $\dperp$ cut that accepts galaxies in a probabilistic way, }
we constructed a model that can handle the color mismatch and probabilistic selection near the $\dperp$ cut all together. 

%Our goal is making galaxy sample that has properties of CMASS sample. 
%Not only we need to consider complicated color relations, we also should reproduce scatters around dperp cut. 
%Therefore we built probabilistic model that can control those things together.

\section{Constructing the Model}
\label{sec:model}

%\notesj{model/membership probability confusing. Use only model probability. }
%In this section, we describe how we construct a model to select BOSS CMASS type galaxies from the DES catalog. 
While BOSS and DES operate in different hemispheres, the survey footprints overlap in an equatorial area of the sky known as Stripe 82. 
%Stripe 82 has been imaged by SDSS over forty times, making the co-added data two times deeper than single epoch SDSS observations \citep{Abazajian2009TheSurvey, Annis2011The82}. 
DES Y1 imaged \stripe of this region, thereby providing a region where data from the two surveys can be matched. 

By using the photometric information in the overlapping region, we build an algorithm for probabilistic target selection that uses density estimation in color and magnitude spaces. 
The general concept of the algorithm is described in Section \ref{sec:3-overview}. The algorithm is trained in half of the overlapping region and validated in the other half. We discuss the training and validation data sets in Section \ref{sec:3-data}. The tools and detailed fitting procedures for training are presented in Section \ref{sec:3-procedures}. The results of validation and application of the algorithm to the target galaxies can be found in Sections \ref{sec:3-validation} and \ref{sec:3-lowprobability}.

\iffalse
\sout{In this section, we describe how we construct a model to select BOSS CMASS type galaxies from the DES catalog. While BOSS and DES operate in different hemispheres, the survey footprints overlap for \stripeGold for DES year 1, in an equatorial area of the sky known as Stripe 82. After masking described in Section \ref{sec:data}, this field contains \trainCMASS CMASS galaxies over the area of \stripe. 
%that are observed by both DES and BOSS. 
We split the galaxies in the overlapping region into training and test samples which we use to develop our model. Once the algorithm is validated with the test sample, we use the full Stripe 82 region and construct the likelihood in the same way as it will be applied to the target galaxies in the DES SPT region. }
\fi

 %that lead us to build a probabilistic model based on Bayes' rule. 
%Next, we describe the construction of the training sample in detail and then describe the model and the extreme deconvolution technique used for the model to resolve spatial variance between Stripe 82(train) and other region.  

\subsection{Overview of the Algorithm}
\label{sec:3-overview}
%Whole paragraph need to be rephrased
%\sout{The basic idea of our selection algorithm is as follows. We wish to classify a set of unlabeled objects as either CMASS or not. We first create samples of CMASS and non-CMASS galaxies that will serve as training sets. For each object in our test set that we wish to classify, we compute its probability of being a CMASS galaxy and its probability of being a non-CMASS galaxy. }

%\notesj{short description of density estimation. Algorithm is based on density estimation }
The probability of being part of the CMASS sample for a source having a property $\boldsymbol{\theta}$ can be written as the combination of the likelihood and the prior according to Bayes' theorem: 
\bea
P(C|\boldsymbol{\theta}) = \frac{P(\boldsymbol{\theta}|C)~P(C)}{P(\boldsymbol{\theta})}~,
\label{eq:bayesian}
\eea
where
\bea
P(\boldsymbol{\theta}) = P(\boldsymbol{\theta}|C)P(C) + P(\boldsymbol{\theta}|N)P(N)~.
%\nonumber
\label{eq:bayesian2}
\eea 
%
%\bea
%P(C|\boldsymbol{a}) = \frac{P(\boldsymbol{a}|C)P(C)}{P(\boldsymbol{a}|C)P(C) + P(\boldsymbol{a}|N)P(N)}~.
%\label{eq:bayesian}
%\eea
The notation $C$ is the class of CMASS, $N$ is the class of non-CMASS galaxies. $P(C)$ is the prior probability that a selected source is part of the CMASS sample, which can be interpreted as the fraction of CMASS in the total galaxy sample. $P(\boldsymbol{\theta}|C)$ is the 
 likelihood of the source under the probability density function (pdf) of the property $\boldsymbol{\theta}$ of CMASS. 
%The pdf can be expressed as a distribution (histogram) of $\boldsymbol{a}$ of CMASS.
The pdf of the property $\boldsymbol{\theta}$ of CMASS can be constructed from a histogram of CMASS as a function of $\boldsymbol{\theta}$. However, since we use noisy quantities such as observed colors and magnitudes, the resulting pdf might be biased by photometric errors that vary by observing conditions. For example, if the training region has a uniquely different observing condition from other regions, the pdf model drawn from the training galaxies will not accurately represent CMASS. 
Therefore, the pdf should take into account measurement errors.

To ensure a uniform selection across the survey, we use the Extreme Deconvolution (XD) technique first proposed in \cite{Bovy2011a}. 
The XD algorithm models the observed distribution of data as a mixture of Gaussians, convolved with a multivariate Gaussian model for the measurement errors on each point. It iterates through expectation and maximization steps to solve for the maximum likelihood estimates of the parameters specifying the underlying mixture model.

%To incorporate measurement errors into the likelihood, we use the Extreme Deconvolution (XD) technique first proposed in \cite{Bovy2011a}. The XD algorithm fits the observed distributions of the input data with a mixture of Gaussians, and obtains the maximum likelihood estimates describing the underlying Gausians through expectation-maximization steps, deconvolving measurement uncertainties given to the input data. 

%\noteemh{This could be a little more clear. Try instead something like: To ensure a uniform selection across the survey, we use the Extreme Devonvolution (XD) technique first proposed in \citealt{Bovy2011a}. The XD algorithm models the observed distribution of data as a mixture of Gaussians, convolved with a multivariate gaussian model for the measurement errors on each point. It uses Expectations Maximization to solve for the maximum likelihood estimates of the parameters specifying the underlying mixture model."}

%\reph{The underlying distribution $P(\boldsymbol{a}_{\rm true}|C)$ obtained from XD is an unbiased, universial distribution free from regional measurement errors in the training region. }

%Assuming the true distribution of CMASS property $\boldsymbol{a}$ is same everywhere, 
The underlying distribution $P(\theta_{\rm true}|C)$ derived from XD is an unbiased pdf free from regional measurement errors in the training set. 
By convolving the underlying distribution back with the measurement uncertainties of the validation sample, 
we can infer the observed distribution that the same kind of galaxies would have in the validation region. For a given observation $\boldsymbol{\theta} = \{ \theta_{\rm obs}, \epsilon \}$, the observed quantity $\theta_{\rm obs}$ with a corresponding measurement uncertainty $\epsilon$, the likelihood of CMASS is written as
%To obtain the observed distribution of the target galaxies, the underlying distribution is convolved back with the measurement uncertainties of the target galaxies as follow:
%We obtain the underlying distribution of CMASS $P(\boldsymbol{a}_{\rm true}|C)$ in the training set from XD and convolve with the measurement uncertainties of the target galaxies given as 
\bea
P(\{\theta_{\rm obs}, \epsilon \} |C) = \int \ud \theta_{\rm true}~p(\{{\theta}_{\rm obs}, {\epsilon} \}|{\theta}_{\rm true})~ P({\theta}_{\rm true}|C)~.
\label{eq:xd}
\eea
The first factor $p(\{{\theta}_{\rm obs}, {\epsilon} \}|{\theta}_{\rm true})$ on the right side stands for the distribution function of measurement uncertainty of ${\theta}$ in the presence of known measurement uncertainty $\epsilon$.  
We assume that the measurement uncertainty distribution of bright galaxies such as CMASS is nearly a Gaussian with a RMS width $\epsilon$. 
%We describe XD in more detail later in the paper.
The same procedure is repeated for non-CMASS galaxies.  
%$P(\{\theta_{\rm obs}, \epsilon \}|N)$.
%, to construct a posterior probability $P(C|\{\theta_{\rm obs}, \epsilon \})$ in Equation \eqref{eq:bayesian}. 
%
%$\boldsymbol{\theta} = \{{\theta}_{\rm obs}, {\epsilon}\}$

Considering all factors, the resulting posterior probability that will be assigned to a target source having a property $\theta_{\rm obs}$ with a measurement uncertainty $\epsilon$ is given as 
\bea
P(C|\{{\theta}_{\rm obs}, {\epsilon}\} ) = \frac{  \int \ud \theta_{\rm true}~p(\{{\theta}_{\rm obs}, {\epsilon} \}|{\theta}_{\rm true}) P({\theta}_{\rm true}|C) P(C)}{   \int \ud \theta_{\rm true}~p(\{{\theta}_{\rm obs}, {\epsilon} \}|{\theta}_{\rm true}) P({\theta}_{\rm true})}~,
\label{eq:bayesian3}
\eea
where
\bea
P({\theta}_{\rm true}) = P({\theta}_{\rm true}|C)P(C) + P({\theta}_{\rm true}|N)P(N)~.
%\nonumber
\label{eq:bayesian4}
\eea

\subsection{The Training and Validation Sets}
\label{sec:3-data}

%\reph{ While BOSS and DES operate in different hemispheres, the survey footprints overlap in an equatorial area of the sky known as Stripe 82. 
%After masking described in Section \ref{sec:data}, this field contains \trainCMASS CMASS galaxies over the area of \stripe. Stripe 82 has been imaged by SDSS over ten times, making the co-added data two times deeper than single epoch SDSS observations \citep{Abazajian2009TheSurvey, Annis2011The82}. DES Y1 imaged \stripeGold of this region, thereby providing a region where data from the two surveys can be matched. }

We use the overlapping area between BOSS and DES to train and validate the algorithm. To label DES galaxies as CMASS or non-CMASS galaxies, we cross-match the refined DES Y1 GOLD catalog (described in Section \ref{sec:data}) to the BOSS CMASS photometric sample\footnote{We do not use the BOSS spectroscopic sample for training. The spectroscopic sample of BOSS CMASS has about $5.8\%$ and $1.8\%$ of missing targets lost by `fiber collision' and `redshift failure', respectively \citep{Reid2016}. Since our probabilistic model is color-based, we utilize the BOSS photometric sample for training in order to include the photometry information from those missing galaxies.} using a $2''$ tolerance.  
The total number of galaxies labelled as CMASS is \trainCMASS 
%, labelled as non-CMASS is \sjcmt{\trainCMASS} 
over the area of \stripe. 

The labelled DES galaxies are split into the training and validation sets. 
%\sout{ We split the overlapping area into the training and validation sets. }
In the overlapping region, the number density of galaxies varies along latitude. Since our probabilistic model assumes that the galaxies are homogeneously distributed in the full sky, we divided the overlapping area into HEALPix\footnote{\url{http://healpix.sourceforge.net}} \citep{HEALPix} pixels of resolution $N_{\rm side} = 64$ in \verb|NEST| ordering and took only even values of HEALPix pixels as the training sets to populate the training regions uniformly. 
%Our probabilistic model assumes galaxies are homogeneously distributed. 
%as $ra > 339~\deg$ for the training and $ra < 339~\deg$ for the validation set. 
The total training set contains \Ntrain CMASS galaxies and \Nno non-CMASS galaxies in \Atrain$\degsqr$. The two samples are used separately to train the algorithm to construct the likelihoods for CMASS and non-CMASS galaxies.
Note that this division is used only to test the algorithm and we will later switch to the full Stripe 82 region as the training set for the DES SPT region.

\subsection{Obtaining True Distributions with the Extreme-Deconvolution Algorithm}
\label{sec:3-procedures}

%\sout{ The true distributions allow us to handle model bias that can happen for the following reasons: 1. Observed distributions on Stripe82 can be biased compared to the SPT region. \sjcmt{2. Observed distributions can vary with changing observing conditions within the DES SPT region. } To overcome these issues, we derived the observed distributions for every target galaxy from the underlying distributions obtained from the Extreme-Deconvolution algorithm (XD) and the corresponding uncertainties of the measurements. }

The Extreme Deconvolution (XD) algorithm developed by \cite{Bovy2011a} is a generalized Gaussian-mixture-model approach to density estimation and is designed to reconstruct the error-deconvolved true distribution function common to all samples, even when noise is significant or there are missing data. Starting from the random initial guess of Gaussian mixtures, the algorithm iteratively calculates the likelihood by varying means and widths of Gaussian components until it finds the best fit of Gaussian mixtures.
%{\cred{ or an improvement in the likelihood is within the tolerance given by a user}}.

We use the Python version of the XD algorithm in the \verb|AstroML|\footnote{\url{http://www.astroml.org}} package \citep{AstroML_XD}.
The following four DES properties are fitted by the XD algorithm:
 %the following four DES properties and their measurement uncertainties are given to the XD algorithm: 
 ($G_{\rm DET}-R_{\rm DET}$), ($R_{\rm DET}-I_{\rm DET}$), $R_{\rm MOD}$, and $I_{\rm MOD}$. 
%$\theta_{\rm obs} = \{ (G_{\rm DET}-R_{\rm DET}), (R_{\rm DET}-I_{\rm DET}), R_{\rm MOD},  I_{\rm MOD} \}$. 
%
%To obtain the true distributions, 
%$P(\boldsymbol{a}_{\rm true}|C)$ and $P(\boldsymbol{a}_{\rm true}|N)$, 
%we choose a total of four DES properties of the training sets: two DES colors $G_{\rm DET}-R_{\rm DET}$ and $R_{\rm DET}-I_{\rm DET}$, and two DES magnitudes $R_{\rm MOD}$ and $I_{\rm MOD}$. 
The two DES colors are selected as they mirror the SDSS information used for the $\dperp$ cut.  The apparent magnitude $I_{\rm MOD}$ is %{\cred{ \sout{selected reflecting the cut}}}
included to extract information induced by the cut given in Equation \eqref{eq:imag_limit}.
There is no $r$ band magnitude cut in the CMASS selection criteria, but we include $R_{\rm MOD}$ in order to provide extra information  %and hope 
to capture the differences between the SDSS and DES filter bands. 
%The distributions of the four aforementioned properties of the training sample are shown in blue in Figure \ref{fig:hist_cmass_dmass}. 
Star-galaxy separation was performed on DES photometry with the flag \verb|MODEST == 1| \citep{Y1GOLD}, therefore we do not apply any further cuts to replace cuts \eqref{eq:sg} and \eqref{eq:sg2}.

%We ran the XD algorithm over the CMASS training set and the non-CMASS training set separately to obtain the underlying distributions of each sets. 
%a model likelihood for CMASS and the model likelihood for non CMASS in the denominator in Equation \ref{eq:model}. Y1 GOLD catalog provides rms errors for different magnitudes as \verb|MAGERR_MODEL| and \verb|MAGERR_DETMODEL| for every source. We derived errors of $G-R$ and $R-I$ colors from the magnitude errors and fed those errors to the XD algorithm with errors of $R_{MOD}$ and $I_{MOD}$ and the chosen input colors in the previous section. 
%
%We use the Python version of the XD algorithm in the \verb|AstroML|\footnote{http://www.astroml.org} package \citep{AstroML_XD} to fit the multi-dimensional distribution of the given properties. 
%
The \verb|AstroML| XD algorithm leaves the initial number of Gaussian mixture components 
%and the tolerance of likelihood 
as a user's choice. 
One of the well-known methods for choosing the correct number of components is to use the Bayesian Information Criterion \citep[BIC;][]{BIC}. 
%We adopt the Bayesian Information Criterion (BIC;\citepp{BIC}) to choose the correct number of Gaussian components to avoid either underfitting or overfitting. 
We use the Gaussian mixture module in the \verb|scikit-learn|\footnote{\url{https://scikit-learn.org/}} package \citep{SCIKIT-LEARN} to compute the BIC scores for a different number of components. The optimal number of
components found by this exercise is 
%components from this module is 
8 components for the CMASS training set and 26 components for the non-CMASS training set. 

%Now, 
The XD algorithm fits the multi-dimensional histogram of the four aforementioned DES properties with the optimal number of Gaussian mixtures and  returns the values of amplitudes, means and widths of the best fit Gaussian mixture model. The resulting best fit model is used as a true distribution $P({\theta}_{\rm true}|C)$  in Equations \eqref{eq:bayesian3}-\eqref{eq:bayesian4}. 
%
%We fitted the observed distributions of the training sets with Gaussian mixture models from the \verb|scikit-learn|\footnote{https://scikit-learn.org/} Gaussian mixture module, increasing the number of mixtures and computing the BIC for every number. We selected the number of mixtures with the smallest BIC. This method gives 8 components for the CMASS training set and 26 components for the non-CMASS training set. 
%
\iffalse
\sout{
For determining the adequate tolerance of likelihood, we test two values: a moderate value of $10^{-5}$ and an extreme choice $10^{-10}$. Since there is no significant difference found between the two results, we stick to a moderate choice of $10^{-5}$.}
\fi

Throughout this work, we assume that there is no correlation between different bands.
% and that the uncertainties in observed magnitudes and colors are nearly Gaussian.

%In the next section, we will discuss CMASS membership probability that the probabilistic model assigns to each galaxy by using the underlying distributions determined as described in this section. 
%that the probabilistic model that the fitted underlying distributions are plugged in produces for each galaxy and how the CMASS membership probability is used to construct the DMASS sample. 

%\cmt{specific method and re-convolution process} 
%The number of gaussian components for the initial guess is decided by BIC \cmt{add details...} 
%We use DES photometric errors of each color and magnitude that to deconvolve the distribution with. We use \verb|magerr_model[_gri]| and \verb|magerr_detmodel[_gri]|. ( \cmt{add later : descriptions about DES photometric error}). The true distribution should be convolved again with the photometric errors in the target region before multiplied by other factors in the full model. The same procedure should be done for non-CMASS sample to obtain $p(a_j| O \notin {\rm CMASS})$.

\subsection{Application to the Target Galaxies}
\label{sec:3-validation}

\begin{figure}
 \includegraphics[width=0.48\textwidth]{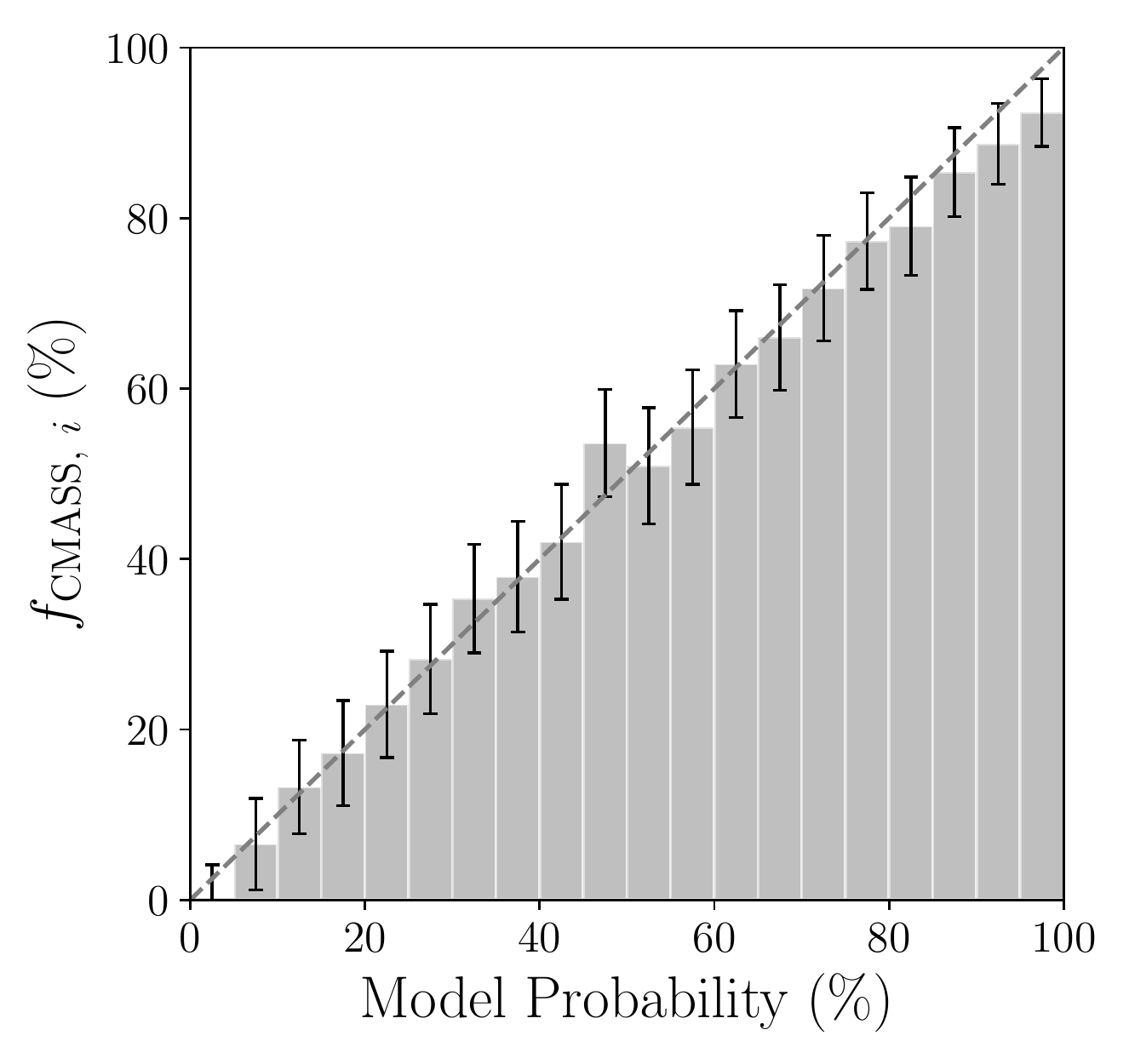}
\caption{
Accuracy of model membership probabilities assigned to the validation set. Galaxies in the validation set are binned based on their probability assigned by the probabilistic model. The x-axis shows 20 bins of the assigned probability, and the y-axis shows the fraction of true CMASS galaxies in each bin. If the model successfully recovers the observed distribution of CMASS in the validation region, the fraction of true CMASS galaxies in each bin should be identical to the assigned model probability. The dashed diagonal line in the figure stands for this ideal case, and the grey bars are given by Poisson errors. 
%\noteemh{I love this figure so much.}
}
\label{fig:pc_calib}
\end{figure}

\begin{figure*}
\centering
\includegraphics[width=\textwidth]{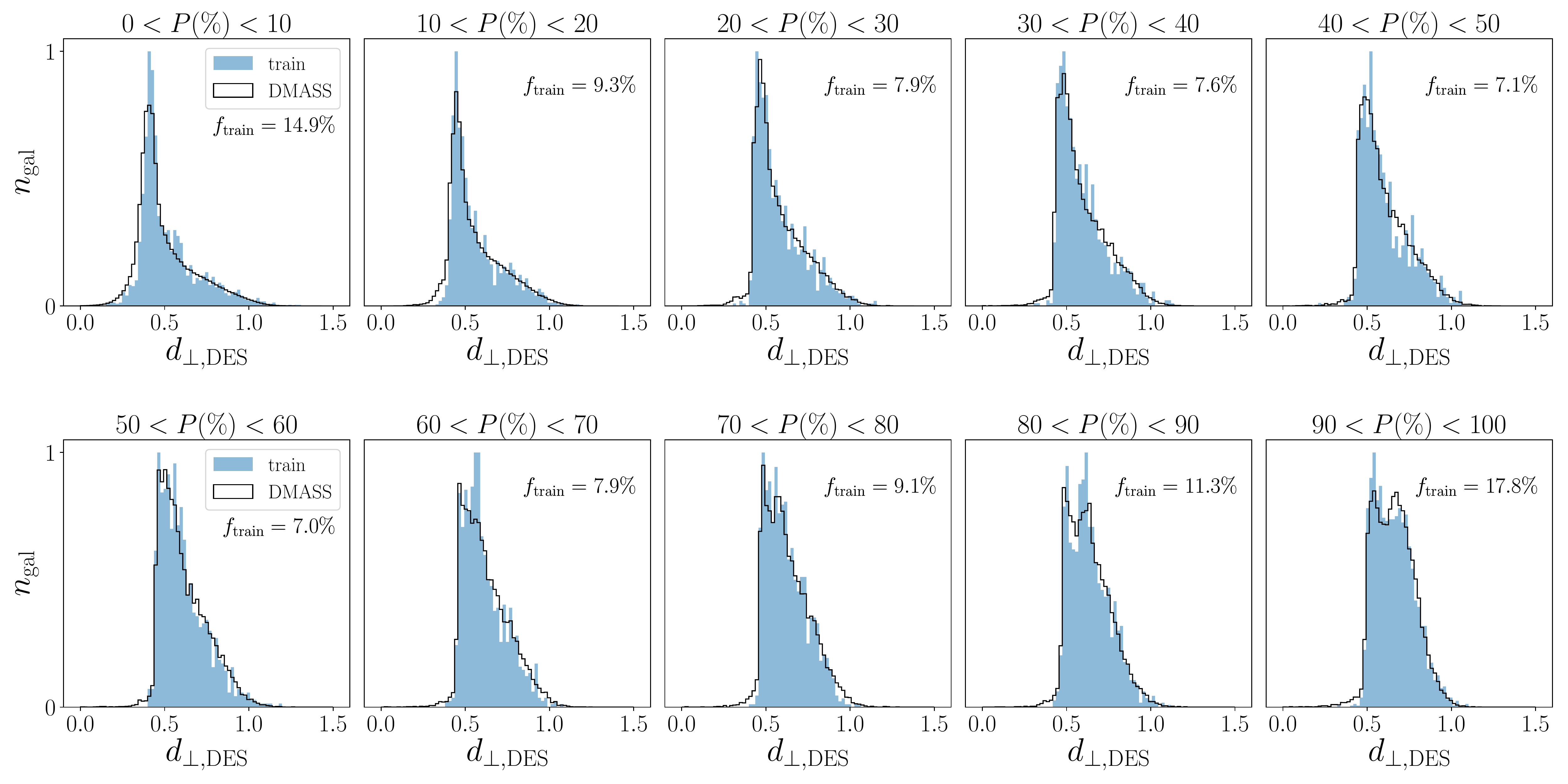}
\cprotect\caption{Histograms of the $\dperp$ color distributions of CMASS (blue) in the training set and DMASS in SPT (solid black line) in $10$ membership probability bins. $\dperp$ on the x-axis consists of only DES quantities ($d_{\perp, \rm DES} = (R_{\rm DET} - I_{\rm DET}) -  (G_{\rm DET} - R_{\rm DET})/8.0$ where the subscript `DET' denotes DES \verb|MAG_DETMODEL| magnitude). $f_{\rm train}$ in the top-right corner of each panel denotes the fraction of training galaxies binned in each probability bin, defined as  $f_{{\rm train},i} = N_{ {\rm train}, i}/N_{ \rm train, total}$ for the $i$th probability bin. 
%\notejep{need to add numbers to panel? \cmtanswer{-- Done} }
}
\label{fig:d_perp_hist}
\end{figure*}

\begin{figure*}
\includegraphics[width=\textwidth]{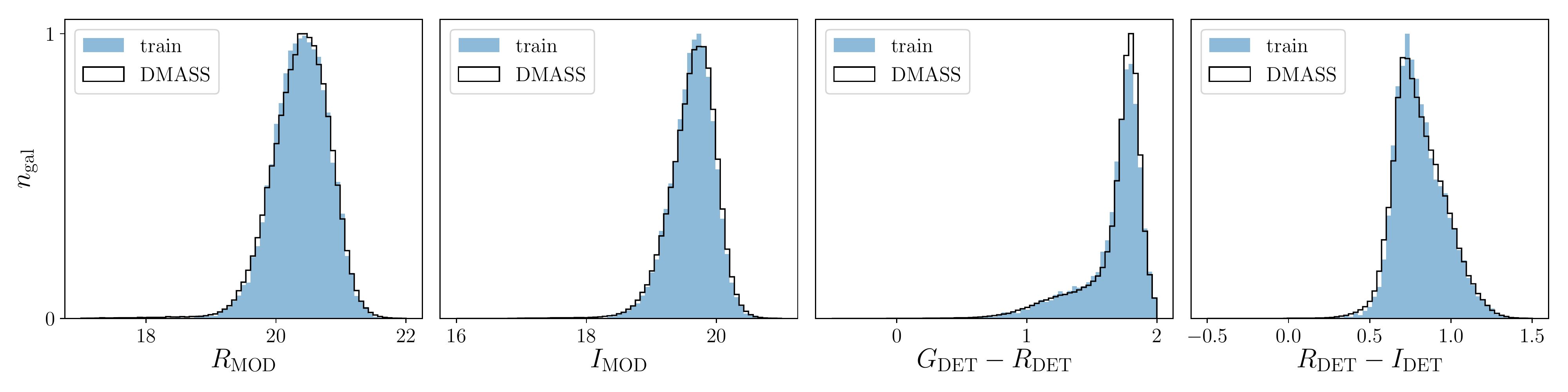}
\caption{Histograms of color and magnitude distributions of CMASS in the training set (blue) and DMASS in SPT (black solid line). The colors and magnitudes on the x-axis are DES quantities. }
%Histograms in stripe82 are deconvolved with photometric errors by extreme deconvolution algorithm to get underlying distributions of CMASS. Then, the obtained underlying distributions are convolved with photometric errors in SPT region and used as a likelihood in our model 
\label{fig:hist_cmass_dmass}
\end{figure*}

%\notesj{Distinguish pdf and likelihood}
In this section, we apply our probabilistic model 
%in Equation \eqref{eq:bayesian3} 
to the validation galaxies in order to validate the algorithm.
The underlying distributions $P({\theta}_{\rm true}|C)$ and $P({\theta}_{\rm true}|N)$ are obtained from the XD algorithm as described in the previous section. 
The Bayesian priors are given as $P(C)=0.018$ and $P(N) = 1 - P(C)$. 
This is based on the fraction of CMASS galaxies in the training set\footnote{The fraction of CMASS may vary depending on the observing condition of the selected area, but we take the value in the training sample as a global prior for simplicity, assuming CMASS galaxies are homogeneously distributed in the Universe.
We will show that this approximation can be justified through validation tests later in this paper.}.
%the fraction of CMASS may vary depending on the observing condition of the selected area, we simply use the priors from the training sample as global priors
%Note that the fraction of CMASS may vary depending on the observing condition of the selected area.
%
%Although we simply use the priors from the training sample as global priors, note that the fraction of CMASS may vary depending on the observing condition of the selected area. We will show that this approximation can be justified through validation tests later in the paper. }

%\sout{Throughout this work, we assume that CMASS galaxies are homogeneously distributed in the Universe. }
%\notesj{ from Niall : However, that does not mean P(C) is constant - the fraction of CMASS galaxies in your selection for a given part of the survey will depend on the observing conditions  in that part of the survey So I think you should just note that this assumption that P(C) is constant is an approximation, which, like the other assumptions in the method will be tested by the validation tests later in the paper }
%\sout{Given a set of 4 observed DES quantities and their photometric errors of a source, the probabilistic model assigns every input galaxy a probability of being a CMASS galaxy between $0 \%$ to $100\%$. }

The probability of being part of CMASS for a given property $\theta$ is analogous to the probability of finding a CMASS galaxy in a group of galaxies having the same property $\theta$.  
%This implies that if the true distributions and Bayesian prior from the training set perfectly represent the full sky, the assigned probability should yield the same fraction of true CMASS in each probability bin.
This implies that in a group of galaxies assigned the same model probability, the assigned probability should be identical to a fraction of galaxies labelled as CMASS.
To confirm this argument, we bin validation galaxies in 20 bins based on their assigned probability. 
In Figure \ref{fig:pc_calib}, the x-axis shows the 20 bins of the assigned probability, and the y-axis shows the fraction of true CMASS galaxies in each bin. The grey bars are the fractions of CMASS-labelled galaxies in the validation set with Poisson error bars. The diagonal dashed line represents the ideal case that the probabilistic model would yield if the model successfully recovers the observed distribution of CMASS in the target region. The computed fractions of CMASS show good agreement with the ideal case within error bars.

Once all galaxies in the target region are assigned a model probability, the widely-accepted next step for classification is dividing target sources into two categories with a threshold probability $> 50 \%$. A similar probabilistic approach was done in \cite{Bovy2011b} to distinguish quasars from stars. However, we take a different approach since we are not interested in classifying individual galaxies accurately, but instead we focus on matching the statistical properties of groups of galaxies.
In order to produce a statistical match, the membership probability we determine must faithfully reflect the probability that an object would be selected into the BOSS CMASS sample based on SDSS imaging.

Figure \ref{fig:d_perp_hist} presents the histograms of the $\dperp$ color of true CMASS in the training set (blue shaded histogram) binned in the different ranges of the membership probability bins. 
% rewrite ----------------------- 
Training galaxies in low probability bins tend to have low $\dperp$ values because of their proximity to the $\dperp$ cut. 
$f_{\rm train}$ in the top-right corner of each panel is the fraction of training galaxies binned in each probability bin, defined as $f_{{\rm train},i} = N_{ {\rm train}, i}/N_{ \rm train, total}$ for the $i$th probability bin. 
Over the 10 probability bins, training galaxies are distributed uniformly, with a relatively high fraction in the lowest and the highest probability bins.  
%Even the lowest probability bins contain a significant portion of training galaxies, 
%The fraction of training galaxies in each bin shows about $10\%$ consistently, 
This indicates 
that galaxies having low membership probabilities contribute to the CMASS sample as significantly as galaxies having high membership probability.

\noteemh{At some point here, we should say that the resampling or weighting is valid because the selection in SDSS is effectively random. If the differences in selection were based on an additional hidden variable not accessible in the photometry, then the reweighting or resampling wouldn't produce a similar sample}

From this, we can infer that 
%This implies that 
in order to generate the same noise level that the original CMASS sample intrinsically has,  galaxies should be populated based on their membership probability in the same way that ones in the CMASS sample are. In this sense, the model probability suggests a natural way of how we should make use of the assigned probabilities.  
We can either sample or weight a galaxy by its assigned probability in order to produce a sample that is a statistical match to the BOSS CMASS sample.  Throughout the rest of the work, we use the membership probability as weights. 

%Note that this division is used only to test the algorithm and we will later switch to the full Stripe 82 region as the training set for the DES SPT region. 
From now on, we apply the validated algorithm to the DES galaxies outside the training area. If not specified otherwise, the DMASS sample only refers to the  DES Y1 GOLD galaxies in the SPT region weighted by the assigned membership probability. 
The black solid lines in Figures \ref{fig:d_perp_hist} and \ref{fig:hist_cmass_dmass} show the various property distributions of the DMASS sample. 
%\sout{The black solid lines in Figures \ref{fig:d_perp_hist} and \ref{fig:hist_cmass_dmass} show the DES Y1 GOLD galaxies in the SPT region weighted by the assigned membership probabilities. }
Figure \ref{fig:d_perp_hist} shows that the weighting scheme successfully reproduces the noisy quantity $\dperp$ by populating each probability bin with the DES galaxies (black solid) as the CMASS galaxies (blue) are distributed. 
Figure \ref{fig:hist_cmass_dmass} shows the distributions of colors and magnitudes that are used to train the algorithm. 
%In Figure \ref{fig:hist_cmass_dmass}, 
The resulting DMASS distributions (black solid) of the colors and magnitudes are in good agreement with the distributions of the training sample (blue). 

%\cmt{kh: check previous sentence}
%The membership probability is stored in the \verb|CMASS_PROB| column.
%Users who want to make use of the DMASS sample should weight the galaxies with the membership probability as default. 

%This demonstrates that We can either sample a galaxy based on or weight a galaxy by its assigned probability in order to produce a sample that is a statistical match to the BOSS CMASS sample. We stored the membership probability for all DES GOLD sources in the \verb|CMASS_PROB| column. Users who want to make use of the DMASS sample should weight the galaxies with the membership probability as default. 

\subsection{ Excluding Low Probability Galaxies }
\label{sec:3-lowprobability}

\begin{figure}
\centering
\includegraphics[width=0.49\textwidth]{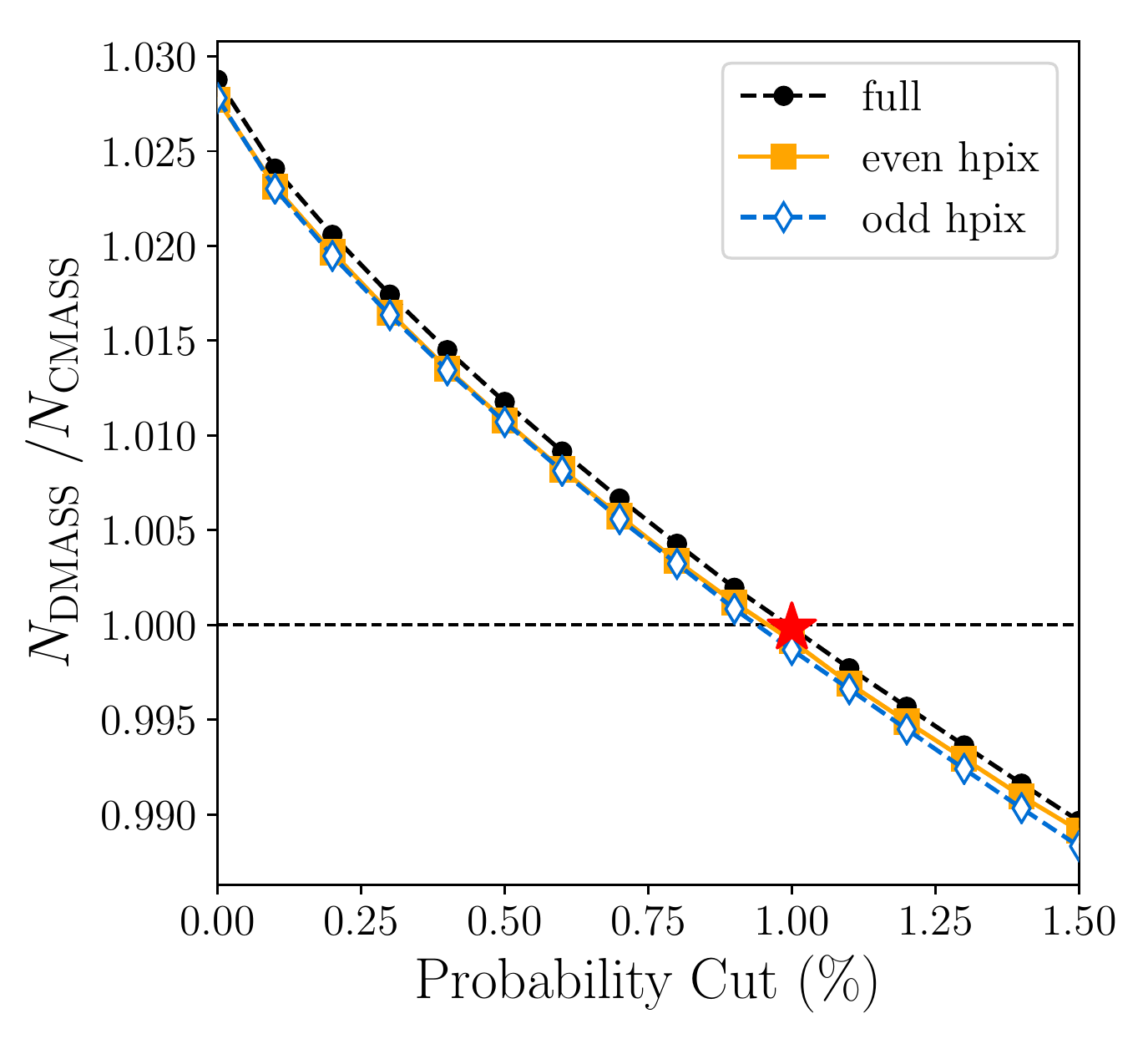}
\caption{ 
Number density of DMASS with respect to the probability cut computed from three diffrent regions - the training region (even HEALPix pixels; square), the validation region (odd HEALPix pixels; circle), and the full region (even + odd HEALPix pixels; black diamond). 
Galaxies below a given probability cut are excluded. 
Number densities are divided by the number density of CMASS in the corresponding regions. 
The extremely similar shape of curves from different samples implies that the model tends to boost the number density of the sample in a predictable way and this tendency can be remedied by cutting out low probability galaxies below a certain threshold. The red star at $P=1\%$ on the black diamond curve is our choice of the probability cut. 
%Derived number density of DMASS in the training sample (even HEALPix pixels; square), the test sample (odd HEALPix pixels; circle), and the full overlapping region (black diamond) divided by the number density of CMASS in the corresponding regions. Galaxies below a given probability cut are excluded. The extremely similar shape of curves from different samples implies that the model tends to boost the number density of the sample in a predictable way and this tendency can be remedied by cutting out low probability galaxies below a certain threshold. The red star at $P=1\%$ on the black diamond curve is our choice of the probability cut. 
}
\label{fig:sys_prob_bias}
\end{figure}

The DES Y1 GOLD catalog contains a lot of galaxies that are much fainter than CMASS.  
%\sout{ In the overlapping region, only $\sim1.8$ per cent of DES galaxies are matched with BOSS CMASS. }
This implies the majority of the DES galaxies have extremely low CMASS probabilities. These galaxies are  likely to only add noise to the sample and potentially bias our measurements and therefore need to be removed. 
%\notejep{we *expect* this to be true. should say are expected to have... \cmtanswer{--this is true because in Stripe82 we exactly know which DES galaxy is CMASS. Reworded the first sentence to make it sound clearer. }}
%
We carefully test how the low probability portion of the training sample (even HEALPix pixels) affects the number density. We remove all galaxies lower than a given probability threshold and compare the number density of each sample with those of CMASS in the training sample. 
Including all sources results in the number density of the resulting sample being about $3\%$ higher than CMASS in the same region, but near a threshold cut $P > 1 \%$, the sample yields a similar number density as CMASS. To validate the threshold cut, we construct a model in the same way but by using only the validation sample (odd HEALPix pixels). Figure \ref{fig:sys_prob_bias} shows that the model from the validation sample produces low probability galaxies that affects the number density of the sample in a very similar way as the training sample.  
%The extremely similar shape of curves from different samples 
The similarity of the curves from different samples 
implies that the model tends to boost the number density of the sample in a predictable way, and this tendency can be remedied by cutting out low probability galaxies below a certain threshold. 
The same procedure is performed for the full Stripe 82 region and yields the same number density as CMASS for a threshold cut $P > 1 \%$ (black points in Figure \ref{fig:sys_prob_bias}). 
%performed the same procedure for the test sample and obtained the same result. 
%\cmt{\sout{ We carefully tested how the low probability portion of the SPT sample affects the recovered galaxy bias. We removed all galaxies lower than a given probability threshold and measured the clustering for each sample. In Figure \ref{fig:sys_prob_bias}, including all sources results in galaxy bias being significantly biased 
%Galaxy bias approaches to the theoretical value by cutting out more low probability objects 
%but after applying a threshold $P>0.4\%$ the derived galaxy bias  becomes stabilized. Throughout this work, We use a threshold $P>1\%$ where the galaxy bias is safely robust. }}
Throughout this work, we use a threshold cut $P>1\%$. This cut excludes $\sim90\%$ of sources in the DMASS catalog, but when considering membership probabilities as weights, the effective portion of galaxies eliminated is $2.96\%$.
After applying the probability cut, we determine the effective sample size of the complete DMASS catalog by summing the weights. We find the sample size of the DMASS sample is 117,293. 

\begin{figure}
%\centering
\includegraphics[width=0.45\textwidth]{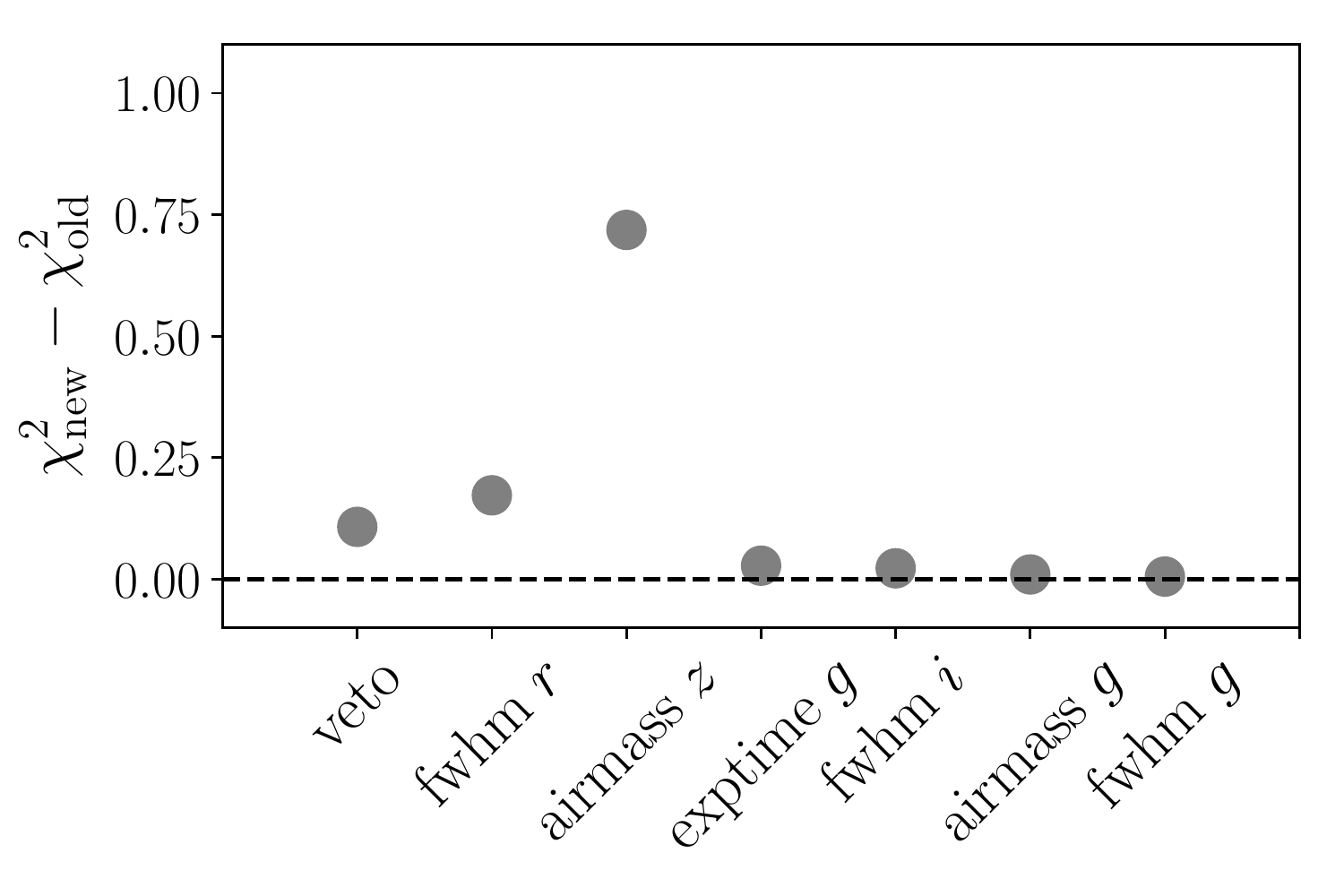}
\caption{ 
%\notesj{Not clear what is chi-square }
The impact of systematic weights. 
Starting from the left, the names of the survey properties are listed on the x-axis in the order that they are corrected. The weight for the particular property is applied on top of the other weights applied earlier. The y-axis shows the $\chi^2$ measured between the correlation function with the new and old weights.
%of the two point function with a new weight against one before the new weight is applied. 
`veto' denotes a veto mask applied to remove regions where fwhm in $r$ band $> 4.5$.
%\notejep{Should say in caption that veto is different to the other points. it is a mask, not a observing map, right? \cmtanswer{ -- Done} }
}
\label{fig:sys_deltachi}
\end{figure}

\begin{figure*}
%exptime_g_fwhm_i_airmass_g_fwhm_g
%\includegraphics[width=0.24\textwidth]{./figures/comparison_systematic_GE_g.pdf}
\includegraphics[width=0.275 \textwidth]{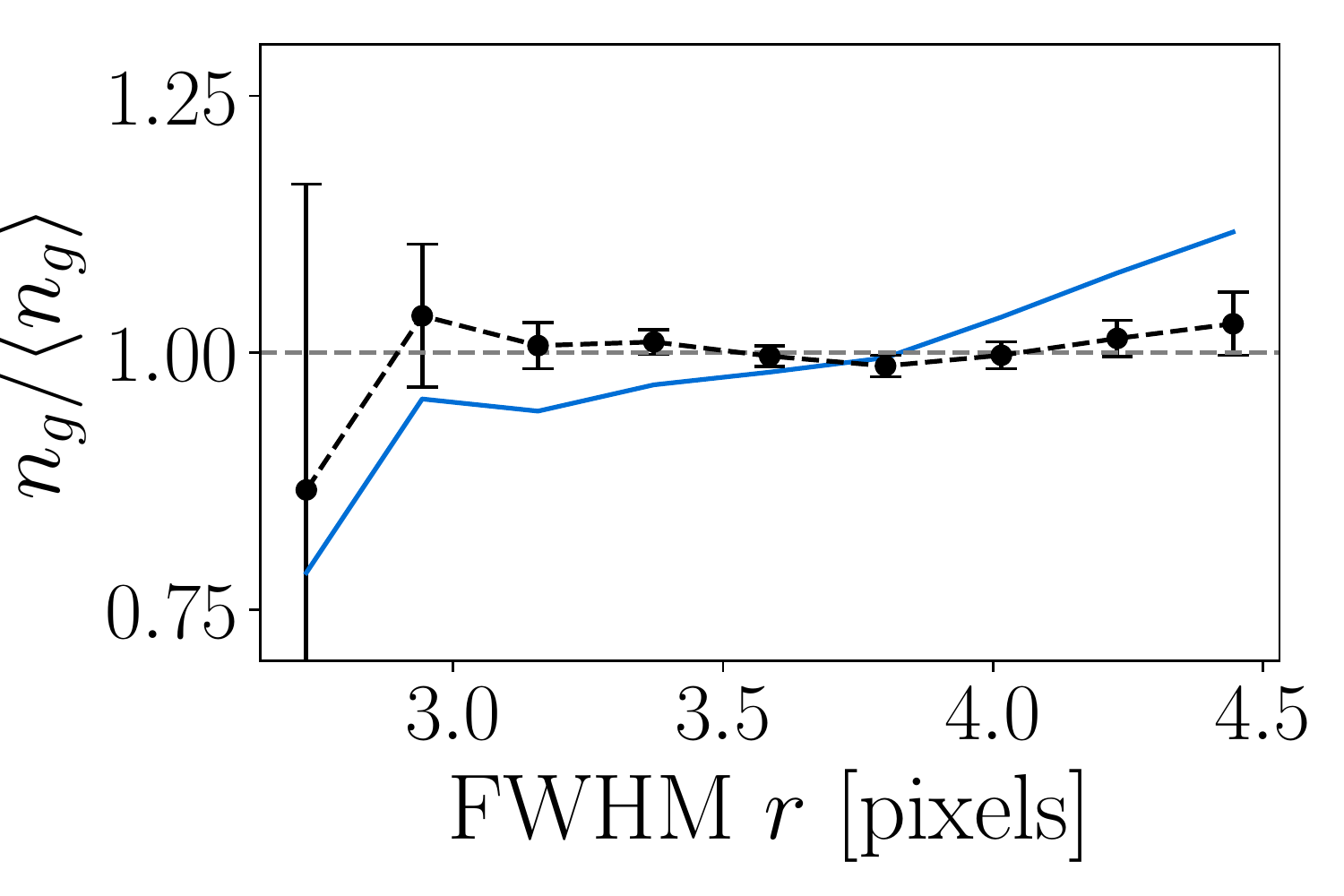}
\includegraphics[width=0.23\textwidth]{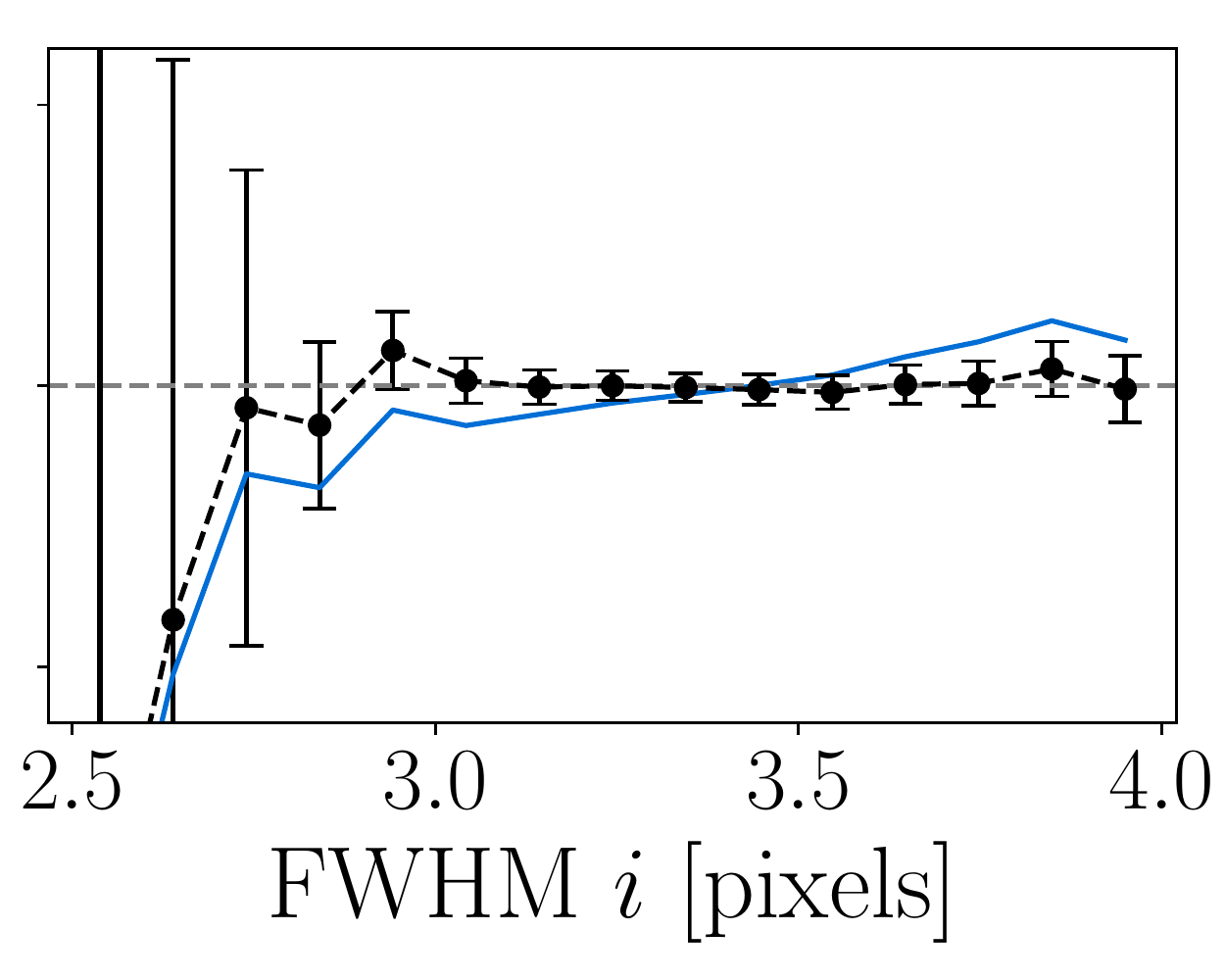}
\includegraphics[width=0.23\textwidth]{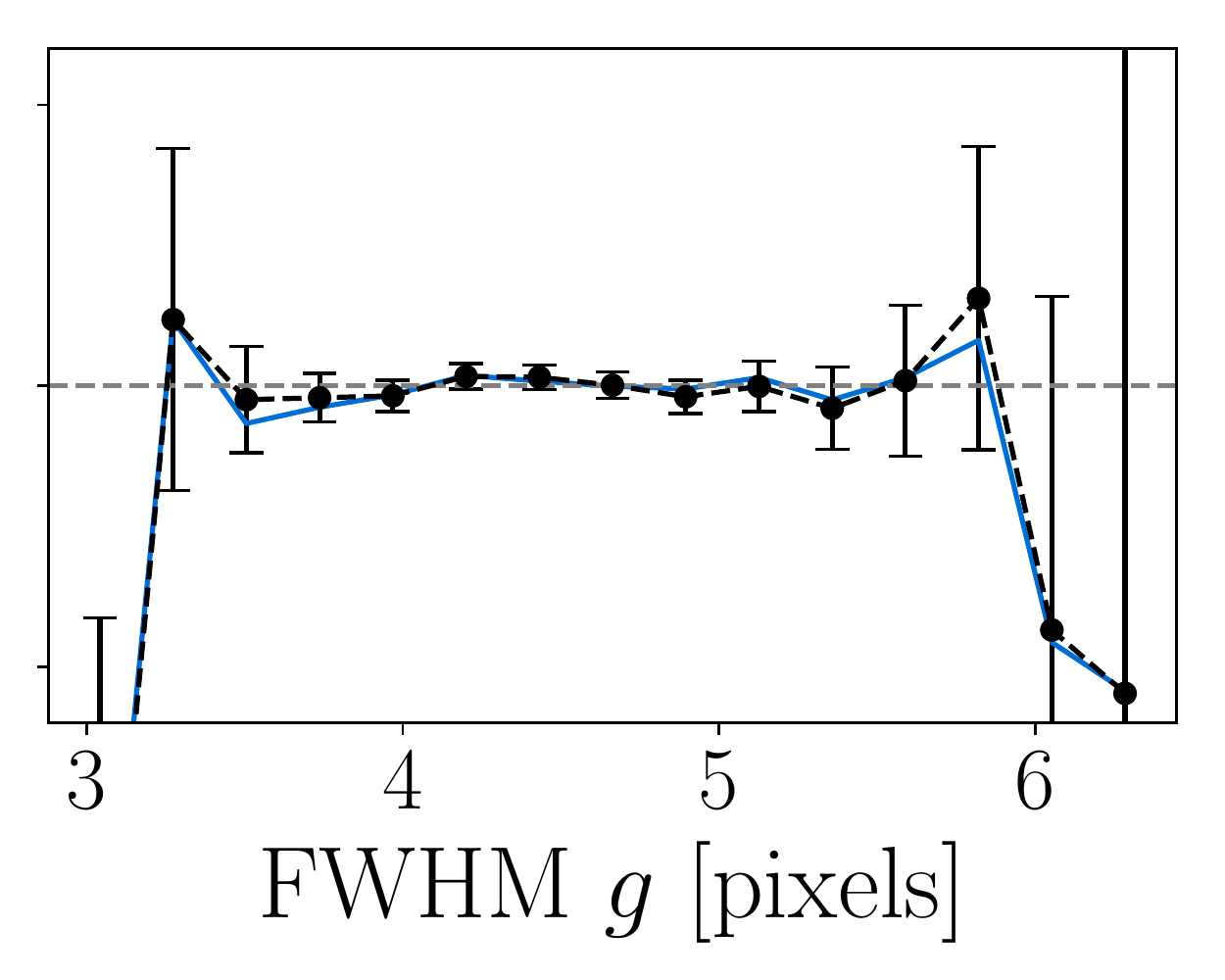}
\includegraphics[width=0.23\textwidth]{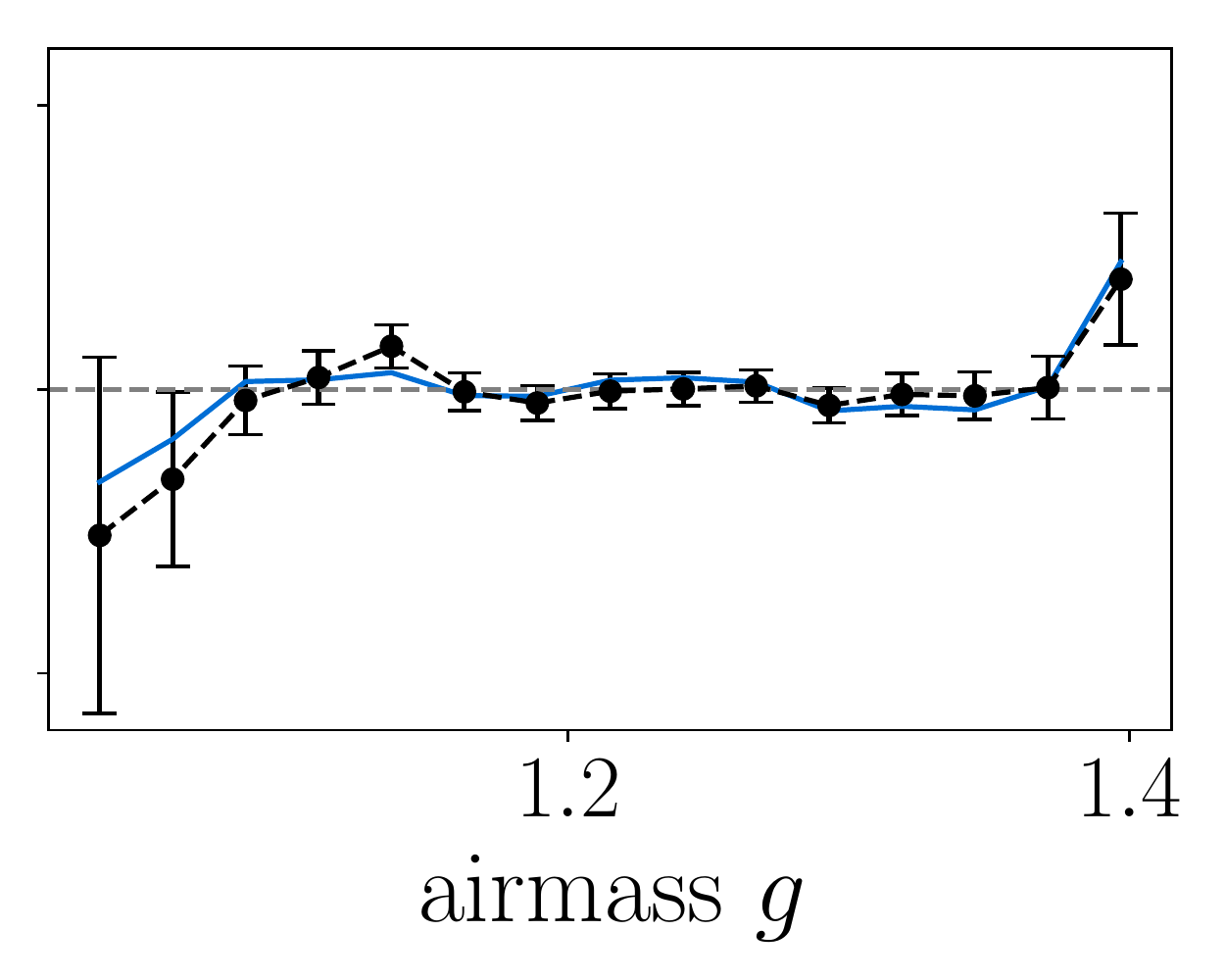} 
\includegraphics[width=0.275\textwidth]{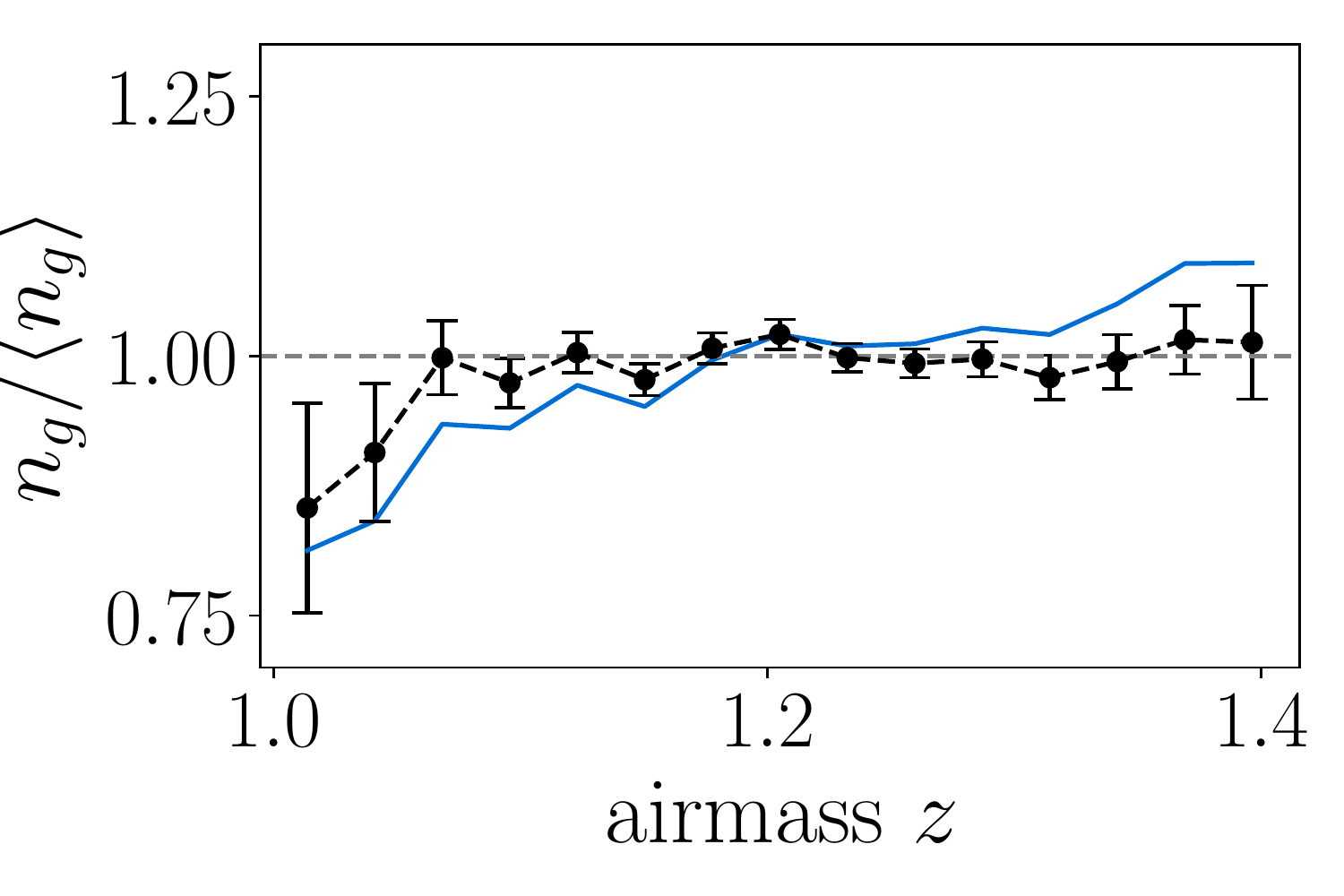}
\includegraphics[width=0.23\textwidth]{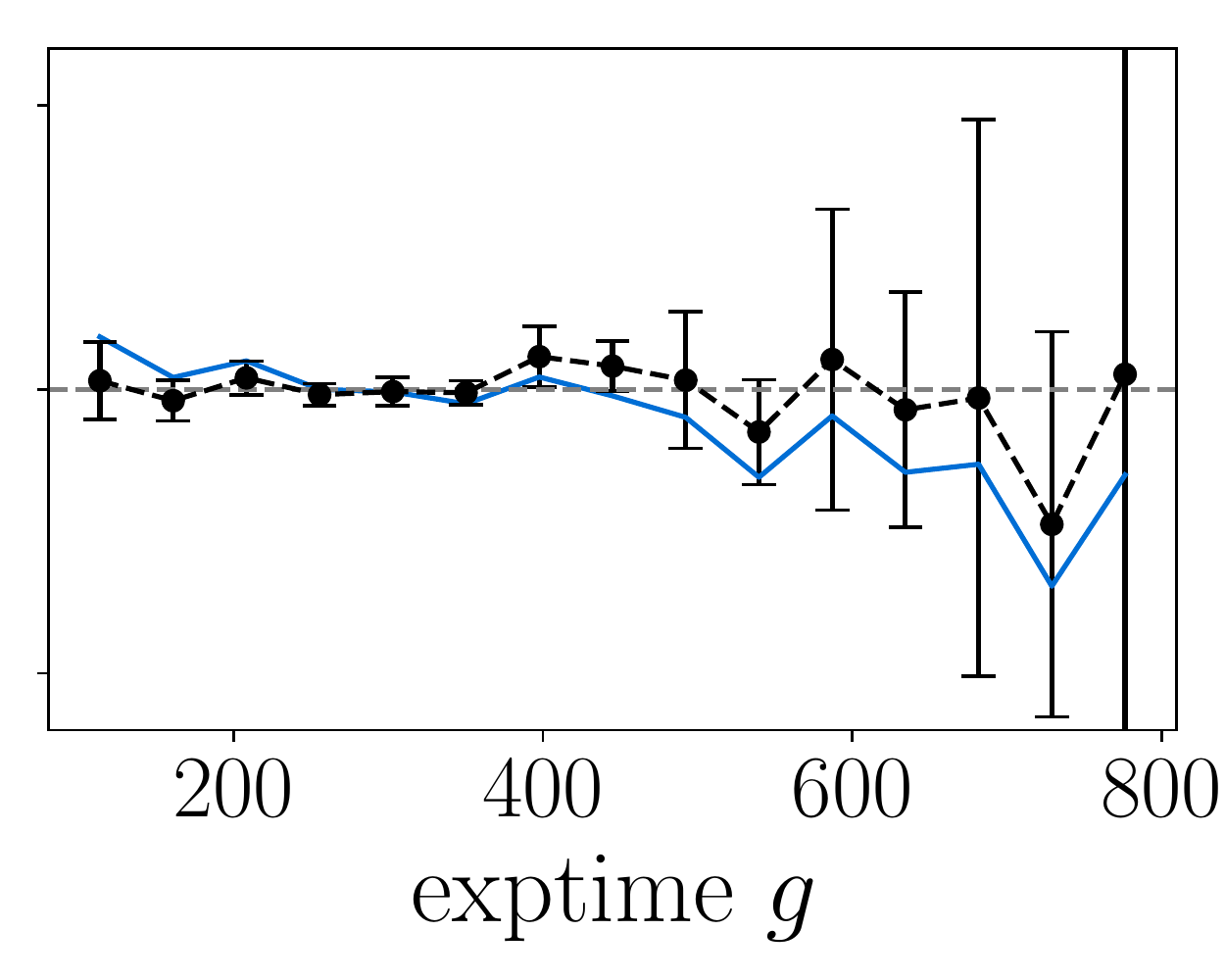}
\includegraphics[width=0.23\textwidth]{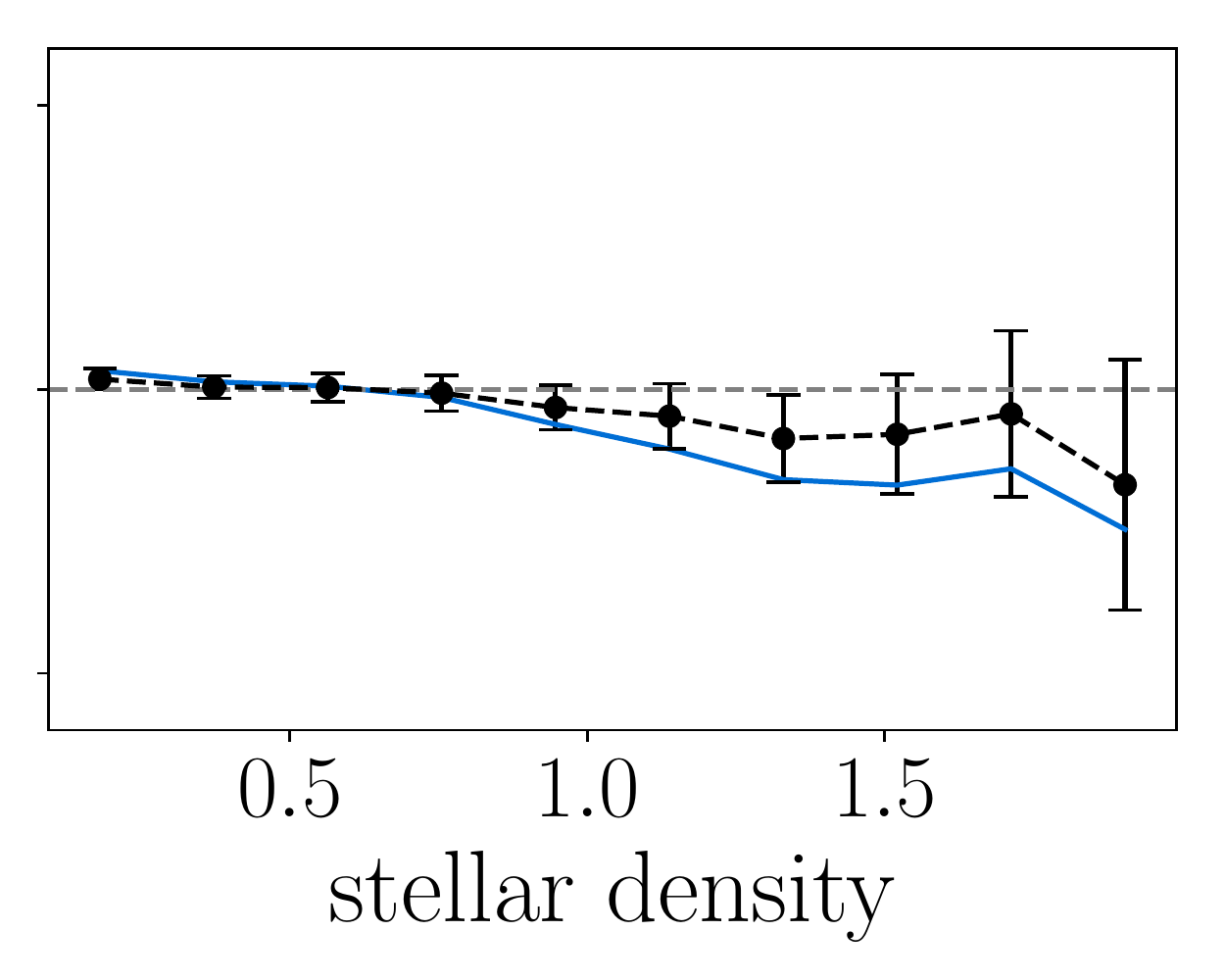}
\caption{
Galaxy number density with respect to survey properties having the top six $\Delta \chi^2$ and stellar density. The solid blue lines are calculated without correction weights. The dashed black lines are calculated with weights. The error bars on the black lines are calculated assuming Poissonian statistics. 
%E(B-V) and stellar density are not directly used to correct systematics, but correcting for other properties automatically corrects for stellar density and keeps galactic extinction same.
}
\label{fig:sys_all}
\end{figure*}

\section{Systematic Error Characterization}
\label{sec:systematics}
%\notesj{Add why this systematics are not considered for training -- ``Because we did not find any trends of systematics in the training sample.''where to add?}

%\cmt{Why we correct systematics} 
Astrophysical foregrounds, observing conditions, and spatially-varying depth are potential sources of systematic uncertainty in galaxy survey analyses. They affect the probability of detecting sources and also their reconstructed properties, and can thereby result in systematic biases in cosmological analyses \citep{Leistedt2015, Crocce2015GalaxyData}.
%\notesj{need to be paraphrased}
%[REF:Leistedt et al] mapped sources of systematics onto the DES footprint. We use these set of survey property maps to quantify and remove the systematic effects that may affect the density of galaxies we select. 

% Leistedt B. et al
%Spatially-varying depth and characteristics of observing conditions, such as seeing, airmass, or sky background, are major sources of systematic uncertainties in modern galaxy survey analyses.
%Spatial fluctuations in the depth or quality of the data (e.g., the properties of the sky noise, the photometry, or galaxy el-lipticity measurements) can impact the galaxy catalogues and lead to systematic biases in cosmological analyses.
%However, it is worth recalling that systematics in clustering and cosmic shear studies are mostly rooted in astrophysical foregrounds (extinction by dust or obscuration by bright stars), ob-serving conditions (e.g., seeing, sky noise, airmass), or processing and calibration (such as the quality of the photometry or the point spread function).
%These affect the probability of detecting sources and also their properties, yielding non-trivial distortions in the re-duced data, in particular the galaxy catalogues.

% Elvin-Poole 
%In order to quantify the extent of these correlations and remove their effect from the two-point function, maps of DES imaging properties were pro-duced using the methods described in Ref. [42].
%We con- sider the possibility that depth, seeing, exposure time, sky brightness and airmass, in each band griz, affect the density of galaxies we select. In

%\cmt{methodology}
We follow the procedures described in \cite{ELVINPOOLE} to identify and correct for these kind of systematic biases on the DMASS sample in the SPT region\footnote{We do not find any systematic biases from CMASS in the training region. Therefore, systematics adderessed in this section were not considered for modeling the probabilistic  model.}. 
To search for potential systematic uncertainties that affect galaxy clustering, we study the correlations between the galaxy number density and survey properties. If the galaxy density is independent from a survey property, we do not consider this property to have an impact on our DMASS sample. 
We use HEALPix maps ($N_{\rm side} = 4096$) of 4 observing conditions (airmass, seeing FWHM, sky brightness, exposure time), $10\sigma$ limiting depth in $griz$ bands, and the 2 astrophysical foregrounds of galactic reddening ($N_{\rm side} = 1024$) and stellar density ($N_{\rm side} = 512$). A detailed description about constructing HEALPix survey property maps can be found in \cite{Leistedt2015}. The construction of stellar density maps is described in \cite{ELVINPOOLE}. The SFD  galactic dust map is available at the \verb|LAMBDA| website\footnote{\url{https://lambda.gsfc.nasa.gov/product/foreground}} \citep{SFD98}. 

%\cmt{methodology - mask} 
We mask HEALPix pixels where the galaxy number density deviates by more than 20 per cent from the mean value (1.0) or changes sharply after some threshold value. We mask HEALPix pixels where seeing FWHM in $r$ band $>$ 4.5 pixels, which removes %$2\%$ of DMASS galaxies with 
$2\%$ of the total area. 

Prior to correcting systematics, we rank survey properties from the most to least significance.  
%\sout{ To avoid over-correcting systematics, we rank survey properties from the most to least significance and test the impact of including each additional correction (i.e., when correcting for the next property, the previously derived corrections are still applied). } 
The survey properties are ranked in order of the value  given by
\bea
\Delta \chi^2 = \chi^2_{\rm null} - \chi^2_{\rm model}~,
\eea
where $\chi^2_{\rm model}$ is the difference in $\chi^2$ between the best fit model of the number density and data points, $\chi^2_{\rm null}$ is $\chi^2$ against a null line $n_{\rm gal}/\langle n_{\rm gal} \rangle = 1$. 
%The best fit model was determined by minimizing $\chi^2_{\rm model}$ with a linear model $N_{\rm gal} \propto A s + B$.
We minimize $\chi^2_{\rm model}$ by fitting a linear model $N_{\rm gal} \propto A s + B$ against the calculated number density with Poisson errors of each data point. 

%{\bf probably can remove this:} \cite{ELVINPOOLE} uses a different value $\Delta \chi^2/\Delta \chi^2(68)$ to determine the significance of survey properties. The denominator $\chi^2(68)$ is the value of $\Delta \chi^2$ measured on the data when it is larger than 68\% of the mocks $\Delta\chi^2$. Since we don't have mock catalogs for DMASS, we simply use $\Delta \chi^2$ as our criteria. 

After ranking properties, 
% based on the $\Delta \chi^2$, 
we correct for them starting from the highest ranked one using the inverse of the best fit model as a weight. Since survey properties are correlated with each other, correcting one survey property can introduce new systematic trends from another survey property. Therefore, the relationship between the galaxy number density and survey properties is re-calculated after applying a weight. Then one moves to the next top-ranked survey property and iterates the procedure. 
 
%every time a new weight is applied. 
%The next weight to be applied is the highest ranked property among the remaining properties.
%This procedure is iterated for the remaining properties. 
%until there is no significant survey properties impacting on galaxy clustering.

%\cmt{This procedure is repeated until there is no significant systematic trend found in the galaxy sample. }
%{\bf What is the definition of significant being used?} \sjcmt{SJ:added. The third sentence in the next paragraph}  

%\cmt{\sout{To investigate any significant impact of systematics on the DMASS sample, we adopt the best fit value of galaxy bias as a diagnostic tool. We derived galaxy biases from angular clustering by correcting the survey property one by one until galaxy bias becomes stable within its error budget. If the derived galaxy bias deviates more than the error budget, we consider the property as one with a significant impact.  }}

The weighting scheme we use assumes that the effects of each survey property are separable. However, there is some correlation between systematic maps  
%. If we correct for all possible systematics, 
that may result in over-correcting the galaxy density field for a large number of systematic maps \citep{Elsner2016}.   
%As demonstrated by [44], correcting all possible systematics may result in over-correcting the galaxy density field. 
To avoid this, we calculate the impact of adding a systematic weight in every iteration to choose the minimum possible number of survey properties to be corrected. 

%\notejep{Mention in this section that each galaxy pair i,j is counted as $w_i w_j$ \cmtanswer{-- done. See the next paragraph} }

To investigate the impact of including additional systematic corrections, we utilize the angular correlation function. 
%We calculated a chisquare of the angular correlation function with a new weight against one before applying the weight, and use the value of chisquare as a diagnostic tool. 
%Figure \ref{fig:sys_deltachi} shows how significant each systematic property is to the angular correlation function. Based on figure \ref{fig:sys_deltachi}, we considered only fwhm r and airmass z as significant systematics that need to be corrected.  
The angular correlation function $w^{\delta_g\delta_g}(\theta)$ is computed with systematic weights using the Landy-Szalay estimator \citep{Landy1993BiasFunctions} as given by 
%We measure the correlation function $\omega(\theta)$ of CMASS and DMASS galaxies using the Landy-Szalay estimator \citep{Landy1993BiasFunctions} as given by 
\bea
w^{\delta_g \delta_g} (\theta) = \frac{DD(\theta) -2~DR(\theta) + RR(\theta)}{RR(\theta)}~,
\eea
where DD, DR and RR are the number of galaxy pairs, galaxy-random pairs, random pairs separated by a distance $\theta$. 
Systematic weights are applied to individual galaxies  as
\bea
DD(\theta) &=& \frac{1}{N_{\rm DD}(\theta)} \sum^{N_{\rm gal}}_i \sum^{N_{\rm gal}}_j w_i w_j \Theta (\theta_i - \theta_j) \\
DR(\theta) &=& \frac{1}{N_{\rm DR}(\theta)} \sum^{N_{\rm gal}}_i \sum^{N_{\rm rand}}_j w_i w_j \Theta (\theta_i - \theta_j) \\
RR(\theta) &=& \frac{1}{N_{\rm RR}(\theta)} \sum^{N_{\rm rand}}_i \sum^{N_{\rm rand}}_j w_i w_j \Theta (\theta_i - \theta_j) 
\eea
where $w_i$ denotes systematic weight ($w_i=1$ for randoms), $N(\theta)$ is the total number of pairs in a given data set in a given angular bin $\theta$, $\Theta(\theta_i - \theta_j)$ is $1$ if a pair lies at an angular distance $\theta$, otherwise zero.
The correlation function is measured in 10 logarithmically spaced angular bins over the range
%\footnote{\sjcmt{just delete to avoid confusion.}The choice of this range of scales is motivated by the scale cuts chosen in the DES analyses \citep{Troxel2018, Krause2017}. \cite{Troxel2018} determined the scale cuts on the shear correlation functions that cover a slightly narrower range than ours, to avoid possible parameter biases in the DES weak lensing cosmology analysis. Since we design DMASS to combine galaxy clustering with weak lensing from DES, matching the clustering signals on scales we care about is a reasonable choice for our ultimate science goals. %matching the clustering here is probably a good proxy for the clustering of the samples matching on scales we care about for our ultimate science goals. %We adopt the same scales for cross-correlation functions with other surveys later in the paper.
%} 
$2.5' < \theta < 250'$.  We adopt the same scales for cross-correlation functions with other surveys later in the paper. All two-point calculations are done with the public code \verb|TreeCorr|\footnote{\url{https://github.com/rmjarvis/TreeCorr}}\citep{TreeCorr}.
%\notesj{Why choose this scale?}
%\noteemh{This isn't a reigorous argument, but this is approximately the range of scales over which we will ultimately measure the weak lensing, so matching the clustering here is probably a good proxy for the clustering of the samples matching on scales we care about for our ultimate science goals.  \cmtanswer{--below}}

Randoms for DMASS are uniformly generated on the surface of a sphere and masked by the same masks described in Section \ref{sec:data}. The number density of randoms is chosen to be $50$ times larger than DMASS, minimizing the impact of any noise from the finite number of randoms and matching the relative number of CMASS randoms. 
%Galaxies in the SPT region are weighted by the CMASS membership probabilities and systematic weights calculated in Section \ref{sec:systematics}. 

To construct a covariance matrix for DMASS, we first compute a covariance matrix for CMASS from the 1000 QPM CMASS mock catalogues used in the BOSS-III analyses  \citep{Alam2015TheSDSS-III}:
\bea
{\bf{C}}(\omega_i, \omega_j) = \frac{1}{N_{\rm mock} - 1} \sum^{N_{\rm mock}}_{k=0} (w_i^k - \bar{w}_i ) (w_j^k -\bar{w}_j) ~,
\eea
where $N_{\rm mock}$ is the total number of mocks, $w_i$ represents the $i$th bin of the angular correlation function, $w_i^{k}$ denotes the $i$th bin of the angular correlation function from the $k$th mock, and $\bar{w}$ is the average value of $w$ over all mocks.

From the resulting CMASS mock covariance matrix, we derive a covariance matrix for DMASS by using the analytic form of the covariance between the angular correlation functions as follows:% \citep{Ross2011a}: 
%\notesj{First citation}
\bea
{\bf{C}}(\theta, \theta') = \frac{(2l+1)^2}{ f_{sky} (4\pi)^2} \sum
_{l=0} P_l (\cos \theta) P_{l}(\cos \theta') \sigma^2(C_{l}) + \frac{\delta_{\theta, \theta'}}{n_{\rm pairs}}
\eea
where $\sigma^2(C_l)$ is the variance of the angular power spectrum $C_l$, $f_{sky}$ is the fraction of the sky, and $n_{\rm pairs}$ is the total number of galaxy pairs. 
Assuming DMASS and CMASS have the same galaxy bias and redshift distribution, 
the first term can be easily adjusted for DMASS by altering the survey area factor. The second term, the shot noise term, can be directly calculated from the data. 
%The first term is proportional to the inverse of survey area and the shot noise term is directly calculable from the data. 
%To construct a DMASS covariance matrix, 
We obtain the first term of the CMASS covariance matrix by subtracting the inverse of pair counts of the CMASS galaxies from the mock covariance, and rescaling it by the ratio of the survey areas. 
The derived form of the covariance matrix for DMASS is 
\bea
{\bf{C}}_{\rm D} = \frac{A_{\rm D}}{A_{\rm C}}\left( {\bf{C}}_{\rm mock,C } - \frac{\delta_{\theta, \theta'}}{n_{\rm pairs, C}} \right) + \frac{\delta_{\theta, \theta'}}{n_{\rm pairs, D}}~,
\eea
where $A$ is the survey area,  ${\bf{C}}_{\rm mock}$ is the mock covariance, and the subscripts C and D stand for CMASS and DMASS respectively.

Figure \ref{fig:sys_deltachi} shows the impact of including additional systematic corrections through the value of $\chi^2$, computed from the re-scaled covariance matrix and residuals between the two measurements - before and after correction. 
%We calculated a chisquare of the angular correlation function with a new weight against one before applying the weight, and use the value of chisquare as a diagnostic tool. 
%The computed values of $\chi^2$ with systematic corrections from the highest order are shown in figure \ref{fig:sys_deltachi}. 
The systematic weights are listed on the x-axis in the order that they are applied to on top of all the previous weights applied. 
%The first value shows chi-square of the angular two point function with veto mask against one without any systematics or masks. 
%Starting from the second point from left, the names of the survey properties we corrected for are listed on the x-axis in the order that they are corrected. 
%The weight for the particular property is applied on top of the weights for the other properties listed earlier.   
Survey properties that show notable impacts are FWHM $r$-band and airmass $z$-band. 
Applying corrections for the rest of the survey properties barely affects the angular correlation function. Therefore, we apply systematic weights only for the top two properties. 
Figure  \ref{fig:sys_all} shows the galaxy number density vs. survey property plots before applying the weights (blue) and after (black).
We additionally find that correcting the top two systematics removes any trend with stellar number density. 
%\notesj{sentence below : this statement needs to be quantitative or should be deleted. 1. Maybe can describe how good DES star-galaxy classifier is ($3\%$ contamination rate for $i<22$). \\ 2. redmagic sample in DES doesn't have correlation with stellar density. CMASS is brighter so should not have that either..?}
Our interpretation is that any trend with the stellar number density is not from pure stellar contamination but from strong correlations between the FWHM and airmass maps\footnote{The lack of correlation with stellar density is consistent with the results for the DES \redmagic  galaxies at similar redshifts \citep{ELVINPOOLE}.}.

%\sout{ since DMASS consists of very bright objects and it is less likely that this level of bright sample contains misclassified stars. }
%This is because there was no evidence found for stellar contamination or obscuration in the DES \redmagic sample in the same footprint \citep{ELVINPOOLE}, and DMASS consists of brighter objects than \redmagic galaxies.}
%\sout{since DMASS consists of very bright objects and the star-galay classifier in the DES imaging pipeline yields only a contamination rate $\leq 3\%$ for $i<22$} %\citep{Y1GOLD}. 

In the later sections, we will apply the systematic weights and veto mask computed in this section to the DMASS sample and report results along with the no systematics case.  

\section{Comparison with the BOSS CMASS Sample}
\label{sec:result}
 
In this section, we compare the properties of the DMASS sample to those of the BOSS CMASS sample. We evaluate the consistency of the overall number density, the amplitude of the auto- and cross-correlation functions, and redshift distribution.

As described in \cite{Ross2011, Ross2012}, the selection functions for BOSS galaxy data in the NGC and the SGC are slightly different due to measurable offsets in the DR8 \citep{SDSSDR8} photometry between the two regions.
% (See appendix \ref{app:ngc_sgc} for details). 
%The offset effectively lowers the $\dperp$ limit in the SGC which results in higher number density and lower clustering amplitude compared to the NGC.
DMASS tends to mimic SGC CMASS as the training set taken is a sub sample of CMASS in the SGC. 
%we use only the CMASS SGC sample for the clustering tests, as \verb|train-CMASS| sample is a sub sample of the CMASS SGC sample.
Therefore, we will specifically compare DMASS with SGC CMASS in addition to comparisons with the full CMASS sample.
%separately from those with the full CMASS sample. 

\begin{figure*}
\includegraphics[width=\textwidth]{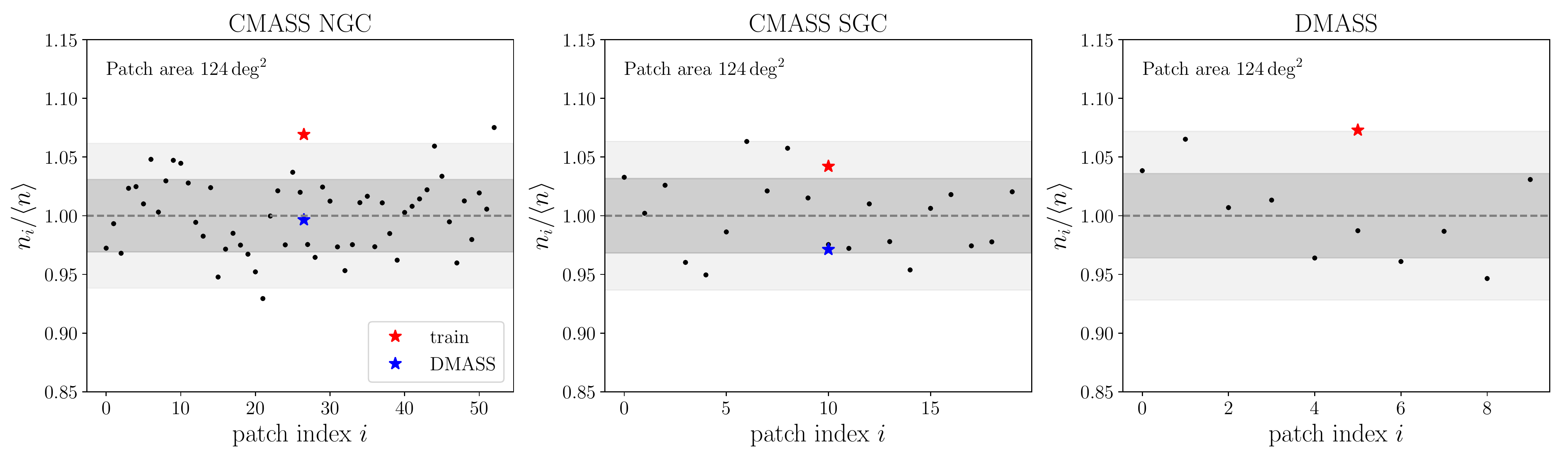}
\caption{Number density and its deviation in the NGC (left), CMASS in the SGC (middle), and DMASS (right). Each region is divided into Stripe 82-size (train region) patches. Red stars represent the number density of BOSS CMASS in the training region and blue stars are the total number density of DMASS.
All values are divided by the mean number density of each CMASS sample.  The dark-grey-shaded region is $1\sigma$, and the light-grey-shaded region is the $2\sigma$ level deviation of the black points in each panel. %The training region has  $\sim~7\%$ higher number density than the mean of the whole CMASS and DMASS shows the similar tendency. The width of shaded region is consistent for all cases. 
 }
\label{fig:boss_num_deviation}
\end{figure*}

\subsection{Number Density}
In this section, 
%we compare the number density of the DMASS sample with the BOSS CMASS sample. 
we will compare the number density of CMASS in the training data, which is from \stripe of Stripe 82 area, to the mean density in three distinct footprints: 1) BOSS CMASS data in the NGC area; 2) BOSS CMASS data in the SGC area; and 3) the DMASS data in the SPT area. 
We divide each of the three regions into many smaller patches that are the size of Stripe 82. This allows us to determine the expected variance between the number density in the training area and the full region.

%Since the model is built on small Stripe 82 that suffers cosmic variance, comparing the number density to the full area can be misleading. To take cosmic variance into account, we divided both CMASS and DES SPT footprint into small patches of stripe82 size and then calculated the deviation in the number densities from all patches.  

The three large footprints are divided as follows: each region is split into HEALPix pixels at resolution $N_{\rm side} = 4096$ where the size of each pixel is $0.72~{\rm arcmin}^2$. Then,  contiguous sets of $\sim606,000$ pixels are combined to make each patch comparable to the size of Stripe 82.
We adopt a slightly larger size for one patch, $124\degsqr$, in order to include all of the HEALPix pixels in the SPT region while keeping the size of all patches the same. The same patch size is applied for the SGC and NGC regions and the remaining HEALPix pixels that cannot be a complete patch are discarded.  The number of patches used for this calculation is 10 patches for the DES SPT region, 53 patches for the NGC region, and 20 patches for the SGC region. 
%The number of pixels combined for one patch is slightly different for each region. The size of patches for each region is written in Figure \ref{fig:boss_num_deviation}. } \sjcmt{now same area}
%
%\cmt{\sout{
%Each region is divided into Healpix pixels at resolution $N_{\rm side} = 16$ where the size of each pixel is $13.4 \degsqr$.
%We combine every nine pixels with their closest neighbors to make each patch comparable to the size of Stripe 82, \stripe.
%HEALPix pixels that overlap with less than $50\%$ of the footprint are excluded. }}

Figure \ref{fig:boss_num_deviation} shows the number density deviation for CMASS in the NGC (left), CMASS in the SGC (middle) and DMASS in the SPT region (right). 
%Since the model is built within S82, the number density calculated from one patch suffers cosmic variance that has to be considered. We divided catalogs into many patches of stripe82 size and then calculated the values of number density for each patch. 
%Figure \ref{fig:boss_num_deviation} shows the values of number density divided by mean value. 
The black dots are the number density values determined in each of the small patches. All values are divided by the mean number density of each panel. The grey-shaded region is the standard deviation of the distribution of the black dots. This represents an estimate for the 1$\sigma$ uncertainty in the number density of a Stripe82-sized patch. All three cases show a similar level of deviations. 
%The slightly larger deviation of DMASS SPT is due to larger number of patches affected by boundaries.
%slightly broader than two CMASS results because the SPT region is smaller and thus it is affected more by patches that overlap with boundaries. 
%{\bf Looks about the same in updated plot?} \sjcmt{SJ: slightly larger but if not clarifying that is better I can delete that sentence.}
%The blue star in the first panel is DMASS sample in Stripe 82. 
The red star in the first and second panels is the number density of the training (BOSS CMASS) galaxies in Stripe 82 and the blue star is the total number density of DMASS. Note that 
%they always have the same number density, so 
the location of the stars in each panel shows the relative number density in each region.
In all panels the red star is consistently $\sim 5-8\%$ away from the total mean value. 
%One can see that the number density in the training region (red star) is considerably greater than the mean number density of DMASS, 
One can see that the number density of DMASS is considerably lower than the number density in the training region (red star), 
but that it is matched to within $1\sigma$ of the overall CMASS number densities. That is, despite the data in the training region having a significantly greater number density than the overall CMASS sample, our model obtains the number density of DMASS that is a fairly good match to both CMASS SGC and CMASS NGC number densities.

%\cmt{ delete cmass and mock plots.... }
%The red star in the first plot is number density of stripe82.  In the first panel, stripe 82 has 5 \% higher number density than mean of the whole CMASS within the shaded region, which implies that the 5\% difference is acceptable. 
%The width of shaded region is consistent with CMASS mock catalog (in the second panel). Figure \ref{fig:number_density_spt} shows the number density and its deviation of DMASS. Likewise, the value of stripe 82 has 5 \% higher number density than mean of the whole sample which is similar to what is shown in CMASS number density plot (Fig. \ref{fig:boss_num_deviation}) . 
%DMASS deviation (shaded region) is broader than CMASS's because the SPT region is smaller than SGC thus it is affected more by patches in boundary than SGC. 

\subsection{Angular Correlation Function  }
\label{sec:wg}

% \usepackage{booktabs}
% \usepackage{graphicx}
\iffalse
\begin{table}
\centering
\caption{My caption}
\label{my-label}
%\resizebox{0.45\textwidth}{!}{%
\begin{tabular}{@{}lcc@{}}
\toprule
SAMPLE    & default    & sys        \\ \midrule
\multicolumn{3}{c}{\textbf{wtheta}} \\
SGC       & 4.94/10    & 2.58/10    \\
FULL      & 10.61/10   & 8.60/10    \\
S-N       & 14.53/10   & -          \\ \midrule
\multicolumn{3}{c}{\textbf{WISE}}   \\
SGC       & 9.04/10    & 9.70/10    \\
FULL      & 12.12/10   & 11.42/10   \\
S-N       & 11.76/10   & -          \\ \midrule
\multicolumn{3}{c}{\textbf{CMB}}    \\
SGC       & 18.50/12   & 12.56/12   \\
FULL      & 28.62/12   & 30.63/12   \\
S-N       & 98.39/12   & -          \\ \bottomrule
\end{tabular}%
%}
\end{table}
\fi

%\notesj{covariance matrix calculated by jksampling, Nk = 100 in the same way as Prat et al 2018}
%\cmt{Why angular clustering?}
We use the angular correlation function as a test to validate that DMASS matches the CMASS sample. Assuming that the number density and redshift distributions are matched, we should expect consistent amplitudes of the correlation functions if we have indeed matched the samples. We can directly compare the amplitude of the correlation functions of DMASS and CMASS and thus test for consistency without any cosmological assumptions.
%Angular clustering is a viable way to validate the DMASS sample as it is completely independent from DES colors and magnitudes used for training the model and also free from any cosmological assumptions such as galaxy bias evolution or redshift distribution. 
Three different probes were chosen for this comparison:  
%In this section we measure 
the galaxy angular auto-correlation and the angular cross-correlation with two full sky surveys - The Wide-field Infrared Survey Explorer all sky survey \citep[WISE;][]{WISE} and CMB lensing from Planck \citep{Planck2015Lensing}.
% - and compare the measurements with the same measurements done with the BOSS CMASS sample.

\subsubsection{Auto-angular Correlation Function}
\label{sec:result-angular}

We measure the correlation function $w^{\delta_g \delta_g}(\theta)$ of CMASS and DMASS galaxies in the same manner as described in Section \ref{sec:systematics}. 
%
%The BOSS CMASS sample is separated into two distinct regions South Galactic Cap (SGC) and North Galactic Cap (NGC). {\bf that might need to come earlier with the number density plots} 
%As described in \REF{ROSS 2011, 2012}, the selection functions for BOSS galaxy data in NGC and SGC are slightly different due to the measurable offsets in the DR8 \REF{Aihara 2011} photometry between two regions. The offset effectively lowers $\dperp$ limit in the SGC which results in higher number density and lower clustering amplitude compared to the NGC.
% For the CMASS sample, the SGC distribution is some-what skewed compared to the NGC selection. The number density is greater at the low redshift end, due to the fact that the offset in photometry effectively lowers d⊥ limit (equation 20) in the SGC compared to the NGC. These differences in n(z) imply that the galaxy populations will be slightly different in the different hemi-spheres and should thus be considered when the results from each hemisphere are combined.
%For this reason, we use only the CMASS SGC sample for the clustering tests, as \verb|train-CMASS| sample is a sub sample of the CMASS SGC sample.
Each galaxy in the CMASS sample is weighted by systematic (systot), close pair (cp), and redshift failure (zp) weights as given by \cite{Reid2016}:
\bea
w_{\rm total} = w_{\rm systot}~(w_{\rm cp} + w_{\rm zp} - 1) ~.
\label{eq:weights}
\eea
%where the subscript `zp' stand for the redshift failure. 
%\notejep{Were the distributions in the CMASS training sample also weighted by any of these quantities? \cmtanswer{-- done. See the next sentence } }
Note that we do not apply these weights in the CMASS training sample because we utilize the BOSS photometric sample for training. The BOSS photometric sample includes all missing galaxies that are dropped from a spectroscopic sample due to fiber collisions and redshift failures. $w_{\rm systot}$ is not considered either as we do not detect any systematic biases from the DES photometry of the CMASS training sample. 
As done in the previous BOSS analyses \citep{Chuang2017, Pellejero-Ibanez2016ThePriors}, we apply the explicit redshift cut $0.43 < z < 0.75$ to the BOSS CMASS sample. This redshift cut is not considered for training because we utilized only matched photometric information in the training sample. 
%Note that we do not use the explicit redshift cut $0.43 < z < 0.75$ for BOSS CMASS as used for the BOSS measurements. This is mainly because we utilized only matched photometric information in the training set.
In Appendix \ref{app:redshift}, we show that the redshift cut negligibly affects the 3D two point functions of BOSS CMASS, which justifies our choice of the CMASS photometric sample as the training set.
%thereby DMASS \sjcmt{is still good.. put later...}
%As our model for selecting DMASS was trained with the photometric CMASS sample without  
%
%from DMASS catalog weighted by the CMASS membership probability and 
For DMASS, 
%\cmt{\sout{only the sample in the SPT region is considered. Randoms for DMASS were uniformly generated on the surface of a sphere and masked by the same masks described in Section \ref{sec:data}. The number density of randoms is chosen to be $50$ times larger than DMASS, minimizing the impact of any noise from the finite number of randoms and matching the relative number of CMASS randoms. G}}
galaxies in the SPT region are weighted by the CMASS membership probabilities and systematic weights calculated in Section \ref{sec:systematics}. 
%
\iffalse
\cmt{\sout{
We use jackknife (JK) covariance matrices to calculate errors as}
\bea
C_{ij}(\omega_i, \omega_j) = \frac{N_{\rm JK} - 1}{N_{\rm JK}} \sum^{N_{\rm JK}}_{k=0} (\omega_i^k - \bar{\omega}_i ) (\omega_j^k -\bar{\omega}_j) ~,
\eea
\sout{
where $N_{\rm JK}$ is the number of Jackknife regions, $\omega_i$ represents $i$th bin of angular clustering $\omega$, $\omega_i^{k}$ denotes the $i$th bin of angular clustering from $k$th realization, and $\bar{\omega}$ is the average value of $\omega$ of all realizations. 
We use $N_{\rm JK} = 100$ for both samples. } }
\fi

\begin{figure}
% (CMH) commented out so this will compile
%\includegraphics[width=0.48\textwidth]{./figures/acf_comparison_cmass_dmass.pdf}
\includegraphics[width=0.5\textwidth]{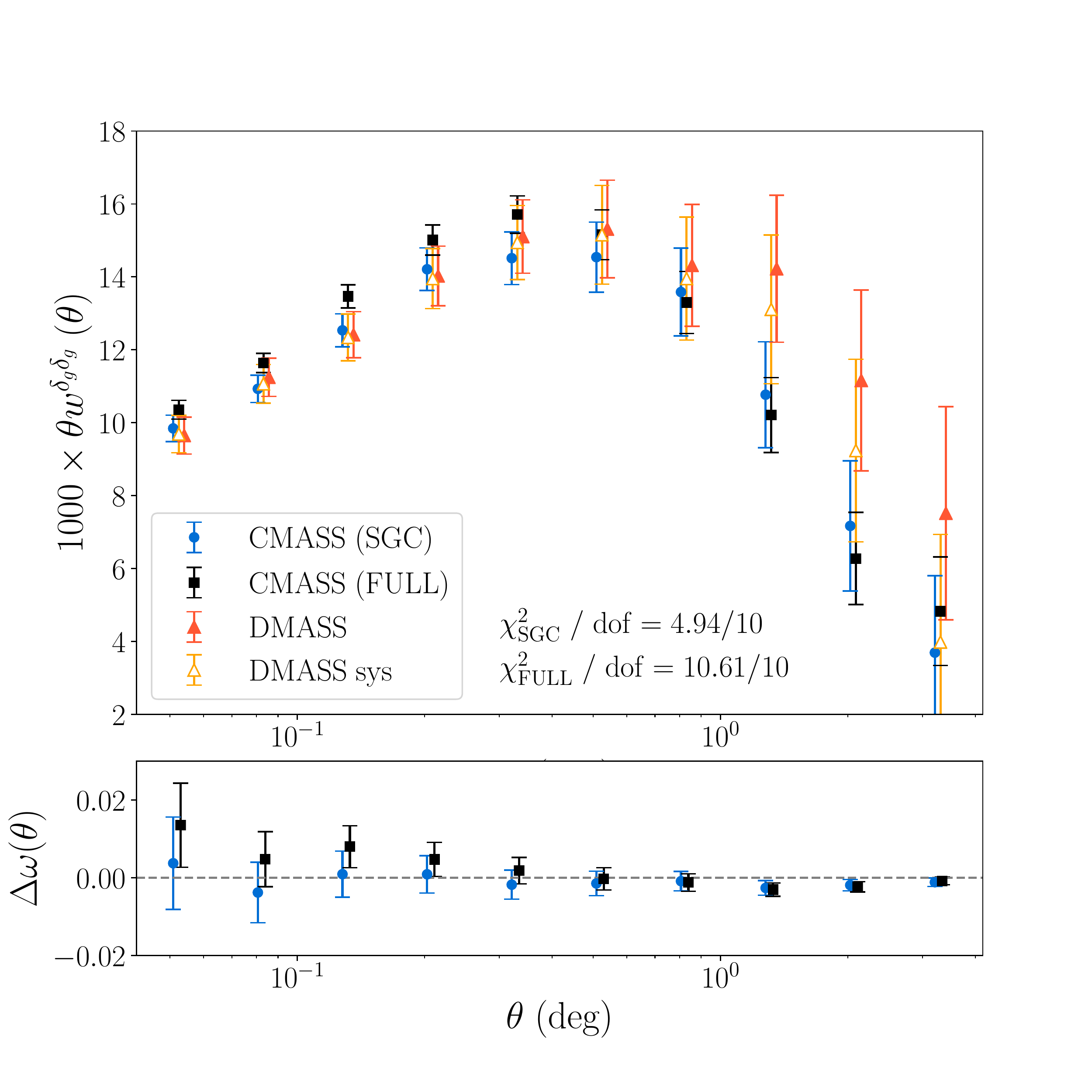}
\caption{The top panel shows the angular correlation function calculated with DMASS (red), DMASS corrected by the systematic weights (orange), CMASS SGC (blue), and full CMASS (black). 
%The black solid line is computed from a theory described in Equation \eqref{eq:wg} with the spectroscopic redshift distribution of CMASS SGC.
%The top panel shows angular clustering, 
The bottom panel shows residuals between DMASS and CMASS SGC (blue) or full CMASS (black). 
%The solid black line is angular clustering of CMASS in SGC, the red line shows  DMASS. 
%We find a better agreement between DMASS and CMASS SGC; the $\chi^2$ obtained when testing the DMASS $\omega(\theta)$ against the CMASS SGC $\omega(\theta)$ is 1.17 for 10 data points. 
$\chi^2_{\rm SGC}$ ($\chi^2_{\rm FULL}$) is the $\chi^2$ of the observed difference of two point functions of DMASS and CMASS SGC (FULL) 
}
\label{fig:acf_comparison_cmass_dmass}
\end{figure}

The result is shown in Figure \ref{fig:acf_comparison_cmass_dmass}. The blue and black data points are the angular correlations of CMASS in the SGC and full CMASS respectively, and the red data points show the DMASS angular correlations. Error bars are obtained from the aforementioned mock covariance matrices in Section \ref{sec:systematics}.  
%We find a good agreement between CMASS in SGC and DMASS; 
We find that the angular correlation function of DMASS has a better agreement with CMASS in SGC than full CMASS. The angular correlation function of full CMASS is slightly higher than the other samples on small scales, as expected from the intrinsic difference between CMASS in the SGC and the NGC. On large scales, DMASS tends to deviate from the two CMASS samples, but adding systematic weights mitigates the difference by suppressing the correlation function of DMASS on large scales.

To quantify consistency between CMASS and DMASS, we use a $\chi^2$ statistic and its associated Probability-To-Exceed (PTE) as our primary metric. 
We take the observed difference of binned two point functions $\Delta {\bf{d}} = w_{C} - w_{D}$ (shown in the bottom panel in Figure \ref{fig:acf_comparison_cmass_dmass}) and its associated covariance as $\boldsymbol{C}_{\rm tot} = \boldsymbol{C}_{C} + \boldsymbol{C}_{D} $. 
%Finally, %we ignore the covariance between these BAO/RSD measurements and those of EDS galaxy clustering and weak lensing
Cross-covariance between the CMASS and DMASS measurements is not considered since the two sets of measurements are carried out on different areas on the sky. 
%We assume the covariance between the CMASS and DMASS measurements is negligible since two sets of measurements are carried out on different areas on the sky. 
Then we calculate the $\chi^2$ of the difference defined by
\bea
\chi^2= \sum_{i,j}^{N_{\rm bins}} \Delta {\bf{d}}^{\rm T}_i (\boldsymbol{C}^{-1}_{\rm tot})_{i,j} \Delta {\bf{d}}_j ~.
\label{eq:chi2}
\eea
and its associated PTE with the degrees of freedom (the number of bins). 
A probability of $(100-{\rm PTE})\% = 68\%~(95\%)$ corresponds to $1 \sigma ~(2\sigma)$ difference.

\iffalse
\reph{
To quantify consistency between CMASS and DMASS, we use a $\chi^2$ statistic and its associated Probability-To-Exceed (PTE) as our primary metric. 
We take the measurements of CMASS ($p_C$) and DMASS ($p_D$) to be consistent with one another if the hypothesis $p_C - p_D = 0$ is acceptable. 
\bea
\chi^2 = (p_C - p_D)^{\rm T} \boldsymbol{C}^{-1}_{\rm tot} (p_C - p_D)
\eea
where $\boldsymbol{C}_{\rm tot} = \boldsymbol{C}_C + \boldsymbol{C}_D$.  We assume cross-covariance between CMASS and DMASS is negligible since two samples cover different part of sky. The associated PTE value $P_{\chi^2}$ is evaluated from the definition: 
\sjcmt{wrong model}
\bea
P_{\chi^2} = {\rm erf} \left(  \frac{\rm No. ~of ~\sigma }{\sqrt{2}} \right)
\eea 
With this definition, a probability of $1-P_{\chi^2} = 68\% ~(95\%)$ corresponds to $1 \sigma~ (2\sigma)$ difference. 
}
\fi

The $\chi^2/{\rm dof}$ obtained between DMASS and CMASS SGC is 4.94/10 (PTE=$90 \%$) in the range $2.5' < \theta < 250'$. For the comparison with the full CMASS sample, we obtain a $\chi^2/{\rm dof}$ of 10.67/10 with PTE=$53 \%$. With the systematic weights, we obtain 2.58/10 (PTE=$99 \%$) for CMASS in the SGC and 8.60/10  (PTE=$47 \%$) for full CMASS. 
%\sout{The $\chi^2$ obtained between DMASS and CMASS SGC is 4.94 for 10 data points in the range $2.5' < \theta < 250'$. For the comparison with the full CMASS sample, we obtain a rather larger $\chi^2$ of 10.67 for 10 data points. With the systematic weights, we obtain 2.58 for CMASS in the SGC and 8.60 for full CMASS. }
%\sout{Note that the result with the systematic weights is more analogous to CMASS, but the reported value of $\chi^2$ implies that adding weights may result in a statistically less robust DMASS sample.  }

\iffalse
\subsubsection{Cross-correlation with WISE and CMB lensing}
\cmt{\sout{
We calculated the cross correlation of the CMASS(DMASS) sample with two external data sets that uniformly cover the entire celestial sphere: the galaxy sample from The Wide-field Infrared Survey Explorer (WISE) all-sky survey} \citep{WISE} \sout{and the Planck CMB lensing map }\citep{Planck2015Lensing}. 
\sout{Those all sky external data sets allow us to validate the DMASS sample independently from any DES or SDSS properties.}}
\fi

%\cmt{wise...} Not only that, external data sets cover different redshift ranges from DMASS or CMASS, therefore weak signal from the low redshift tail at $z<0.43$ is not smeared by strong signals from the high redshift range. 

%high redshift having high signal to noise. 
%Also good way to check that our extreme deconvolution model is successfully restoring dperp cut that is isolating high redshift galaxies in the CMASS criteria.

\subsubsection{Cross-correlation with WISE Galaxies}

\begin{figure*}
\includegraphics[width=0.49\textwidth]{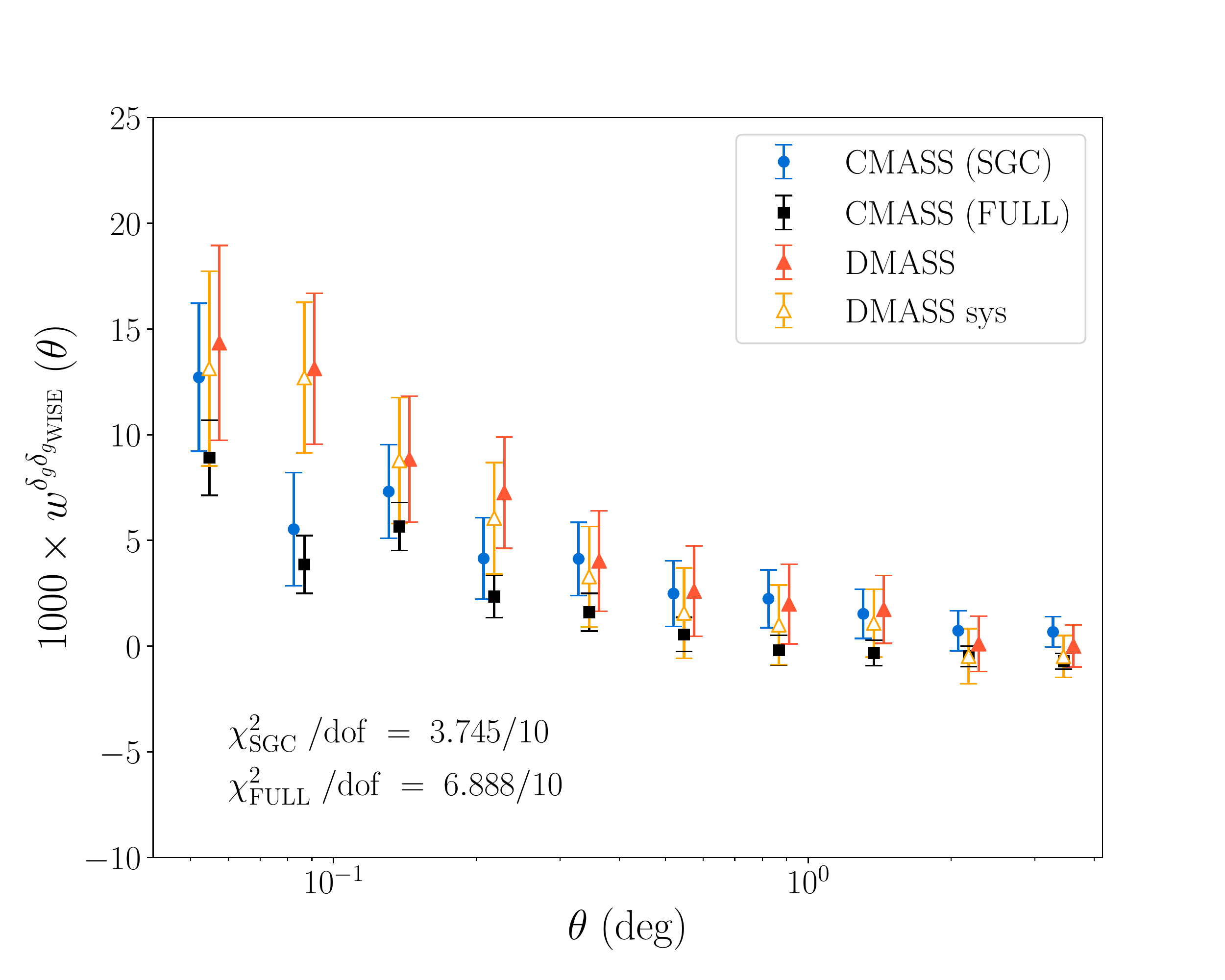}
\includegraphics[width=0.49\textwidth]{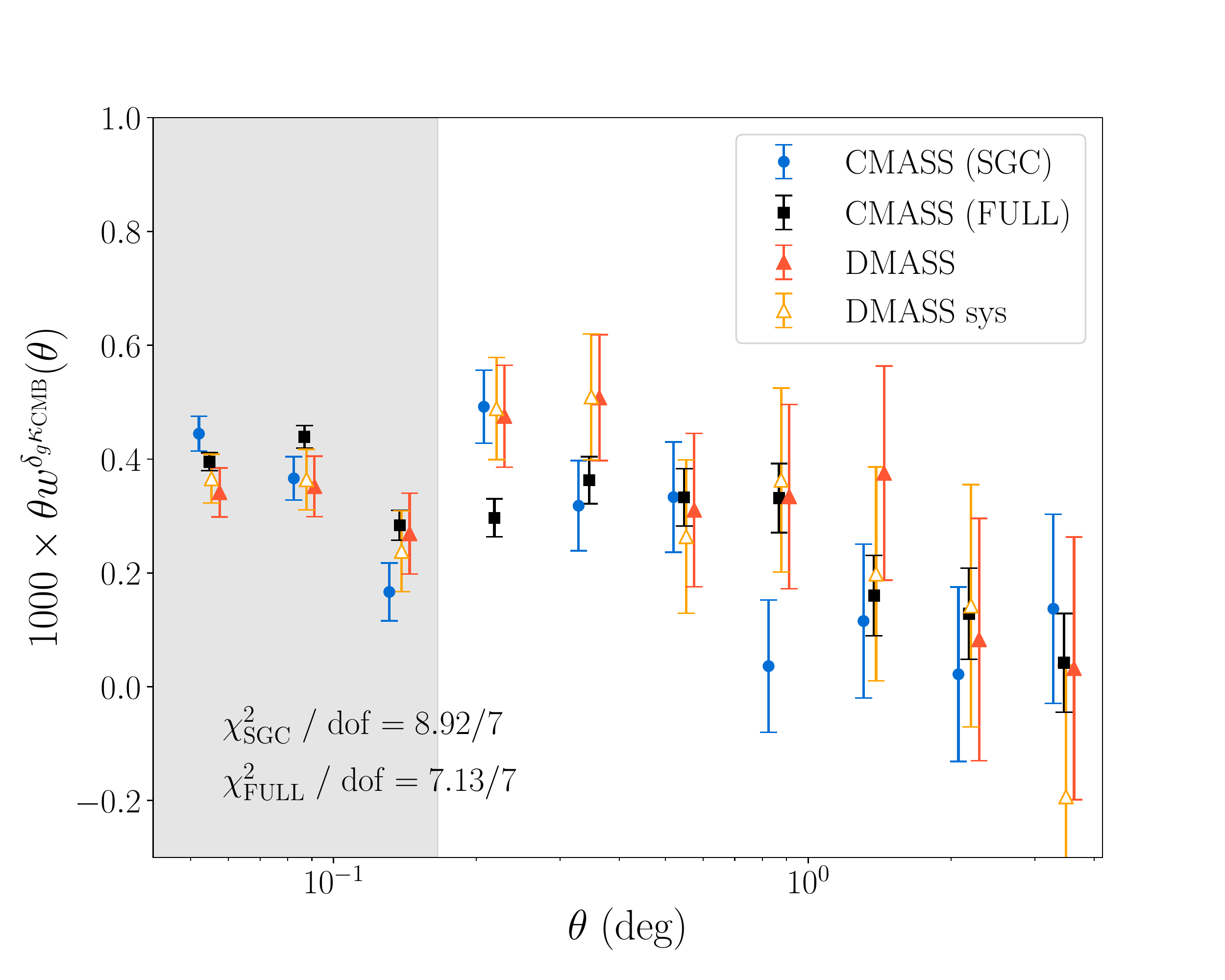}
\caption{
Cross-correlation measurements of the DMASS (red), DMASS with systematic weights (orange), CMASS SGC (blue) and full CMASS (black) samples with WISE galaxies (left) and CMB convergence map (right). We dropped the first three data points of the cross-correlation with CMB lensing (in grey shaded region) from the measurements of $\chi^2$ to include only reliable scales where the analytic covariance matrices are valid. 
%A good agreement in cross-correlation with WISE implies that all three samples have the similar fraction of galaxies at $z < 0.4$ where the noisy CMASS $\dperp$ cut is located at. 
%\notesj{description for cmblensing}
%Cross-correlation with CMB convergence map shows a better agreement with full CMASS on small scales. 
%The black solid line is computed from theory with assuming galaxy bias $b_g = 2$.
	 }
\label{fig:wise}
\end{figure*}

%a. The WISE satellite surveys the entire sky at wavelengths of 3.4, 4.6, 12, and 22 µm (W1 through W4), with the point source sensitivities of 0.08, 0.11, 1, and 6 mJy, respectively.\\ 
%b. We use the first data release issued on April 14, 2011, containing about half the sky. \\
%1. We removed stars with the color cuts [(According to Goto et al. (2012) the majority of stars near to the galactic plane have a W3.4 −W4.6 ? 0.2 mag color. Moreover, it was found that a W4.6 −W12 ? 2.9 mag selection reduced the stellar contamination.)]. 2. To select a uniform galaxy sample, we only use galaxies with W13.6µm < 15.2 mag. Following Goto et al. (2012) we select sources to a flux limit ofW1 ? 15.2 mag to have a uniform dataset. 3. Remove Stripe like features

%This analysis uses the WISE All-Sky Release Source Catalog  taken by the the Wide-field Infrared Survey Explorer. 
%The approximate coverage of the All-Sky Release Source Catalog is $42195 \degsqr$, or $99.86\%$ of the entire sky. 
The WISE satellite surveys  $99.86\%$ of the entire sky at wavelengths of $3.4$, $4.6$, $12$, and $22~\rm \mu m$ (W1 through W4). To have a uniform galaxy dataset, we select sources to a flux limit of ${\rm W1 < 15.2}$ and remove stars with the cuts $W1-W2 < 0.2$ and $W2-W3 < 2.9$, following \cite{Goto2011Cross-correlationBackground}. %We select only high luminosity galaxies from the WISE galaxy catalog by following procedures in \REF{WISE}. 
Regions contaminated by scattered moonlight are excluded by the `moonlev' flag if at least one of the bands has a value higher than $3$ 
%in at least one of the bands 
\citep{Kovacs2012Cross-correlationRelease}. We also remove regions having the extreme level of galactic extinction, $0.367 \times\rm E(B-V)_{\rm SFD} > 0.05$.

The resulting WISE galaxy sample approximately spans the redshift range from 0 to 0.4 with median redshift $z\sim0.15$ \citep{Goto2011Cross-correlationBackground, Kovacs2012Cross-correlationRelease}. CMASS in the SGC is known to have $5.24\%$ of galaxies and CMASS in the NGC has $3.73\%$ in the low redshift tail $z < 0.43$. If the probabilistic model effectively reproduces the  $\dperp$ cut in the DES photometry, the DMASS sample would have a similar fraction of galaxies in the low redshift tail and this would result in the same cross-correlation signal.

We adopt the Landy-Szalay estimator for the cross-correlation given as
\bea
w^{\delta_g \delta_{g_{\rm WISE }}}(\theta) = \frac{D D_W - D R_W - D_W R + R R_W}{R R_W}~,
\eea
where $D_W$ and $R_W$ stand for WISE galaxies and WISE randoms. WISE randoms are uniformly generated on the surface of a sphere within the masked region, with a size 50 times larger than the WISE galaxy sample.

Errors are derived from analytic covariance matrices. We calculate the covariance matrices of the cross-correlation as the sum of the Gaussian covariance and non-Gaussian covariance, and the super-sample covariance as detailed in \cite{COSMOLIKE}. We adopt the measurement of galaxy bias $b_{\rm WISE} = 1.06$ and the spectroscopic redshift distribution shown in Figure 3 in \cite{Goto2011Cross-correlationBackground}.

With the same angular binning choice as the auto-correlation function, we measure the cross-correlation function between  WISE galaxies and SGC CMASS, full CMASS, and DMASS as shown in Figure \ref{fig:wise}. The cross-correlation of full CMASS shows a slightly lower amplitude than CMASS in the SGC and DMASS on all scales which is expected because CMASS in the NGC has a smaller number density than CMASS in the SGC at the low redshift end.
We find that the $\chi^2/{\rm dof}$ of DMASS computed with respect to CMASS in the SGC is $9.04/10$ (PTE=$53\%$), and the one computed with respect to full CMASS is $12.12/10$ (PTE=$28\%$). With the systematic weights, the value is 9.70/10 (PTE=$47\%$) with SGC CMASS and 11.42/10 (PTE=$33\%$) with full CMASS. 
%\reph{We find that the $\chi^2$ of DMASS computed with respect to CMASS in the SGC is $9.038$ for 10 data points, and the one computed with respect to full CMASS is $12.124$ for 10 data points. With the systematic weights, the value is 9.70 with SGC CMASS and 11.42 with full CMASS. }
From the results, we do not find strong discrepancies between any of the CMASS samples and DMASS sample. The result also implies %two things; First, the effect of $4\%$ of CMASS sample at the low redshift tail is negligible.
that the probabilistic model is successfully reproduces the $\dperp$ cut in the DES system that excludes low redshift objects. 
\notejep{So the final decision was to include all scales in chi2? Does this require a comment somewhere? }

%The CMB photons released from the time of last scattering ($z\sim 1100$) are gravitationally deflected by the mass distribution as they travel through the large-scale structure(LSS). The imprint on CMB anisotropies by this deflection of photons is called CMB lensing \REF{CMB lensing}. 

%The intervening large-scale structure (LSS) of the Universe can alter the energies and paths of the CMB photons, producing a range of effects beyond the primary CMB power spectrum; these are collectively known as secondary CMB anisotropies.

%Finally, as they travel through the LSS, the CMB photons are gravitation-ally deflected by the mass distribution along their way, distorting the image we eventually observe. Here we focus on this last effect,CMB lensing.

%The large-scale structure (LSS) leaves an imprint on CMB anisotropies by gravitationally deflecting CMB photons during their journey from the last scattering surface to us. 

%In the standard structure formation scenario, galaxies reside in dark matter halos (Mo et al. 2010), the most massive of which are the signposts of larger scale structures that act as lenses for CMB photons. 

\subsubsection{Cross-correlation with the CMB Lensing Map}

The CMB photons released from the time of last scattering ($z\sim 1100$) are gravitationally deflected by the foreground mass distribution as they travel through the large-scale structure. The imprint on CMB anisotropies by this deflection of photons is called CMB lensing \citep{Planck2015Lensing}. 
The cross-correlation of galaxy positions and CMB lensing has two advantages for this work. First, CMB lensing is extremely homogeneous compared to galaxy catalogs. All information from the CMB departs from the same redshift $z=1100$ (considered as a very thin redshift bin) and travels the same distance until today regardless of the northern or southern part of the sky. Any difference found between the cross-correlation signals between different galaxy samples and the CMB would originate from differences between the galaxy samples themselves. Second, the galaxy bias is tied to the matter-matter correlation function in a different way that might give us complementary information. 

This analysis uses the 2015 CMB convergence map provided by the Planck collaboration \citep{Planck2015Lensing}. 
We use the lensing multipole range of $8 < l < 2048$ and apply a Gaussian smoothing of $\theta_{\rm FWHM} = 1.71'$ to the map. 
%We evaluated the cross-correlation function with CMASS in SGC and full CMASS separately as done in the previous sections. %photometric offsets in two regions affect the amplitude of two point function differently {\bf previously it was stated that only the SGC would be used}. 
%We followed the general methodology described in \cite{Omori2018}. 
The cross-correlation function is calculated in $10$ logarithmically spaced bins between $\rm 2.5' < \theta < 250'$  using the estimator  \citep{Omori2018}: 
\bea
w^{\delta_g \kappa_{\rm CMB}} (\theta) = D \kappa (\theta) - R \kappa(\theta)~,   
\eea
with 
\bea
D\kappa (\theta) = \frac{1}{N^{D\kappa}} \sum^{N_{\rm gal}}_{i=1}\sum^{N_{\rm pix}}_{j=1} w^D_i w^{\kappa}_j \kappa_{,j} \Theta (\theta_i - \theta_j)\\
R\kappa (\theta) = \frac{1}{N^{R\kappa}} \sum^{N_{\rm gal}}_{i=1}\sum^{N_{\rm pix}}_{j=1}  w^R_i w^{\kappa}_j \kappa_{j} \Theta (\theta_i - \theta_j)
\eea
where $D$ and $R$ stand for galaxies and randoms respectively, $w^D$ and $w^R$ are weights for galaxies and galaxy randoms, $N$ in the denominator is the total number of pairs, and $\kappa_j$ represents the value of convergence at the $j$th pixel.  
 
The measurements are shown in Figure \ref{fig:wise}. 
%Data points with error bars are our measurement with real galaxy samples. 
Error bars are from Gaussian covariance matrices computed by cosmoSIS \citep{COSMOSIS}.  
%For full details we refer readers to read \cite{Omori2018} and  \cite{Baxter2019}. 
%\notejep{I think the covariance in Baxter and Omori included NG terms (though they would likely have been negligible). Cosmosis is just Gaussian \cmtanswer{ -- done. deleted references and added a description in the next paragraph that the absent of NG term may cause high chi-square values } }
%
With the Gaussian covariance matrices and measured cross correlation functions, we estimate the values of $\chi^2$ between CMASS and DMASS. We find that the $\chi^2/{\rm dof}$ value between CMASS SGC and DMASS is $24.38/10$ (PTE $<1\%$) and the value between full CMASS and DMASS is $21.56/10$ (PTE=$2\%$). The value between CMASS SGC and NGC is $101.48/10$ (PTE $<1\%$), which is even more extreme than the former two cases. 
%\reph{With the Gaussian covariance matrices and measured cross correlation functions, we estimate the values of $\chi^2$ between CMASS and DMASS. We find that the $\chi^2$ value between CMASS SGC and DMASS is 18.50 and the value between full CMASS and DMASS is 28.62 for 10 data points. The value between CMASS SGC and NGC is 98.39, which is even more extreme than the former two cases. }
This implies that the large $\chi^2$ values between CMASS and DMASS are not from the difference between CMASS and DMASS.
Since our analytic covariance matrix is Gaussian, 
we believe that these large values of $\chi^2$ are due to the lack of the non-linear contributions on small scales. 
%the non-Gaussian contribution begins to dominate on small scales, 
Therefore, we exclude data points on the scales $\theta < 10 \arcmin$ and re-calculated $\chi^2$. The minimum angular cut is motivated by the measurement of the angular correlation function in Section \ref{sec:result-angular}. We compare the mock covariance for the auto-correlation function with the analytic calculation and find that the analytic calculation underestimates uncertainties by more than $20\%$ at $\theta < 10 \arcmin$. We simply utilize this scale to cut out unreliable information, expecting the non-linear contribution to be dominant on a similar scale in this case.
The $\chi^2/{\rm dof}$ values with the minimum scale cut are improved to $8.92/7$ (PTE=$26\%$) between CMASS SGC and DMASS, and $7.13/7$ (PTE=$42\%$) between full CMASS and DMASS. These values of $\chi^2/{\rm dof}$ are smaller than the $\chi^2/{\rm dof}$ between CMASS NGC and CMASS SGC (shown in Appendix \ref{app:ngc_sgc}). 
\notesj{with systematic weights, why PTE becomes bad? 6\% is still 1sigma level difference though. }

\subsection{Redshift Distribution}
\label{sec:clustering-z}
%\notesj{have not checked yet}
%\cmt{Why redshift distribution is important}
%Photometric redshift is inaccurate. DES photometric redshift is derived from DES colors. Therefore, it conveys various of combination of DES photometric errors. Its error reaches [..]. and DMASS photometric redshift is biased by ( ) compared to CMASS spectroscopic redshift. Our extreme deconvolution model is selecting galaxies based on their photometric color sets, thus even if the resulted galaxy sample has the same photometric redshift as the train sample it may not have the same true redshift distribution. 

%\cmt{introducing clustering redshift distribution}
In this section, we evaluate the redshift distribution of the DMASS sample by cross-correlating DMASS galaxies with the DES \redmagic sample \citep{Rozo2016RedMaGiC:Data, ELVINPOOLE}.
The concept of this technique called `clustering-z' is to recover redshift distributions of an unknown sample by cross-correlating it with a galaxy sample whose redshift distribution is known and accurate.
The technique was first demonstrated in %\cite{Schneider2006CrossCorrelation}, 
\cite{Newman2008}, and has been developed and applied to various cosmological analyses including DES \citep{2016MNRAS.460..163R,2016MNRAS.463.3737C,2017MNRAS.465.4118J,2017MNRAS.467.3576M,2018MNRAS.474.3921S}. 
%This method is referred to 'clustering-z' and 
%originally intended to cross-correlate a spectroscopic sample with an unknown sample that one wants to evaluate redshift distributions as \cite{Cawthon2017DarkCross-Correlations}
%One can also use photometric redshifts in the same manner, given that they are accurate and the redshift distribution that one wishes to characterize is broad \citep{Gatti2018}.
%using spectroscopic sample as a reference sample, 
\cite{Gatti2018} and \cite{Davis2017} calibrated redshift distributions of the DES Y1 source samples by using the DES \redmagic sample \citep{ELVINPOOLE} as a reference sample. 
\cite{Cawthon2017DarkCross-Correlations} calibrated the DES \redmagic sample by cross-correlating with the BOSS spectroscopic galaxy samples. For further details about the clustering-z method, we refer interested readers to references in \cite{Cawthon2017DarkCross-Correlations}. 
%The reference sample in their work is the DES redmagic high luminosity sample \REF{REF:ROZO et al}
%as a reference sample to calibrate the redshift distributions of the DES Y1 source sample. 

We utilize the \redmagic sample as a reference sample and follow the general procedures described in \cite{Davis2017}. 
The \redmagic galaxies are red luminous galaxies selected by the redMaPPer algorithm \citep{Rykoff2014RedMaPPer.Catalog} above three different luminosity threshold cuts ($L/L_*>0.5$, $L/L_*>1.0$, and $L/L_*>1.5$).
%The \redmagic sample is made of red luminous galaxies above some minimal luminosity threshold selected by the redMaPPer algorithm \citep{Rykoff2014RedMaPPer.Catalog}. 
These galaxies have excellent photometric redshifts with an approximately Gaussian scatter of $\sigma_z/(1+z) < 0.02$ and cover 
the entire redshift range of DMASS within the full DES Y1 footprint.
%the entire DES Y1 footprint. 
This makes them suitable as a reference sample to evaluate the redshift distribution of the DMASS sample.
%We choose the higher luminosity \redmagic sample selected above a luminosity threshold of $L > 1.5L_*$.
We opt for the higher luminosity \redmagic sample selected above a luminosity threshold of $L > 1.5L_*$ because the sample's redshifts reach up to $0.9$.

\begin{figure}
\centering
~\\
~\\
\includegraphics[width=0.45\textwidth]{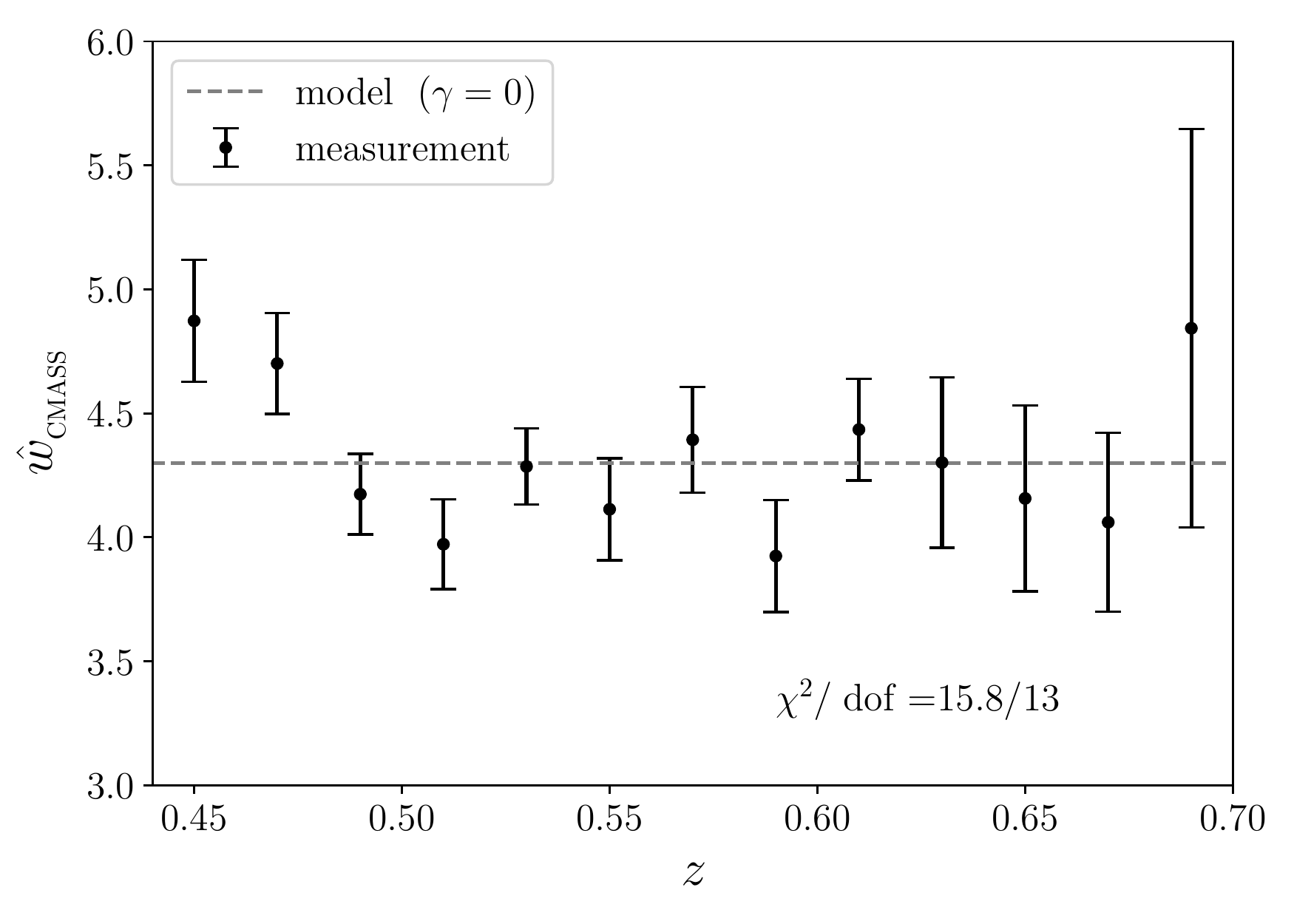}
\caption{ 
Integrated auto-correlations (Equation \ref{eq:w_ur_estimator}) of the CMASS SGC sample (black points).  
The grey-dashed line is the redshift evolution model $(1+z)^{\gamma}$ with $\gamma = 0$ (Equation \ref{eq:redshift_evolution_model}).
%We fit the auto-correlations by the redshift evolution model $(1+z)^{\gamma}$ in Equation \eqref{eq:redshift_evolution_model}. 
The value of $\chi^2$ between the model and the measurement is 15.8 for 13 data points, which indicates that the measurement is well consistent with the model. }
\label{fig:w_ur_cmass}
\end{figure}

To obtain the redshift distribution of the unknown sample, we split the reference sample in narrow redshift bins, $\Delta z = 0.02$, and measure cross-correlations between the galaxies in each redshift bin and the unknown sample. The cross-correlation for the $i$th redshift bin measures the quantity:
%for each subsample i we measure its angular or spa- tial correlations with the unknown population
\bea
w_{ur}(z_i) = \int^{R_{\rm max}}_{R_{\rm min}} \ud R~ \int \ud z' ~b_u(R,z') b_r(R,z_i) 
\label{eq:wur} \\
\times ~ n_u(z') n_r(z_i) \xi_m (R, z_i, z') 
\nonumber
\eea
where $b$ denotes galaxy bias of the unknown(`u') and reference(`r') samples, $n(z)$ stands for a normalized redshift distribution, $\xi_m$ stands for the matter-matter correlation function. $z_i$ is the $i$th redshift bin of the reference sample. $R$ is the comoving distance, $R = (1+z) D_A(z) \theta$. 
%\reph{The method counts unknown galaxies in annuli around each reference galaxy bounded by comoving distance $R = (1+z) D_A \theta$ from $R_{\rm min}$ to $R_{\rm max}$. } 
We adopt $R_{\rm min} = 500 \kpc$ and $R_{\rm max} = 1500 \kpc$ based on \cite{Gatti2018} and \cite{Davis2017}. %\cite{Gatti2018} demonstrated that in simulation clustering-z is largely scale-independent, with a decrease in signal-to-noise at larger scales. 
We also assume that the galaxy biases of both the reference and unknown sample are scale-independent on these scales.  \cite{Schmidt2013RecoveringBoundaries} demonstrated that implementing a linear bias model in the clustering-z does not significantly affect the methodology, even if these scales are non-linear. The same point was made in \cite{Gatti2018} and \cite{Cawthon2017DarkCross-Correlations} as well. For narrow redshift bins $n_r(z_i) = \delta(z-z_i)$, Equation \eqref{eq:wur} is simplified as 
\bea
w_{ur}(z_i) = n_u(z_i) b_u(z_i)  b_r(z_i) w_m(z_i)
\eea
where $w_m$ is the integrated matter-matter correlation function between $R_{\rm min}$ and $R_{\rm max}$. 
The cross-correlation was measured with the estimator from \cite{DavisPeebles1983} as follows:
%We use the estimator from \REF{DavisPeebles1983} to measure cross-correlations as follow:
\bea
\hat{w}_{ur}(z) = \frac{N_{R_r}}{{N_{D_r}}} \frac{ \int^{R_{\rm max}}_{R_{\rm min}}  \ud R ~W(R) [D_u D_r(R,z)] }{ \int^{R_{\rm max}}_{R_{\rm min}} \ud R~ W(R) [D_u R_r(R,z)] }-1 
\label{eq:w_ur_estimator}
\eea
where $W(R)$ is a weighting function, $D_u D_r$ and $D_u R_r$ stand for the number of galaxy-galaxy and galaxy-random pairs, $N_{R_r}$ and $N_{D_r}$ stand for the total number of randoms and galaxies of the reference sample. With the measured cross-correlation, the redshift distribution of the unknown sample is given as 
\bea
n_u (z) \propto \frac{\hat{w}_{ur} (z) }{ b_u(z) b_r (z) \hat{w}_m(z) }.
\label{eq:nz_unknown}
\eea
%\sout{The galaxy bias evolution of the reference sample $b_r$ can be obtained from the measured auto-correlation of the reference sample as follows: }
%The galaxy bias evolution of the reference sample $b_r$ can be obtained from the measured auto-correlation of the reference sample. 
If the redshift bins are sufficiently narrow so the biases and matter-matter correlations can be considered to be constant in each bin, the auto-correlation of the reference and unknown samples are given as
\bea
\hat{w}_{rr}(z) = b_r(z)^2 \hat{w}_m (z)~, \\
\hat{w}_{uu}(z) = b_u(z)^2 \hat{w}_m (z)~.
\eea
Then Equation \eqref{eq:nz_unknown} is re-written as 
\bea
n_u (z) \propto \frac{\hat{w}_{ur} (z) }{ \sqrt{\hat{w}_{uu}(z) \hat{w}_{rr}(z) }}~.
\eea
For the redshift evolution of $\hat{w}_{uu}$, we adopt a power law parametrization \citep{Cawthon2017DarkCross-Correlations}:
\bea
\sqrt{\hat{w}_{uu}(z)} \propto (1 + z)^{\gamma}~.
\label{eq:redshift_evolution_model}
\eea
Since we do not have access to the true redshifts of the DMASS galaxies, we infer the redshift evolution of $\hat{w}_{uu}$ from the auto-correlations of the CMASS galaxies using their spectroscopic redshfits (see Figure \ref{fig:w_ur_cmass}). 
Based on the nearly constant $\hat{w}$ of CMASS, we adopt $\gamma = 0$ for DMASS. 

\begin{figure}
%\includegraphics[width=0.6\textwidth]{./figures/dmass_spt_deviation}
% (CMH) commented out so this will compile
\centering
\includegraphics[width=0.5\textwidth]{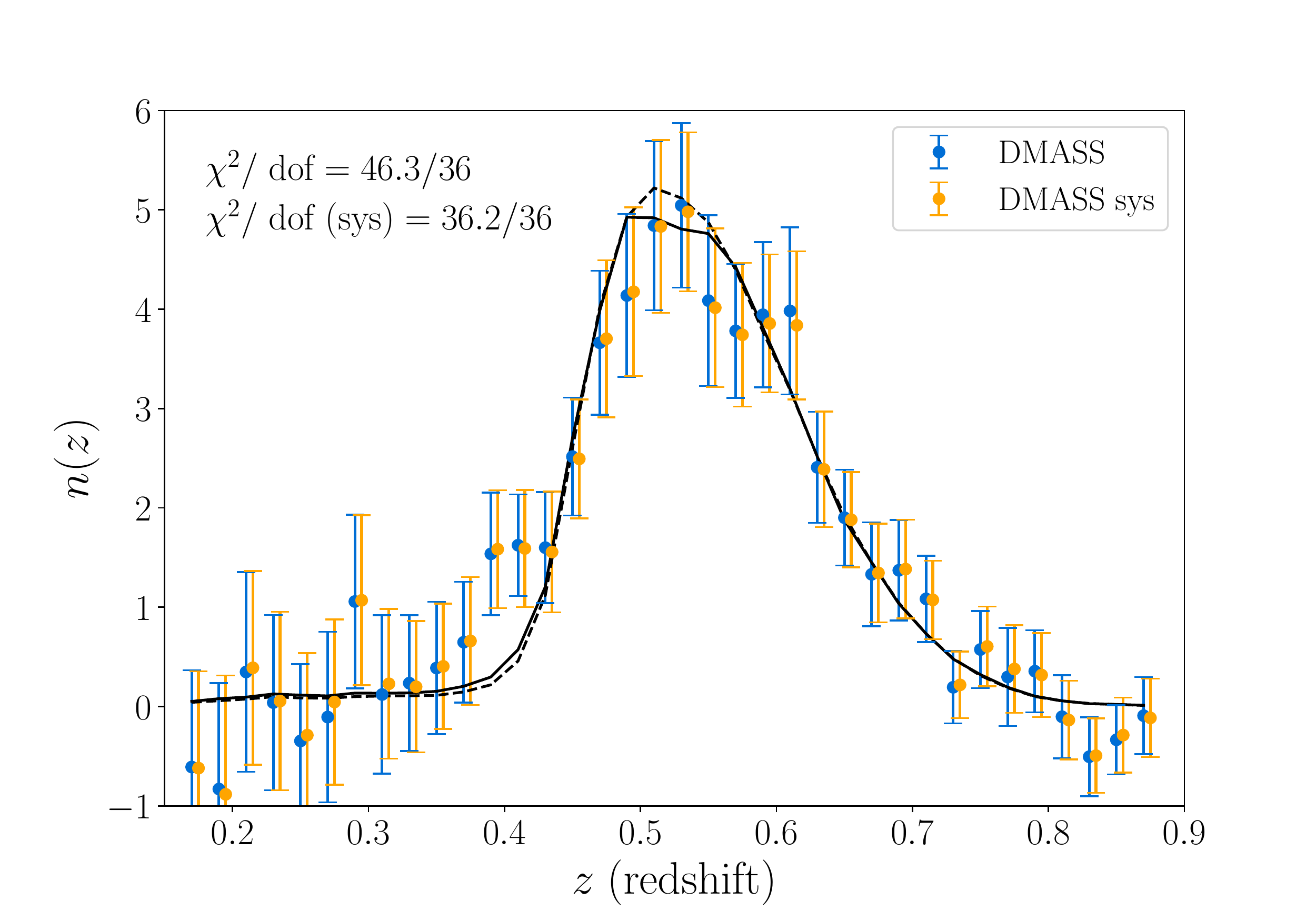}
\caption{ Redshift distribution of DMASS (blue) and DMASS with systematic weights (orange) recovered by the  clustering-z method with redMaGiC. The solid black and dashed lines show the spectroscopic redshift distribution of CMASS SGC and full CMASS. 
%The value of $\chi^2/{\rm dof}$ obtained against CMASS SGC is less than 1.0.
	 }
\label{fig:dmass_nz}
\end{figure}

Figure \ref{fig:dmass_nz} shows the result obtained from clustering-z. We find an excellent agreement between the clustering-z distribution of DMASS (blue points with error bars) and the spectroscopic redshift of CMASS SGC (solid black curve). The $\chi^2$ obtained when comparing the two is $46.3$ for 36 data points. With the systematic weights, the value is $36.2$ for 36 data points. 
We conclude that the clustering-z method returns a $n(z)$ for DMASS that is consistent with the BOSS SGC $n(z)$.

\subsection{ Difference in Galaxy Bias}
\label{sec:result-bias}
%\notesj{better title?}
\notesj{galaxy bias - mean galaxy bias. See Salazar-Albornoz et al. }
%\notesj{Why we constrain galaxy bias?}
%\cmt{Why we do this} 
In this section, we present the constraint on the mean galaxy bias difference between DMASS and CMASS derived from the combination of different probes aforementioned. 
Due to the weaker constraining power of the cross-correlation functions compared to the auto-correlation function, we utilize only the auto-correlation function (Section~\ref{sec:result-angular}) and the clustering-z distribution (Section~\ref{sec:clustering-z}) in this section. 
\notesj{ quantitative numbers }

To constrain the shifts in galaxy bias and redshift distribution compared to CMASS, we model the angular correlation function as follows: 
\begin{flalign}
&w^{\delta_g \delta_g}(\theta, b, \Delta b, \Delta z ) \nonumber \\
&~~~~~ = \int \ud z~f(z, b, \Delta b) \int \ud z'~ f(z', b, \Delta b) ~\xi_m (R, z, z')
\label{eq:modelw}
\end{flalign}
with
\bea
f(z, b, \Delta b) = (b + \Delta b)~ n(z + \Delta z) 
\eea
where $\xi_m$ is the matter angular correlation function, $R$ is the comoving distance defined as $R = (1+z) D_A(z) \theta$, $n(z)$ is the normalized redshift distribution, $b$ is galaxy bias. The galaxy bias of CMASS is known to be nearly a constant within the redshift $0.43 < z < 0.7$, so we do not consider redshift evolution of the galaxy bias \citep{Salazar-albornoz2018TheImplications}.
$\Delta b$ and $\Delta z$ are shifts in the galaxy bias and the redshift distribution from fiducial quantities. For CMASS, $\Delta b$ and $\Delta z$ are set to zero. 
Then, the residuals of the angular correlations of CMASS and DMASS is defined as
\begin{flalign}
& \Delta w^{\delta_g \delta_g} (\theta, b, \Delta b, \Delta z) 
\nonumber \\
&~~~~~= w^{\delta_g \delta_g} (\theta, b, \Delta b, \Delta z) - w^{\delta_g \delta_g} (\theta,b,0,0)~.
\label{eq:residual}
\end{flalign}

%To constrain the shifts in galaxy bias and redshift distribution from CMASS, we model residuals of angular correlation functions of CMASS and DMASS as follows: 
%\bea
%\Delta w^{\delta_g \delta_g} (\theta, \Delta b, \Delta z) = w^{\delta_g \delta_g} (\theta, \Delta b, \Delta z) - w^{\delta_g \delta_g} (\theta,0,0)~,
%\label{eq:residual}
%\eea
%with the angular correlation function defined as 
%\bea
%\omega_g(\theta, \Delta b, \Delta z ) = \int \ud z ~(b_g+\Delta b_g)^2 ~ n_g(z + \Delta z)^2 ~\xi_m (\chi\theta, z)~,
%\eea
%where $\xi_m$ is the matter angular correlation function, $\Delta b$ and $\Delta z$ imply a shift in galaxy bias and the redshift distribution from fiducial quantities. 
%The subscripts `C' and `D' in Equation \ref{eq:residual} denote CMASS and DMASS respectively. 
%For CMASS, $\Delta b$ and $\Delta z$ are set to zero. }
%

We also model the residuals of the redshift distributions to constrain the redshift shift $\Delta z$ independently with the clustering-z measurement in the previous section. The residual model is given as 
\bea
\Delta n(z, \Delta z) =n(z + \Delta z) -  n(z)
\label{eq:residual_nz}
\eea
%where the subscripts `C' and `D' represent CMASS and DMASS, and 
where $\Delta z$ is the same parameter shown in Equation \eqref{eq:modelw}. We use the spectroscopic redshift distribution of CMASS as the true distribution.  

Using a combination of the residuals of the angular correlation (Section \ref{sec:wg}) and clustering-z (Section \ref{sec:clustering-z}) measurements, we perform 
Markov Chain Monte-Carlo likelihood analyses %a maximum likelihood analysis 
to constrain the parameter set of $\{b, \Delta b, \Delta z\}$.
%to constrain parameters $b$, $\Delta b$ and $\Delta z$. 
The likelihood of the combined cosmological probe is given by the sum of individual log likelihoods given as
\bea
\ln \mathcal{L}(\boldsymbol{p}) = - \frac{1}{2} \left( \chi^2_{w^{\delta_g \delta_g}} (\boldsymbol{p}) + \chi^2_{\rm n(z)} (\boldsymbol{p}) \right)
\eea
where $\boldsymbol{p}$ is the set of varied parameters. 
We estimate $\chi^2$ defined in Equation \eqref{eq:chi2}. The data vector $\Delta {\bf{d}}$ is the difference between the measurement and theoretical prediction given as $\Delta {\bf{d}} = {\bf{d}}_{\rm true }- {\bf{d}}$. Equations \eqref{eq:residual} and \eqref{eq:residual_nz} are adopted as the true data vector $\bf{d}_{\rm true }$ for $\chi^2_{w^{\delta_g \delta_g}}$ and $\chi^2_{\rm n(z)}$, respectively. Residuals of the measurements between CMASS and DMASS are used as an input data vector $\bf{d}$ for a corresponding probe as well. 
%
\iffalse
$\chi^2$ is defined as 
\bea
\chi^2 = \sum_{i,j}^{N_{\rm bins}}( {\bf{d}}_{\rm true} - {\bf{d}} )_i^{T}  ({\bf{C}}^{-1})_{ij} ( {\bf{d}}_{\rm true} - {\bf{d}})_j. 
\label{eq:chi2b}
\eea
Equations \eqref{eq:residual} and \eqref{eq:residual_nz} are adopted as the true data vector ${\bf{d}}^T$ for $\chi^2_{w^{\delta_g \delta_g}}$ and $\chi^2_{\rm n(z)}$, respectively. Residuals of the measurements between CMASS and DMASS are used as an input data vector $\bf{d}$ for a corresponding probe as well. 
\fi
%
The covariance matrix for the angular correlation probe is given as the sum of the CMASS and DMASS covariance matrices:
\bea
{\bf{C}}_{w^{\delta_g \delta_g}} = {\bf{C}}_{ \rm DMASS} + {\bf{C}}_{\rm CMASS}~, 
\eea
and the covariance matrix for the clustering-z probe $\bf{C}_{\rm n(z)}$ is obtained from the clustering-z calculation in Section \ref{sec:clustering-z}. 
%The diagonal components of this covariance matrix is shown in Figure \ref{fig:dmass_nz}.
To evaluate the likelihood values and matter power spectrum for a given cosmology, we use the DES analysis pipeline in CosmoSIS \citep{COSMOSIS}. 
Further details of the likelihood framework are described in \cite{Krause2017}. 

\begin{figure}
%\includegraphics[width=0.6\textwidth]{./figures/dmass_spt_deviation}
% (CMH) commented out so this will compile
%\left
%\includegraphics[width=0.45\textwidth]{./figures/emcee_bias_nz.pdf}
%\includegraphics[width=0.45\textwidth]{./figures/emcee_nz_b_db.pdf}
\includegraphics[width=0.45\textwidth]{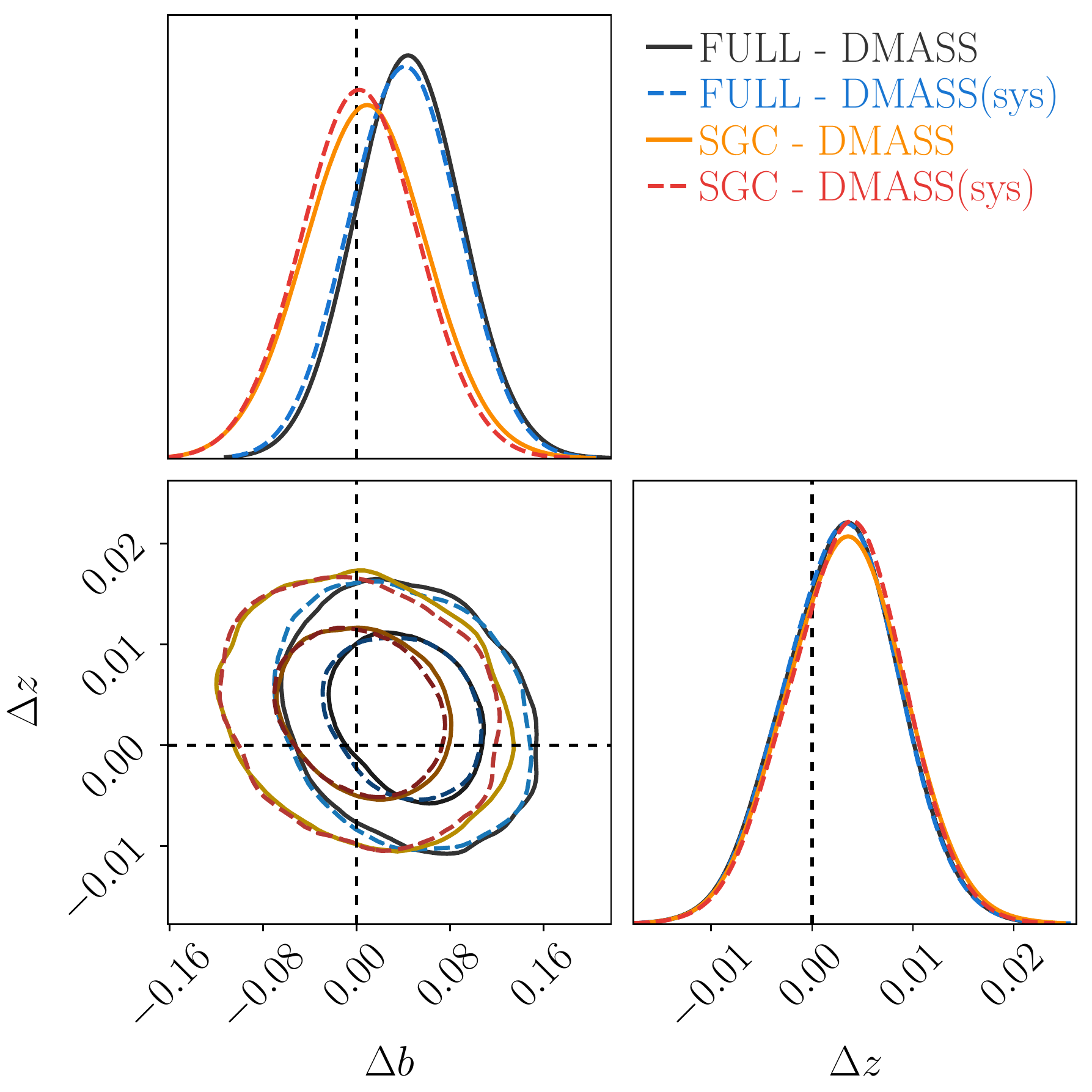}
\caption{ 
Constraints on the galaxy bias shift $\Delta b$ and redshift distribution shift $\Delta z$ from combination of the auto-angular correlation function and clustering-z. 
The dashed vertical and horizontal lines show the ideal case where DMASS is perfectly matched with CMASS. 
Orange-solid and red-dashed contours show shifts from the values of CMASS SGC. Black-solid and Blue-dashed contours show shifts from the values of full CMASS. The DMASS systematic weights are added for dashed contours. Adding systematic weights has very little impact on galaxy bias and redshift distributions.
%Because of the strong degeneracy between galaxy bias and redshift distribution, angular correlation function only shows extremely stretched contours in the diagonal direction. Adding clustering-z breaks the degeneracy and results in 
%$\Delta b = \dbsgc$ for CMASS SGC (orange) and $\Delta b = \dbfull$ for full CMASS (black). 
%The uncertainty of the galaxy bias shift for CMASS SGC is $\sigma_{\Delta b} \sim 0.08$, which is about $4\%$ of the fiducial value.
%Dashed contours are obtained with the DMASS systematic weights. 
}
\label{fig:bias_final}
\end{figure}

Figure \ref{fig:bias_final} shows the constraints of the bias shift $\Delta b$ and redshift shift $\Delta z$.
% from the combination of the angular correlation function and clustering-z. 
%The blue and black contours show the constraints of galaxy bias shift from CMASS SGC and full CMASS respectively when the redshift distribution is marginalized. The extremely stretched contours imply that uncertainties of galaxy bias and redshift distribution are strongly degenerate, therefore, extra information is needed to break the degeneracy. The green contour is only constraining $\Delta z$ with DMASS clustering-z. The results shows $\Delta z = \dzonly$ for CMASS SGC as a fiducial sample. 
The orange-solid and red-dashed contours show shifts $\Delta b$ and $\Delta z$ of DMASS when the values of CMASS SGC are fiducial. The black-solid and blue-dashed contours present shifts $\Delta b$ and $\Delta z$ of DMASS when full CMASS is used as fiducial. Dashed contours of both cases are obtained with the systematic weights of DMASS. 
The resulting numbers are $\Delta b = \dbsgc$ and $\Delta z = \dzsgc$  between CMASS SGC and DMASS, and $\Delta b = \dbfull$ and $\Delta z = \dzfull$ between full CMASS and DMASS. Since adding systematic weights has a negligible effect on numbers as shown in Figure \ref{fig:bias_final}, we do not report the results separately. As expected, DMASS has a better agreement with CMASS in SGC.
% but achieves $1\sigma$ deviations for both cases. }
%The red and orange contours show the combination of the two probes, angular correlation function and clustering-z. The resulting numbers for DMASS are $\Delta b = \dbsgc$ and $\Delta z = \dzsgc$ against CMASS SGC, and $\Delta b = \dbfull$ and $\Delta z = \dzfull$ against full CMASS. As expected, DMASS has a better agreement with CMASS in SGC but achieves $1\sigma$ deviations for both cases. \cmt{ \sout{ , despite the correlation function comparison with full CMASS is the one showing the least agreement among all comparison tests performed in this paper. }}
%\cmt{Note that this result with full CMASS is a pessimistic case because other than the clustering-z distributions, we only adopt the auto-correlation functions that shows the least agreement among all comparison tests with full CMASS performed in this paper. 
The resulting constraints of $\Delta b$ show that the mean galaxy bias of DMASS is consistent with both CMASS samples within $1\sigma$.
Moreover, $\Delta b$ between DMASS and full CMASS is  comparable to $2.6\%$ of the intrinsic difference in CMASS between the SGC and NGC shown in Appendix \ref{app:ngc_sgc}. 
%This indicates that the resulting DMASS sample has a strong potential as lenses for an alternative of CMASS where BOSS CMASS is not available. 

%The resulting constraints of $\Delta b$ is within $1\sigma$ of the true value for both cases which is safely below the error budget of CMASS galaxy bias $10\%$ with a combination of the structure growth rate $f$ \citep{Chuang2017} and also comparable to the $2.6\%$ intrinsic difference in CMASS between the SGC and NGC shown in the appendix \ref{app:ngc_sgc}. 
In this work, 
%we do not consider the redshift uncertainties of the \redmagic samples, which are known to be 
we do not consider the redshift bin biases and their uncertainties of the \redmagic samples, which are known to be  
%$\Delta z = ( 0.008, -0.005, 0.006, 0.0, 0.0 )$ 
$\Delta z = ( 0.010, -0.004, -0.004 )$ 
and   
%$\sigma_{\Delta z} = (0.007, 0.007, 0.006, 0.010, 0.010)$ 
$\sigma_{\Delta z} = (0.011, 0.010, 0.008)$
for three redshift bins from $z = 0.15$ to $z=0.6$ \citep{Cawthon2017DarkCross-Correlations}.
%These uncertainties are obtained from the clustering-z method relying on only $\sim10\%$ of the DES \redmagic sample that overlaps BOSS CMASS. Since we utilize the full \redmagic sample, we expect the impact of the redshift uncertainties are \sout{negligible} \cmt{rather smaller than the reported uncertainties.  }
Including the redshift uncertainties as priors would widen the final contours, but still keep the final constraints consistent with CMASS as all the biases are within $1\sigma$. Future analyses using DMASS will likely need to use a similar prior on the redshift bias for DMASS as used for DES \redmagic in DES Y1 \citep{ELVINPOOLE}.
%\notesj{ Negligible? How much negligible? proportional to the survey area?  }
\notejep{Should be $\sigma_{\Delta z}$? \cmtanswer{-- Done} }
\notejep{Should make it clear that these are from clustering-z and the values depend on the spectra available. This is not a quantity intrinsic to the redmagic sample. In reality, redmagic redshifts might be better than this. \cmtanswer{--done. See the paragraph above}}

%{\bf AJR didn't do anything in this section, will wait until it is more complete}

%Goal : bias of DMASS sample (obtained from ggl) should be within error budget of CMASS cl + Planck when the uncertainty of lowz fraction is marginalized.
%\bea
%b_{CMASS,cl} - b_{DMASS,ggl} < \sigma_{b, planck + CMASS}
%\eea

%\input{redshift}
\section{Conclusion}
\label{sec:conclusion}
%\notesj{flow is bad. rewrite}
In this paper, we constructed a catalog of DES galaxies from the full footprint of DES, whose statistical properties match those of the BOSS DR12 CMASS galaxy sample. We developed an algorithm for probabilistic target selection that uses density estimation in color and magnitude spaces. 
%We used the DES photometry from the overlapping area between the DES and BOSS footprints to train and validate the algorithm. 
The algorithm was trained and validated by the DES photometry from the overlapping area between the DES and BOSS footprints. 
From the distribution of the input DES galaxies in the overlapping region, the algorithm 
%derives the underlying distribution  and 
predicts an observed distribution that the same kind of galaxies would have in the target region.
% based on the measurement uncertainties of the area. 
A membership probability calculated based on the predicted observation was assigned to each source in the DES Y1 GOLD catalog. By weighting galaxies by their assigned probability, the resulting DMASS sample mimics the noise level the original CMASS sample has.

We showed that the resulting DMASS catalog matches well with both the SGC subset of CMASS as well as the full CMASS sample in various aspects: the number density, auto-angular correlation function, cross-angular correlation function with other full sky surveys and redshift distribution. 
%We also quote the 
%The final galaxy bias constraints we obtained by combining the angular correlation function and clustering-z are $\Delta b = \dbsgc$ for CMASS SGC and $\Delta b = \dbfull$ for full CMASS. 
%The final galaxy bias constraints we obtained by combining the angular correlation function and clustering-z
We determined differences in galaxy bias and shifts in the redshift distribution between DMASS and other CMASS samples by combining the angular correlation function and redshift distribution from the clustering-z method.   
The resulting constraints of $\Delta b$ show that the galaxy bias of the DMASS sample is consistent with both CMASS samples within $1\sigma$.
Furthermore, $\Delta b$ between DMASS and full CMASS is  comparable to the $2.6\%$ intrinsic difference of CMASS between the SGC and NGC regions.

The resulting DMASS sample can be used in cosmological analyses in various ways. The most promising application is using DMASS as a lens sample for galaxy-galaxy lensing.
\iffalse
\sjcmt{put more detail or delete }
\cmt{The sample can also be used as a reference sample for photometric redshift training. %Since weak gravitational lensing became a principal probe for cosmological analysis, computing precise photometric redshift has become more important for the various image-based surveys that are ongoing or planned in the near future.
Current photometric redshift algorithms depend on a small subset of spectroscopic galaxies from a minimal overlap area. Having a reference sample in the full DES footprint gives enormous statistical power to compute photometric redshifts. }
\fi
Beyond the sample used in our work, 
%Not only for the sample, 
the probabilistic technique used for this work can be easily applied to other image-based and spectroscopic surveys to identify another CMASS-like sample or other specific types of samples. 
%The model only uses a few color sets of galaxies and takes into account for spatial variance automatically.
%
%\cmt{\sout{For future surveys that will have vast observing areas such as DESI and LSST, this method will be a promising way to utilize complementary information between imaging and spectroscopic surveys.}}
Future surveys such as LSST \citep{LSST} can be a great application for this novel approach as the survey footprint of LSST occupies the entire southern sky but has only a small overlapping area with spectroscopic surveys such as eBOSS or DESI \citep{DESICollaboration2016} that view the northern sky. Producing a spectroscopic galaxy sample with the LSST imaging will enable us to utilize almost the entire sky and yield a wealth of information on the accelerated expansion of the Universe.

\section*{Acknowledgements}

The figures in this work are produced with plotting routines from matplotlib \citep{matplotlib} and ChainConsumer \citep{Hinton2016}.
Some of the results in this paper have been derived using the healpy and HEALPix package \citep{healpy}.

We thank D. H. Weinberg and C. Chuang for useful conversations in the course of preparing this work. 

% Personal ack
AC acknowledges support from NASA grant 15-WFIRST15-0008. During the preparation of this paper, C.H.\ was supported by the Simons Foundation, NASA, and the US Department of Energy.

% DES ack
Funding for the DES Projects has been provided by the U.S. Department of Energy, the U.S. National Science Foundation, the Ministry of Science and Education of Spain, 
the Science and Technology Facilities Council of the United Kingdom, the Higher Education Funding Council for England, the National Center for Supercomputing 
Applications at the University of Illinois at Urbana-Champaign, the Kavli Institute of Cosmological Physics at the University of Chicago, 
the Center for Cosmology and Astro-Particle Physics at the Ohio State University,
the Mitchell Institute for Fundamental Physics and Astronomy at Texas A\&M University, Financiadora de Estudos e Projetos, 
Funda{\c c}{\~a}o Carlos Chagas Filho de Amparo {\`a} Pesquisa do Estado do Rio de Janeiro, Conselho Nacional de Desenvolvimento Cient{\'i}fico e Tecnol{\'o}gico and 
the Minist{\'e}rio da Ci{\^e}ncia, Tecnologia e Inova{\c c}{\~a}o, the Deutsche Forschungsgemeinschaft and the Collaborating Institutions in the Dark Energy Survey. 

The Collaborating Institutions are Argonne National Laboratory, the University of California at Santa Cruz, the University of Cambridge, Centro de Investigaciones Energ{\'e}ticas, 
Medioambientales y Tecnol{\'o}gicas-Madrid, the University of Chicago, University College London, the DES-Brazil Consortium, the University of Edinburgh, 
the Eidgen{\"o}ssische Technische Hochschule (ETH) Z{\"u}rich, 
Fermi National Accelerator Laboratory, the University of Illinois at Urbana-Champaign, the Institut de Ci{\`e}ncies de l'Espai (IEEC/CSIC), 
the Institut de F{\'i}sica d'Altes Energies, Lawrence Berkeley National Laboratory, the Ludwig-Maximilians Universit{\"a}t M{\"u}nchen and the associated Excellence Cluster Universe, 
the University of Michigan, the National Optical Astronomy Observatory, the University of Nottingham, The Ohio State University, the University of Pennsylvania, the University of Portsmouth, 
SLAC National Accelerator Laboratory, Stanford University, the University of Sussex, Texas A\&M University, and the OzDES Membership Consortium.

Based in part on observations at Cerro Tololo Inter-American Observatory, National Optical Astronomy Observatory, which is operated by the Association of 
Universities for Research in Astronomy (AURA) under a cooperative agreement with the National Science Foundation.

The DES data management system is supported by the National Science Foundation under Grant Numbers AST-1138766 and AST-1536171.
The DES participants from Spanish institutions are partially supported by MINECO under grants AYA2015-71825, ESP2015-66861, FPA2015-68048, SEV-2016-0588, SEV-2016-0597, and MDM-2015-0509, 
some of which include ERDF funds from the European Union. IFAE is partially funded by the CERCA program of the Generalitat de Catalunya.
Research leading to these results has received funding from the European Research
Council under the European Union's Seventh Framework Program (FP7/2007-2013) including ERC grant agreements 240672, 291329, and 306478.
We  acknowledge support from the Brazilian Instituto Nacional de Ci\^encia
e Tecnologia (INCT) e-Universe (CNPq grant 465376/2014-2).

This manuscript has been authored by Fermi Research Alliance, LLC under Contract No. DE-AC02-07CH11359 with the U.S. Department of Energy, Office of Science, Office of High Energy Physics. The United States Government retains and the publisher, by accepting the article for publication, acknowledges that the United States Government retains a non-exclusive, paid-up, irrevocable, world-wide license to publish or reproduce the published form of this manuscript, or allow others to do so, for United States Government purposes.

% SDSS-III ack
Funding for SDSS-III has been provided by the Alfred P. Sloan Foundation, the Participating Institutions, the National Science Foundation, and the U.S. Department of Energy Office of Science. The SDSS-III web site is \url{http://www.sdss3.org/}.

SDSS-III is managed by the Astrophysical Research Consortium for the Participating Institutions of the SDSS-III Collaboration including the University of Arizona, the Brazilian Participation Group, Brookhaven National Laboratory, Carnegie Mellon University, University of Florida, the French Participation Group, the German Participation Group, Harvard University, the Instituto de Astrofisica de Canarias, the Michigan State/Notre Dame/JINA Participation Group, Johns Hopkins University, Lawrence Berkeley National Laboratory, Max Planck Institute for Astrophysics, Max Planck Institute for Extraterrestrial Physics, New Mexico State University, New York University, Ohio State University, Pennsylvania State University, University of Portsmouth, Princeton University, the Spanish Participation Group, University of Tokyo, University of Utah, Vanderbilt University, University of Virginia, University of Washington, and Yale University.

This work used resources at the Owens Cluster at the Ohio Supercomputer Center  \citep{OSC}.

\bibliography{main.bib}

\begin{thebibliography}{}
\makeatletter
\relax
\def\mn@urlcharsother{\let\do\@makeother \do\$\do\&\do\#\do\^\do\_\do\%\do\~}
\def\mn@doi{\begingroup\mn@urlcharsother \@ifnextchar [ {\mn@doi@}
  {\mn@doi@[]}}
\def\mn@doi@[#1]#2{\def\@tempa{#1}\ifx\@tempa\@empty \href
  {http://dx.doi.org/#2} {doi:#2}\else \href {http://dx.doi.org/#2} {#1}\fi
  \endgroup}
\def\mn@eprint#1#2{\mn@eprint@#1:#2::\@nil}
\def\mn@eprint@arXiv#1{\href {http://arxiv.org/abs/#1} {{\tt arXiv:#1}}}
\def\mn@eprint@dblp#1{\href {http://dblp.uni-trier.de/rec/bibtex/#1.xml}
  {dblp:#1}}
\def\mn@eprint@#1:#2:#3:#4\@nil{\def\@tempa {#1}\def\@tempb {#2}\def\@tempc
  {#3}\ifx \@tempc \@empty \let \@tempc \@tempb \let \@tempb \@tempa \fi \ifx
  \@tempb \@empty \def\@tempb {arXiv}\fi \@ifundefined
  {mn@eprint@\@tempb}{\@tempb:\@tempc}{\expandafter \expandafter \csname
  mn@eprint@\@tempb\endcsname \expandafter{\@tempc}}}

\bibitem[\protect\citeauthoryear{{Abbott} et~al.,}{{Abbott}
  et~al.}{2018a}]{DESCollaboration2017}
{Abbott} T.~M.~C.,  et~al., 2018a, \mn@doi [\prd] {10.1103/PhysRevD.98.043526},
  \href {https://ui.adsabs.harvard.edu/\#abs/2018PhRvD..98d3526A} {98, 043526}

\bibitem[\protect\citeauthoryear{{Abbott} et~al.,}{{Abbott}
  et~al.}{2018b}]{DESDR1}
{Abbott} T.~M.~C.,  et~al., 2018b, \mn@doi [The Astrophysical Journal
  Supplement Series] {10.3847/1538-4365/aae9f0}, \href
  {https://ui.adsabs.harvard.edu/\#abs/2018ApJS..239...18A} {239, 18}

\bibitem[\protect\citeauthoryear{{Aihara} et~al.,}{{Aihara}
  et~al.}{2011}]{SDSSDR8}
{Aihara} H.,  et~al., 2011, \mn@doi [The Astrophysical Journal Supplement
  Series] {10.1088/0067-0049/193/2/29}, \href
  {https://ui.adsabs.harvard.edu/\#abs/2011ApJS..193...29A} {193, 29}

\bibitem[\protect\citeauthoryear{{Alam} et~al.,}{{Alam}
  et~al.}{2015}]{Alam2015TheSDSS-III}
{Alam} S.,  et~al., 2015, \mn@doi [The Astrophysical Journal Supplement Series]
  {10.1088/0067-0049/219/1/12}, \href
  {https://ui.adsabs.harvard.edu/\#abs/2015ApJS..219...12A} {219, 12}

\bibitem[\protect\citeauthoryear{{Alam}, {Miyatake}, {More}, {Ho}  \&
  {Mandelbaum}}{{Alam} et~al.}{2017}]{Alam2017TestingCMASS}
{Alam} S.,  {Miyatake} H.,  {More} S.,  {Ho} S.,   {Mandelbaum} R.,  2017,
  \mn@doi [\mnras] {10.1093/mnras/stw3056}, \href
  {https://ui.adsabs.harvard.edu/\#abs/2017MNRAS.465.4853A} {465, 4853}

\bibitem[\protect\citeauthoryear{{Amon} et~al.,}{{Amon}
  et~al.}{2018}]{Amon2018}
{Amon} A.,  et~al., 2018, \mn@doi [\mnras] {10.1093/mnras/sty1624}, \href
  {https://ui.adsabs.harvard.edu/\#abs/2018MNRAS.479.3422A} {479, 3422}

\bibitem[\protect\citeauthoryear{{Annis} et~al.,}{{Annis}
  et~al.}{2014}]{Annis2011The82}
{Annis} J.,  et~al., 2014, \mn@doi [\apj] {10.1088/0004-637X/794/2/120}, \href
  {https://ui.adsabs.harvard.edu/\#abs/2014ApJ...794..120A} {794, 120}

\bibitem[\protect\citeauthoryear{{Baldauf}, {Smith}, {Seljak}  \& {Mand
  elbaum}}{{Baldauf} et~al.}{2010}]{Baldauf2010}
{Baldauf} T.,  {Smith} R.~E.,  {Seljak} U.,   {Mand elbaum} R.,  2010, \mn@doi
  [\prd] {10.1103/PhysRevD.81.063531}, \href
  {https://ui.adsabs.harvard.edu/\#abs/2010PhRvD..81f3531B} {81, 063531}

\bibitem[\protect\citeauthoryear{{Bertin} \& {Arnouts}}{{Bertin} \&
  {Arnouts}}{1996}]{SEXTRACTOR}
{Bertin} E.,  {Arnouts} S.,  1996, \mn@doi [Astronomy and Astrophysics
  Supplement Series] {10.1051/aas:1996164}, \href
  {https://ui.adsabs.harvard.edu/\#abs/1996A&AS..117..393B} {117, 393}

\bibitem[\protect\citeauthoryear{{Bolton} et~al.,}{{Bolton}
  et~al.}{2012}]{Bolton2012SpectralSurvey}
{Bolton} A.~S.,  et~al., 2012, \mn@doi [\aj] {10.1088/0004-6256/144/5/144},
  \href {https://ui.adsabs.harvard.edu/\#abs/2012AJ....144..144B} {144, 144}

\bibitem[\protect\citeauthoryear{{Bovy}, {Hogg}  \& {Roweis}}{{Bovy}
  et~al.}{2011a}]{Bovy2011a}
{Bovy} J.,  {Hogg} D.~W.,   {Roweis} S.~T.,  2011a, \mn@doi [Annals of Applied
  Statistics] {10.1214/10-AOAS439}, \href
  {https://ui.adsabs.harvard.edu/\#abs/2011AnApS...5.1657B} {5, 1657}

\bibitem[\protect\citeauthoryear{{Bovy} et~al.,}{{Bovy}
  et~al.}{2011b}]{Bovy2011b}
{Bovy} J.,  et~al., 2011b, \mn@doi [\apj] {10.1088/0004-637X/729/2/141}, \href
  {https://ui.adsabs.harvard.edu/abs/2011ApJ...729..141B} {729, 141}

\bibitem[\protect\citeauthoryear{{Burke} et~al.,}{{Burke}
  et~al.}{2018}]{Burke2017ForwardSurvey}
{Burke} D.~L.,  et~al., 2018, \mn@doi [\aj] {10.3847/1538-3881/aa9f22}, \href
  {https://ui.adsabs.harvard.edu/\#abs/2018AJ....155...41B} {155, 41}

\bibitem[\protect\citeauthoryear{{Carlstrom} et~al.,}{{Carlstrom}
  et~al.}{2011}]{SPT}
{Carlstrom} J.~E.,  et~al., 2011, \mn@doi [Publications of the Astronomical
  Society of the Pacific] {10.1086/659879}, \href
  {https://ui.adsabs.harvard.edu/\#abs/2011PASP..123..568C} {123, 568}

\bibitem[\protect\citeauthoryear{{Cawthon} et~al.,}{{Cawthon}
  et~al.}{2018}]{Cawthon2017DarkCross-Correlations}
{Cawthon} R.,  et~al., 2018, \mn@doi [\mnras] {10.1093/mnras/sty2424}, \href
  {https://ui.adsabs.harvard.edu/\#abs/2018MNRAS.481.2427C} {481, 2427}

\bibitem[\protect\citeauthoryear{{Choi}, {Tyson}, {Morrison}, {Jee}, {Schmidt},
  {Margoniner}  \& {Wittman}}{{Choi} et~al.}{2012}]{Choi2012}
{Choi} A.,  {Tyson} J.~A.,  {Morrison} C.~B.,  {Jee} M.~J.,  {Schmidt} S.~J.,
  {Margoniner} V.~E.,   {Wittman} D.~M.,  2012, \mn@doi [\apj]
  {10.1088/0004-637X/759/2/101}, \href
  {http://adsabs.harvard.edu/abs/2012ApJ...759..101C} {759, 101}

\bibitem[\protect\citeauthoryear{{Choi} et~al.,}{{Choi}
  et~al.}{2016}]{2016MNRAS.463.3737C}
{Choi} A.,  et~al., 2016, \mn@doi [\mnras] {10.1093/mnras/stw2241}, \href
  {http://adsabs.harvard.edu/abs/2016MNRAS.463.3737C} {463, 3737}

\bibitem[\protect\citeauthoryear{{Chuang} et~al.,}{{Chuang}
  et~al.}{2017}]{Chuang2017}
{Chuang} C.-H.,  et~al., 2017, \mn@doi [\mnras] {10.1093/mnras/stx1641}, \href
  {https://ui.adsabs.harvard.edu/\#abs/2017MNRAS.471.2370C} {471, 2370}

\bibitem[\protect\citeauthoryear{{Crocce} et~al.,}{{Crocce}
  et~al.}{2016}]{Crocce2015GalaxyData}
{Crocce} M.,  et~al., 2016, \mn@doi [\mnras] {10.1093/mnras/stv2590}, \href
  {https://ui.adsabs.harvard.edu/\#abs/2016MNRAS.455.4301C} {455, 4301}

\bibitem[\protect\citeauthoryear{{Crocce} et~al.,}{{Crocce}
  et~al.}{2019}]{Y1BAO}
{Crocce} M.,  et~al., 2019, \mn@doi [\mnras] {10.1093/mnras/sty2522}, \href
  {https://ui.adsabs.harvard.edu/\#abs/2019MNRAS.482.2807C} {482, 2807}

\bibitem[\protect\citeauthoryear{{Cuesta} et~al.,}{{Cuesta}
  et~al.}{2016}]{Cuesta2016}
{Cuesta} A.~J.,  et~al., 2016, \mn@doi [\mnras] {10.1093/mnras/stw066}, \href
  {https://ui.adsabs.harvard.edu/abs/2016MNRAS.457.1770C} {457, 1770}

\bibitem[\protect\citeauthoryear{{DESI Collaboration} et~al.,}{{DESI
  Collaboration} et~al.}{2016}]{DESICollaboration2016}
{DESI Collaboration} et~al., 2016, arXiv e-prints, \href
  {https://ui.adsabs.harvard.edu/\#abs/2016arXiv161100036D} {p.
  arXiv:1611.00036}

\bibitem[\protect\citeauthoryear{{Dark Energy Survey Collaboration}
  et~al.,}{{Dark Energy Survey Collaboration} et~al.}{2016}]{DESOverview}
{Dark Energy Survey Collaboration} et~al., 2016, \mn@doi [\mnras]
  {10.1093/mnras/stw641}, \href
  {https://ui.adsabs.harvard.edu/\#abs/2016MNRAS.460.1270D} {460, 1270}

\bibitem[\protect\citeauthoryear{{Davis} \& {Peebles}}{{Davis} \&
  {Peebles}}{1983}]{DavisPeebles1983}
{Davis} M.,  {Peebles} P.~J.~E.,  1983, \mn@doi [\apj] {10.1086/160884}, \href
  {https://ui.adsabs.harvard.edu/\#abs/1983ApJ...267..465D} {267, 465}

\bibitem[\protect\citeauthoryear{{Davis} et~al.,}{{Davis}
  et~al.}{2017}]{Davis2017}
{Davis} C.,  et~al., 2017, arXiv e-prints, \href
  {https://ui.adsabs.harvard.edu/\#abs/2017arXiv171002517D} {p.
  arXiv:1710.02517}

\bibitem[\protect\citeauthoryear{{Dawson} et~al.,}{{Dawson}
  et~al.}{2013}]{Dawson2013TheSDSS-III}
{Dawson} K.~S.,  et~al., 2013, \mn@doi [\aj] {10.1088/0004-6256/145/1/10},
  \href {https://ui.adsabs.harvard.edu/\#abs/2013AJ....145...10D} {145, 10}

\bibitem[\protect\citeauthoryear{{Drlica-Wagner} et~al.,}{{Drlica-Wagner}
  et~al.}{2018}]{Y1GOLD}
{Drlica-Wagner} A.,  et~al., 2018, \mn@doi [The Astrophysical Journal
  Supplement Series] {10.3847/1538-4365/aab4f5}, \href
  {https://ui.adsabs.harvard.edu/\#abs/2018ApJS..235...33D} {235, 33}

\bibitem[\protect\citeauthoryear{{Eisenstein} et~al.,}{{Eisenstein}
  et~al.}{2001}]{Eisenstein2001LRG}
{Eisenstein} D.~J.,  et~al., 2001, \mn@doi [\aj] {10.1086/323717}, \href
  {https://ui.adsabs.harvard.edu/\#abs/2001AJ....122.2267E} {122, 2267}

\bibitem[\protect\citeauthoryear{{Eisenstein} et~al.,}{{Eisenstein}
  et~al.}{2011}]{Eisenstein2011BOSS}
{Eisenstein} D.~J.,  et~al., 2011, \mn@doi [\aj] {10.1088/0004-6256/142/3/72},
  \href {https://ui.adsabs.harvard.edu/\#abs/2011AJ....142...72E} {142, 72}

\bibitem[\protect\citeauthoryear{{Elsner}, {Leistedt}  \& {Peiris}}{{Elsner}
  et~al.}{2016}]{Elsner2016}
{Elsner} F.,  {Leistedt} B.,   {Peiris} H.~V.,  2016, \mn@doi [\mnras]
  {10.1093/mnras/stv2777}, \href
  {https://ui.adsabs.harvard.edu/abs/2016MNRAS.456.2095E} {456, 2095}

\bibitem[\protect\citeauthoryear{{Elvin-Poole} et~al.,}{{Elvin-Poole}
  et~al.}{2018}]{ELVINPOOLE}
{Elvin-Poole} J.,  et~al., 2018, \mn@doi [\prd] {10.1103/PhysRevD.98.042006},
  \href {https://ui.adsabs.harvard.edu/\#abs/2018PhRvD..98d2006E} {98, 042006}

\bibitem[\protect\citeauthoryear{{Flaugher} et~al.,}{{Flaugher}
  et~al.}{2015}]{Flaugher2015THECAMERA}
{Flaugher} B.,  et~al., 2015, \mn@doi [\aj] {10.1088/0004-6256/150/5/150},
  \href {https://ui.adsabs.harvard.edu/\#abs/2015AJ....150..150F} {150, 150}

\bibitem[\protect\citeauthoryear{{Foreman-Mackey}, {Hogg}, {Lang}  \&
  {Goodman}}{{Foreman-Mackey} et~al.}{2013}]{EMCEE}
{Foreman-Mackey} D.,  {Hogg} D.~W.,  {Lang} D.,   {Goodman} J.,  2013, \mn@doi
  [Publications of the Astronomical Society of the Pacific] {10.1086/670067},
  \href {https://ui.adsabs.harvard.edu/abs/2013PASP..125..306F} {125, 306}

\bibitem[\protect\citeauthoryear{{Frieman}, {Turner}  \& {Huterer}}{{Frieman}
  et~al.}{2008}]{FriemanReview08}
{Frieman} J.~A.,  {Turner} M.~S.,   {Huterer} D.,  2008, \mn@doi [Annual Review
  of Astronomy and Astrophysics] {10.1146/annurev.astro.46.060407.145243},
  \href {https://ui.adsabs.harvard.edu/\#abs/2008ARA&A..46..385F} {46, 385}

\bibitem[\protect\citeauthoryear{{Gatti} et~al.,}{{Gatti}
  et~al.}{2018}]{Gatti2018}
{Gatti} M.,  et~al., 2018, \mn@doi [\mnras] {10.1093/mnras/sty466}, \href
  {https://ui.adsabs.harvard.edu/\#abs/2018MNRAS.477.1664G} {477, 1664}

\bibitem[\protect\citeauthoryear{{Gil-Mar{\'\i}n} et~al.,}{{Gil-Mar{\'\i}n}
  et~al.}{2016a}]{Gil-Marin2016a}
{Gil-Mar{\'\i}n} H.,  et~al., 2016a, \mn@doi [\mnras] {10.1093/mnras/stw1096},
  \href {https://ui.adsabs.harvard.edu/abs/2016MNRAS.460.4188G} {460, 4188}

\bibitem[\protect\citeauthoryear{{Gil-Mar{\'\i}n} et~al.,}{{Gil-Mar{\'\i}n}
  et~al.}{2016b}]{Gil-Marin2016b}
{Gil-Mar{\'\i}n} H.,  et~al., 2016b, \mn@doi [\mnras] {10.1093/mnras/stw1264},
  \href {https://ui.adsabs.harvard.edu/abs/2016MNRAS.460.4210G} {460, 4210}

\bibitem[\protect\citeauthoryear{{G{\'o}rski}, {Hivon}, {Banday}, {Wandelt},
  {Hansen}, {Reinecke}  \& {Bartelmann}}{{G{\'o}rski} et~al.}{2005}]{HEALPix}
{G{\'o}rski} K.~M.,  {Hivon} E.,  {Banday} A.~J.,  {Wandelt} B.~D.,  {Hansen}
  F.~K.,  {Reinecke} M.,   {Bartelmann} M.,  2005, \mn@doi [\apj]
  {10.1086/427976}, \href {http://adsabs.harvard.edu/abs/2005ApJ...622..759G}
  {622, 759}

\bibitem[\protect\citeauthoryear{{Goto}, {Szapudi}  \& {Granett}}{{Goto}
  et~al.}{2012}]{Goto2011Cross-correlationBackground}
{Goto} T.,  {Szapudi} I.,   {Granett} B.~R.,  2012, \mn@doi [\mnras]
  {10.1111/j.1745-3933.2012.01240.x}, \href
  {https://ui.adsabs.harvard.edu/\#abs/2012MNRAS.422L..77G} {422, L77}

\bibitem[\protect\citeauthoryear{{Gunn} et~al.,}{{Gunn}
  et~al.}{1998}]{Gunn1998TheCamera}
{Gunn} J.~E.,  et~al., 1998, \mn@doi [\aj] {10.1086/300645}, \href
  {https://ui.adsabs.harvard.edu/\#abs/1998AJ....116.3040G} {116, 3040}

\bibitem[\protect\citeauthoryear{{Gunn} et~al.,}{{Gunn}
  et~al.}{2006}]{Gunn2006TheSurvey}
{Gunn} J.~E.,  et~al., 2006, \mn@doi [\aj] {10.1086/500975}, \href
  {https://ui.adsabs.harvard.edu/\#abs/2006AJ....131.2332G} {131, 2332}

\bibitem[\protect\citeauthoryear{{Heymans} et~al.,}{{Heymans}
  et~al.}{2012}]{CFHTLenS}
{Heymans} C.,  et~al., 2012, \mn@doi [\mnras]
  {10.1111/j.1365-2966.2012.21952.x}, \href
  {https://ui.adsabs.harvard.edu/\#abs/2012MNRAS.427..146H} {427, 146}

\bibitem[\protect\citeauthoryear{{Hinton}}{{Hinton}}{2016}]{Hinton2016}
{Hinton} S.~R.,  2016, \mn@doi [The Journal of Open Source Software]
  {10.21105/joss.00045}, \href
  {https://ui.adsabs.harvard.edu/abs/2016JOSS....1...45H} {1, 00045}

\bibitem[\protect\citeauthoryear{{Hirata} et~al.,}{{Hirata}
  et~al.}{2004}]{Hirata2004}
{Hirata} C.~M.,  et~al., 2004, \mn@doi [\mnras]
  {10.1111/j.1365-2966.2004.08090.x}, \href
  {https://ui.adsabs.harvard.edu/abs/2004MNRAS.353..529H} {353, 529}

\bibitem[\protect\citeauthoryear{{Hunter}}{{Hunter}}{2007}]{matplotlib}
{Hunter} J.~D.,  2007, \mn@doi [Computing in Science and Engineering]
  {10.1109/MCSE.2007.55}, \href
  {https://ui.adsabs.harvard.edu/abs/2007CSE.....9...90H} {9, 90}

\bibitem[\protect\citeauthoryear{{Huterer} \& {Shafer}}{{Huterer} \&
  {Shafer}}{2018}]{HutererReview181}
{Huterer} D.,  {Shafer} D.~L.,  2018, \mn@doi [Reports on Progress in Physics]
  {10.1088/1361-6633/aa997e}, \href
  {https://ui.adsabs.harvard.edu/\#abs/2018RPPh...81a6901H} {81, 016901}

\bibitem[\protect\citeauthoryear{{Jarvis}}{{Jarvis}}{2015}]{TreeCorr}
{Jarvis} M.,  2015, {TreeCorr: Two-point correlation functions} (\mn@eprint
  {ascl} {1508.007})

\bibitem[\protect\citeauthoryear{{Johnson} et~al.,}{{Johnson}
  et~al.}{2017}]{2017MNRAS.465.4118J}
{Johnson} A.,  et~al., 2017, \mn@doi [\mnras] {10.1093/mnras/stw3033}, \href
  {http://adsabs.harvard.edu/abs/2017MNRAS.465.4118J} {465, 4118}

\bibitem[\protect\citeauthoryear{{Jullo} et~al.,}{{Jullo}
  et~al.}{2019}]{Jullo2019}
{Jullo} E.,  et~al., 2019, arXiv e-prints, \href
  {https://ui.adsabs.harvard.edu/\#abs/2019arXiv190307160J} {p.
  arXiv:1903.07160}

\bibitem[\protect\citeauthoryear{{Kaiser}}{{Kaiser}}{1984}]{Kaiser1984}
{Kaiser} N.,  1984, \mn@doi [\apj] {10.1086/184341}, \href
  {https://ui.adsabs.harvard.edu/\#abs/1984ApJ...284L...9K} {284, L9}

\bibitem[\protect\citeauthoryear{{Kitaura} et~al.,}{{Kitaura}
  et~al.}{2016}]{Kitaura2016TheRelease}
{Kitaura} F.-S.,  et~al., 2016, \mn@doi [\mnras] {10.1093/mnras/stv2826}, \href
  {https://ui.adsabs.harvard.edu/\#abs/2016MNRAS.456.4156K} {456, 4156}

\bibitem[\protect\citeauthoryear{{Kovacs}, {Szapudi}, {Granett}  \&
  {Frei}}{{Kovacs} et~al.}{2013}]{Kovacs2012Cross-correlationRelease}
{Kovacs} A.,  {Szapudi} I.,  {Granett} B.~R.,   {Frei} Z.,  2013, \mn@doi
  [\mnras] {10.1093/mnrasl/slt002}, \href
  {https://ui.adsabs.harvard.edu/\#abs/2013MNRAS.431L..28K} {431, L28}

\bibitem[\protect\citeauthoryear{{Krause} \& {Eifler}}{{Krause} \&
  {Eifler}}{2017}]{COSMOLIKE}
{Krause} E.,  {Eifler} T.,  2017, \mn@doi [\mnras] {10.1093/mnras/stx1261},
  \href {https://ui.adsabs.harvard.edu/abs/2017MNRAS.470.2100K} {470, 2100}

\bibitem[\protect\citeauthoryear{{Krause} et~al.,}{{Krause}
  et~al.}{2017}]{Krause2017}
{Krause} E.,  et~al., 2017, arXiv e-prints, \href
  {https://ui.adsabs.harvard.edu/abs/2017arXiv170609359K} {p. arXiv:1706.09359}

\bibitem[\protect\citeauthoryear{{LSST Science Collaboration} et~al.,}{{LSST
  Science Collaboration} et~al.}{2009}]{LSST}
{LSST Science Collaboration} et~al., 2009, arXiv e-prints, \href
  {https://ui.adsabs.harvard.edu/\#abs/2009arXiv0912.0201L} {p.
  arXiv:0912.0201}

\bibitem[\protect\citeauthoryear{{Landy} \& {Szalay}}{{Landy} \&
  {Szalay}}{1993}]{Landy1993BiasFunctions}
{Landy} S.~D.,  {Szalay} A.~S.,  1993, \mn@doi [\apj] {10.1086/172900}, \href
  {https://ui.adsabs.harvard.edu/\#abs/1993ApJ...412...64L} {412, 64}

\bibitem[\protect\citeauthoryear{{Leistedt} et~al.,}{{Leistedt}
  et~al.}{2016}]{Leistedt2015}
{Leistedt} B.,  et~al., 2016, \mn@doi [The Astrophysical Journal Supplement
  Series] {10.3847/0067-0049/226/2/24}, \href
  {https://ui.adsabs.harvard.edu/\#abs/2016ApJS..226...24L} {226, 24}

\bibitem[\protect\citeauthoryear{{Lewis}, {Challinor}  \& {Lasenby}}{{Lewis}
  et~al.}{2000}]{CAMB}
{Lewis} A.,  {Challinor} A.,   {Lasenby} A.,  2000, \mn@doi [\apj]
  {10.1086/309179}, \href
  {https://ui.adsabs.harvard.edu/abs/2000ApJ...538..473L} {538, 473}

\bibitem[\protect\citeauthoryear{{Li} et~al.,}{{Li}
  et~al.}{2016}]{Li2016ASSESSMENTSURVEYS}
{Li} T.~S.,  et~al., 2016, \mn@doi [\aj] {10.3847/0004-6256/151/6/157}, \href
  {https://ui.adsabs.harvard.edu/\#abs/2016AJ....151..157L} {151, 157}

\bibitem[\protect\citeauthoryear{{Mandelbaum}}{{Mandelbaum}}{2018}]{Mandelbaum2017WeakCosmology}
{Mandelbaum} R.,  2018, \mn@doi [Annual Review of Astronomy and Astrophysics]
  {10.1146/annurev-astro-081817-051928}, \href
  {https://ui.adsabs.harvard.edu/\#abs/2018ARA&A..56..393M} {56, 393}

\bibitem[\protect\citeauthoryear{{Mandelbaum}, {Slosar}, {Baldauf}, {Seljak},
  {Hirata}, {Nakajima}, {Reyes}  \& {Smith}}{{Mandelbaum}
  et~al.}{2013}]{Mandelbaum2013}
{Mandelbaum} R.,  {Slosar} A.,  {Baldauf} T.,  {Seljak} U.,  {Hirata} C.~M.,
  {Nakajima} R.,  {Reyes} R.,   {Smith} R.~E.,  2013, \mn@doi [\mnras]
  {10.1093/mnras/stt572}, \href
  {https://ui.adsabs.harvard.edu/\#abs/2013MNRAS.432.1544M} {432, 1544}

\bibitem[\protect\citeauthoryear{{Maraston}, {Str{\"o}mb{\"a}ck}, {Thomas},
  {Wake}  \& {Nichol}}{{Maraston} et~al.}{2009}]{Maraston2009}
{Maraston} C.,  {Str{\"o}mb{\"a}ck} G.,  {Thomas} D.,  {Wake} D.~A.,   {Nichol}
  R.~C.,  2009, \mn@doi [\mnras] {10.1111/j.1745-3933.2009.00621.x}, \href
  {https://ui.adsabs.harvard.edu/\#abs/2009MNRAS.394L.107M} {394, L107}

\bibitem[\protect\citeauthoryear{{Miyatake} et~al.,}{{Miyatake}
  et~al.}{2015}]{Miyatake2015}
{Miyatake} H.,  et~al., 2015, \mn@doi [\apj] {10.1088/0004-637X/806/1/1}, \href
  {https://ui.adsabs.harvard.edu/\#abs/2015ApJ...806....1M} {806, 1}

\bibitem[\protect\citeauthoryear{{Moraes} et~al.,}{{Moraes}
  et~al.}{2014}]{CFHT-S82}
{Moraes} B.,  et~al., 2014, in Revista Mexicana de Astronomia y Astrofisica
  Conference Series. pp 202--203

\bibitem[\protect\citeauthoryear{{More}, {Miyatake}, {Mandelbaum}, {Takada},
  {Spergel}, {Brownstein}  \& {Schneider}}{{More} et~al.}{2015}]{More2015}
{More} S.,  {Miyatake} H.,  {Mandelbaum} R.,  {Takada} M.,  {Spergel} D.~N.,
  {Brownstein} J.~R.,   {Schneider} D.~P.,  2015, \mn@doi [\apj]
  {10.1088/0004-637X/806/1/2}, \href
  {https://ui.adsabs.harvard.edu/\#abs/2015ApJ...806....2M} {806, 2}

\bibitem[\protect\citeauthoryear{{Morrison}, {Hildebrandt}, {Schmidt},
  {Baldry}, {Bilicki}, {Choi}, {Erben}  \& {Schneider}}{{Morrison}
  et~al.}{2017}]{2017MNRAS.467.3576M}
{Morrison} C.~B.,  {Hildebrandt} H.,  {Schmidt} S.~J.,  {Baldry} I.~K.,
  {Bilicki} M.,  {Choi} A.,  {Erben} T.,   {Schneider} P.,  2017, \mn@doi
  [\mnras] {10.1093/mnras/stx342}, \href
  {http://adsabs.harvard.edu/abs/2017MNRAS.467.3576M} {467, 3576}

\bibitem[\protect\citeauthoryear{{Newman}}{{Newman}}{2008}]{Newman2008}
{Newman} J.~A.,  2008, \mn@doi [\apj] {10.1086/589982}, \href
  {https://ui.adsabs.harvard.edu/abs/2008ApJ...684...88N} {684, 88}

\bibitem[\protect\citeauthoryear{OSC}{OSC}{1987}]{OSC}
OSC 1987, Ohio Supercomputer Center, \url {http://osc.edu/ark:/19495/f5s1ph73}

\bibitem[\protect\citeauthoryear{{Omori} et~al.,}{{Omori}
  et~al.}{2018}]{Omori2018}
{Omori} Y.,  et~al., 2018, arXiv e-prints, \href
  {https://ui.adsabs.harvard.edu/\#abs/2018arXiv181002342O} {p.
  arXiv:1810.02342}

\bibitem[\protect\citeauthoryear{{Padmanabhan} et~al.,}{{Padmanabhan}
  et~al.}{2007}]{Padmanabhan2007}
{Padmanabhan} N.,  et~al., 2007, \mn@doi [\mnras]
  {10.1111/j.1365-2966.2007.11593.x}, \href
  {https://ui.adsabs.harvard.edu/\#abs/2007MNRAS.378..852P} {378, 852}

\bibitem[\protect\citeauthoryear{{Park} et~al.,}{{Park}
  et~al.}{2016}]{Park2016CombiningClustering}
{Park} Y.,  et~al., 2016, \mn@doi [\prd] {10.1103/PhysRevD.94.063533}, \href
  {https://ui.adsabs.harvard.edu/abs/2016PhRvD..94f3533P} {94, 063533}

\bibitem[\protect\citeauthoryear{{Pedregosa} et~al.,}{{Pedregosa}
  et~al.}{2012}]{SCIKIT-LEARN}
{Pedregosa} F.,  et~al., 2012, arXiv e-prints, \href
  {https://ui.adsabs.harvard.edu/abs/2012arXiv1201.0490P} {p. arXiv:1201.0490}

\bibitem[\protect\citeauthoryear{{Pellejero-Ibanez} et~al.,}{{Pellejero-Ibanez}
  et~al.}{2017}]{Pellejero-Ibanez2016ThePriors}
{Pellejero-Ibanez} M.,  et~al., 2017, \mn@doi [\mnras] {10.1093/mnras/stx751},
  \href {https://ui.adsabs.harvard.edu/\#abs/2017MNRAS.468.4116P} {468, 4116}

\bibitem[\protect\citeauthoryear{{Perlmutter}, {Aldering}, {Goldhaber}  \&
  {Knop}}{{Perlmutter} et~al.}{1999}]{Perlmutter99}
{Perlmutter} S.,  {Aldering} G.,  {Goldhaber} G.,   {Knop} R.~A.,  1999,
  \mn@doi [\apj] {10.1086/307221}, \href
  {https://ui.adsabs.harvard.edu/\#abs/1999ApJ...517..565P} {517, 565}

\bibitem[\protect\citeauthoryear{{Planck Collaboration} et~al.,}{{Planck
  Collaboration} et~al.}{2016}]{Planck2015Lensing}
{Planck Collaboration} et~al., 2016, \mn@doi [\aap]
  {10.1051/0004-6361/201525941}, \href
  {https://ui.adsabs.harvard.edu/\#abs/2016A&A...594A..15P} {594, A15}

\bibitem[\protect\citeauthoryear{{Rahman}, {Mendez}, {M{\'e}nard}, {Scranton},
  {Schmidt}, {Morrison}  \& {Budav{\'a}ri}}{{Rahman}
  et~al.}{2016}]{2016MNRAS.460..163R}
{Rahman} M.,  {Mendez} A.~J.,  {M{\'e}nard} B.,  {Scranton} R.,  {Schmidt}
  S.~J.,  {Morrison} C.~B.,   {Budav{\'a}ri} T.,  2016, \mn@doi [\mnras]
  {10.1093/mnras/stw981}, \href
  {https://ui.adsabs.harvard.edu/abs/2016MNRAS.460..163R} {460, 163}

\bibitem[\protect\citeauthoryear{{Reid} et~al.,}{{Reid}
  et~al.}{2016}]{Reid2016}
{Reid} B.,  et~al., 2016, \mn@doi [\mnras] {10.1093/mnras/stv2382}, \href
  {https://ui.adsabs.harvard.edu/\#abs/2016MNRAS.455.1553R} {455, 1553}

\bibitem[\protect\citeauthoryear{{Riess}, {Filippenko}  \& {Challis}}{{Riess}
  et~al.}{1998}]{Riess98}
{Riess} A.~G.,  {Filippenko} A.~V.,   {Challis} P.,  1998, \mn@doi [\aj]
  {10.1086/300499}, \href
  {https://ui.adsabs.harvard.edu/\#abs/1998AJ....116.1009R} {116, 1009}

\bibitem[\protect\citeauthoryear{{Ross} et~al.,}{{Ross}
  et~al.}{2011}]{Ross2011}
{Ross} A.~J.,  et~al., 2011, \mn@doi [\mnras]
  {10.1111/j.1365-2966.2011.19351.x}, \href
  {https://ui.adsabs.harvard.edu/\#abs/2011MNRAS.417.1350R} {417, 1350}

\bibitem[\protect\citeauthoryear{{Ross} et~al.,}{{Ross}
  et~al.}{2012}]{Ross2012}
{Ross} A.~J.,  et~al., 2012, \mn@doi [\mnras]
  {10.1111/j.1365-2966.2012.21235.x}, \href
  {https://ui.adsabs.harvard.edu/abs/2012MNRAS.424..564R} {424, 564}

\bibitem[\protect\citeauthoryear{{Rozo} et~al.,}{{Rozo}
  et~al.}{2016}]{Rozo2016RedMaGiC:Data}
{Rozo} E.,  et~al., 2016, \mn@doi [\mnras] {10.1093/mnras/stw1281}, \href
  {https://ui.adsabs.harvard.edu/\#abs/2016MNRAS.461.1431R} {461, 1431}

\bibitem[\protect\citeauthoryear{{Rykoff} et~al.,}{{Rykoff}
  et~al.}{2014}]{Rykoff2014RedMaPPer.Catalog}
{Rykoff} E.~S.,  et~al., 2014, \mn@doi [\apj] {10.1088/0004-637X/785/2/104},
  \href {https://ui.adsabs.harvard.edu/\#abs/2014ApJ...785..104R} {785, 104}

\bibitem[\protect\citeauthoryear{{Salazar-Albornoz} et~al.,}{{Salazar-Albornoz}
  et~al.}{2017}]{Salazar-albornoz2018TheImplications}
{Salazar-Albornoz} S.,  et~al., 2017, \mn@doi [\mnras] {10.1093/mnras/stx633},
  \href {https://ui.adsabs.harvard.edu/\#abs/2017MNRAS.468.2938S} {468, 2938}

\bibitem[\protect\citeauthoryear{{Schlafly} \& {Finkbeiner}}{{Schlafly} \&
  {Finkbeiner}}{2011}]{Schlafly2011MEASURINGSFD}
{Schlafly} E.~F.,  {Finkbeiner} D.~P.,  2011, \mn@doi [\apj]
  {10.1088/0004-637X/737/2/103}, \href
  {https://ui.adsabs.harvard.edu/\#abs/2011ApJ...737..103S} {737, 103}

\bibitem[\protect\citeauthoryear{{Schlafly}, {Finkbeiner}, {Schlegel},
  {Juri{\'c}}, {Ivezi{\'c}}, {Gibson}, {Knapp}  \& {Weaver}}{{Schlafly}
  et~al.}{2010}]{Schlafly2010THESURVEY}
{Schlafly} E.~F.,  {Finkbeiner} D.~P.,  {Schlegel} D.~J.,  {Juri{\'c}} M.,
  {Ivezi{\'c}} {\v{Z}}.,  {Gibson} R.~R.,  {Knapp} G.~R.,   {Weaver} B.~A.,
  2010, \mn@doi [\apj] {10.1088/0004-637X/725/1/1175}, \href
  {https://ui.adsabs.harvard.edu/\#abs/2010ApJ...725.1175S} {725, 1175}

\bibitem[\protect\citeauthoryear{{Schlegel}, {Finkbeiner}  \&
  {Davis}}{{Schlegel} et~al.}{1998}]{SFD98}
{Schlegel} D.~J.,  {Finkbeiner} D.~P.,   {Davis} M.,  1998, \mn@doi [\apj]
  {10.1086/305772}, \href
  {https://ui.adsabs.harvard.edu/\#abs/1998ApJ...500..525S} {500, 525}

\bibitem[\protect\citeauthoryear{{Schmidt}, {M{\'e}nard}, {Scranton},
  {Morrison}  \& {McBride}}{{Schmidt}
  et~al.}{2013}]{Schmidt2013RecoveringBoundaries}
{Schmidt} S.~J.,  {M{\'e}nard} B.,  {Scranton} R.,  {Morrison} C.,   {McBride}
  C.~K.,  2013, \mn@doi [\mnras] {10.1093/mnras/stt410}, \href
  {https://ui.adsabs.harvard.edu/\#abs/2013MNRAS.431.3307S} {431, 3307}

\bibitem[\protect\citeauthoryear{{Schwarz}}{{Schwarz}}{1978}]{BIC}
{Schwarz} G.,  1978, Annals of Statistics, \href
  {https://ui.adsabs.harvard.edu/abs/1978AnSta...6..461S} {6, 461}

\bibitem[\protect\citeauthoryear{{Scottez}, {Benoit-L{\'e}vy}, {Coupon},
  {Ilbert}  \& {Mellier}}{{Scottez} et~al.}{2018}]{2018MNRAS.474.3921S}
{Scottez} V.,  {Benoit-L{\'e}vy} A.,  {Coupon} J.,  {Ilbert} O.,   {Mellier}
  Y.,  2018, \mn@doi [\mnras] {10.1093/mnras/stx3056}, \href
  {https://ui.adsabs.harvard.edu/abs/2018MNRAS.474.3921S} {474, 3921}

\bibitem[\protect\citeauthoryear{{Seljak} et~al.,}{{Seljak}
  et~al.}{2005}]{Seljak2003SDSSImplications}
{Seljak} U.,  et~al., 2005, \mn@doi [\prd] {10.1103/PhysRevD.71.043511}, \href
  {https://ui.adsabs.harvard.edu/\#abs/2005PhRvD..71d3511S} {71, 043511}

\bibitem[\protect\citeauthoryear{{Singh}, {Mandelbaum}, {Seljak},
  {Rodr{\'\i}guez-Torres}  \& {Slosar}}{{Singh} et~al.}{2018}]{Singh2018}
{Singh} S.,  {Mandelbaum} R.,  {Seljak} U.,  {Rodr{\'\i}guez-Torres} S.,
  {Slosar} A.,  2018, arXiv e-prints, \href
  {https://ui.adsabs.harvard.edu/\#abs/2018arXiv181106499S} {p.
  arXiv:1811.06499}

\bibitem[\protect\citeauthoryear{{Singh}, {Alam}, {Mandelbaum}, {Seljak},
  {Rodriguez-Torres}  \& {Ho}}{{Singh} et~al.}{2019}]{Singh2019}
{Singh} S.,  {Alam} S.,  {Mandelbaum} R.,  {Seljak} U.,  {Rodriguez-Torres} S.,
    {Ho} S.,  2019, \mn@doi [\mnras] {10.1093/mnras/sty2681}, \href
  {https://ui.adsabs.harvard.edu/\#abs/2019MNRAS.482..785S} {482, 785}

\bibitem[\protect\citeauthoryear{{The Dark Energy Survey Collaboration}}{{The
  Dark Energy Survey Collaboration}}{2005}]{DESCollaboration2006}
{The Dark Energy Survey Collaboration} 2005, arXiv e-prints, \href
  {https://ui.adsabs.harvard.edu/\#abs/2005astro.ph.10346T} {pp
  astro--ph/0510346}

\bibitem[\protect\citeauthoryear{{VanderPlas}, {Connolly}, {Ivezic}  \&
  {Gray}}{{VanderPlas} et~al.}{2012}]{AstroML_XD}
{VanderPlas} J.,  {Connolly} A.~J.,  {Ivezic} Z.,   {Gray} A.,  2012, in
  Proceedings of Conference on Intelligent Data Understanding (CIDU. pp 47--54
  (\mn@eprint {arXiv} {1411.5039}), \mn@doi{10.1109/CIDU.2012.6382200}

\bibitem[\protect\citeauthoryear{{Weinberg}, {Mortonson}, {Eisenstein},
  {Hirata}, {Riess}  \& {Rozo}}{{Weinberg} et~al.}{2013}]{Weinberg13}
{Weinberg} D.~H.,  {Mortonson} M.~J.,  {Eisenstein} D.~J.,  {Hirata} C.,
  {Riess} A.~G.,   {Rozo} E.,  2013, \mn@doi [\physrep]
  {10.1016/j.physrep.2013.05.001}, \href
  {http://adsabs.harvard.edu/abs/2013PhR...530...87W} {530, 87}

\bibitem[\protect\citeauthoryear{{Wright} et~al.,}{{Wright}
  et~al.}{2010}]{WISE}
{Wright} E.~L.,  et~al., 2010, \mn@doi [\aj] {10.1088/0004-6256/140/6/1868},
  \href {https://ui.adsabs.harvard.edu/\#abs/2010AJ....140.1868W} {140, 1868}

\bibitem[\protect\citeauthoryear{{Yoo} \& {Seljak}}{{Yoo} \&
  {Seljak}}{2012}]{Yoo2012}
{Yoo} J.,  {Seljak} U.,  2012, \mn@doi [\prd] {10.1103/PhysRevD.86.083504},
  \href {https://ui.adsabs.harvard.edu/\#abs/2012PhRvD..86h3504Y} {86, 083504}

\bibitem[\protect\citeauthoryear{{York} et~al.,}{{York}
  et~al.}{2000}]{York2000SDSS}
{York} D.~G.,  et~al., 2000, \mn@doi [\aj] {10.1086/301513}, \href
  {https://ui.adsabs.harvard.edu/\#abs/2000AJ....120.1579Y} {120, 1579}

\bibitem[\protect\citeauthoryear{Zonca, Singer, Lenz, Reinecke, Rosset, Hivon
  \& Gorski}{Zonca et~al.}{2019}]{healpy}
Zonca A.,  Singer L.,  Lenz D.,  Reinecke M.,  Rosset C.,  Hivon E.,   Gorski
  K.,  2019, \mn@doi [Journal of Open Source Software] {10.21105/joss.01298},
  4, 1298

\bibitem[\protect\citeauthoryear{{Zuntz} et~al.,}{{Zuntz}
  et~al.}{2015}]{COSMOSIS}
{Zuntz} J.,  et~al., 2015, \mn@doi [Astronomy and Computing]
  {10.1016/j.ascom.2015.05.005}, \href
  {https://ui.adsabs.harvard.edu/\#abs/2015A&C....12...45Z} {12, 45}

\bibitem[\protect\citeauthoryear{{de Jong}, {Verdoes Kleijn}, {Kuijken}  \&
  {Valentijn}}{{de Jong} et~al.}{2013}]{KiDS}
{de Jong} J. T.~A.,  {Verdoes Kleijn} G.~A.,  {Kuijken} K.~H.,   {Valentijn}
  E.~A.,  2013, \mn@doi [Experimental Astronomy] {10.1007/s10686-012-9306-1},
  \href {https://ui.adsabs.harvard.edu/\#abs/2013ExA....35...25D} {35, 25}

\bibitem[\protect\citeauthoryear{{de Vaucouleurs}}{{de
  Vaucouleurs}}{1948}]{DeVaucouleurs}
{de Vaucouleurs} G.,  1948, Annales d'Astrophysique, \href
  {http://adsabs.harvard.edu/abs/1948AnAp...11..247D} {11, 247}

\bibitem[\protect\citeauthoryear{{van den Bosch}, {More}, {Cacciato}, {Mo}  \&
  {Yang}}{{van den Bosch} et~al.}{2013}]{Vandenbosch2013}
{van den Bosch} F.~C.,  {More} S.,  {Cacciato} M.,  {Mo} H.,   {Yang} X.,
  2013, \mn@doi [\mnras] {10.1093/mnras/sts006}, \href
  {https://ui.adsabs.harvard.edu/\#abs/2013MNRAS.430..725V} {430, 725}

\makeatother
\end{thebibliography}

\appendix
\section{ The impact of redshift tails in BOSS CMASS on galaxy bias }
\label{app:redshift}

%\notesj{grammar check required}
\noteac{I've gone through all of the appendices and fixed all of the grammar and typo issues I found. 04/12/19 4:30pm EST.}
The BOSS analyses use the CMASS galaxies only within the redshift range $( 0.43 < z < 0.75 )$, by applying the spectroscopic redshift cuts on the CMASS targets selected by photometric cuts in Equations (\ref{eq:imag_limit}) - (\ref{eq:dperp}) \citep{Chuang2017,  Pellejero-Ibanez2016ThePriors}. Through the redshift cuts, nearly $10\%$ of sources are discarded from the photometric targets. As the DMASS algorithm only utilizes the photometric information of galaxies, the resulting DMASS sample includes sources at the low end ($z < 0.43$) or high end ($z > 0.75$), and they cannot be excluded as done in the BOSS CMASS sample.   
%
%The DMASS algorithm uses the photometric redshift sample that includes sources in the low and high redshift tails as well. 
To combine the BOSS measurements with the weak lensing measurements of DMASS, the effect of the redshift tails on galaxy clustering should be examined. 
Here, we test the impact of the redshift tails on the galaxy clustering of BOSS CMASS, specifically on galaxy bias, by computing the correlation function monopole $\xi_0 (r)$ and quadrupole  $\xi_2 (r)$. 

We use the three dimensional, two point correlation function estimator given by \cite{Landy1993BiasFunctions}:
\bea
\xi(s, \mu) = \frac{ DD(s, \mu) -2 DR (s, \mu) + RR ( s, \mu) }{ RR ( s, \mu) }
\eea
where $s$ is the separation of a pair of objects and $\mu$ is the cosine of the angle between the directions between the line of sight (LOS) and the line connecting the pair of objects. $DD$, $DR$, and $RR$ represent the normalized galaxy-galaxy, galaxy-random, and random-random pair counts, for a given separation $s$ and $\mu$. The weights described in Equation (\ref{eq:weights}) are applied.

To derive the monopole and quadrupole, the two-point correlation function $\xi(s, \mu)$ is integrated over a spherical shell with radius $s$: % to obtain monopole (l=0) and quadrupole (l=2). 
\bea
\xi_l (s) = \frac{1}{N_{\mu}} \sum_{i=0}^{N_{\mu}}  (2 l + 1) ~\xi(s, \mu_i) ~P_l (\mu_i)  
\eea
where $N_\mu$ is the number of $\mu$ bins, $P_l(\mu)$ is the Legendre Polynomial.

\begin{figure}
	\centering
	\includegraphics[width=0.45\textwidth]{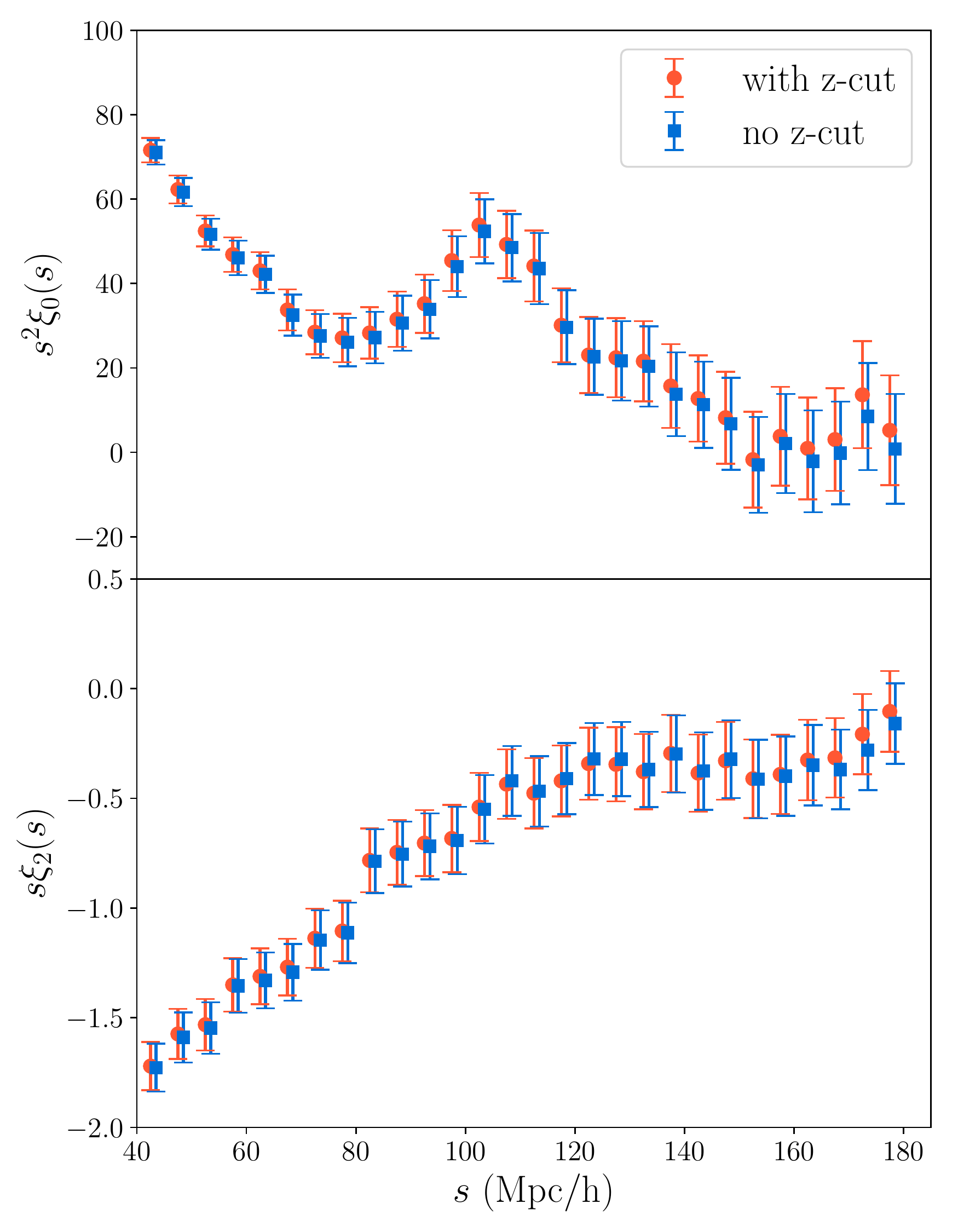}
	\caption{ Monopole (top) and quadrupole (bottom) correlation function of the CMASS sample before (red) and after (blue) applying redshift cuts at $40 \hMpc < s < 180 \hMpc$. 
	}
\label{fig:cfz_comparison_zcut}
\end{figure}

Figure \ref{fig:cfz_comparison_zcut} shows the monopole and quadrupole of CMASS with the redshift cuts (red) and without the redshift cuts (blue). 
These multipoles are computed in the scale range $40 \hMpc< s <180 \hMpc $ with the bin size $5 \hMpc$, the same scales and bin size adopted in the previous BOSS analyses \citep{Chuang2017,  Pellejero-Ibanez2016ThePriors}. 
%The redshift distribution is specifically impact on small scales, we only plotted small scales here. 
Error bars are computed from the MultiDark-PATCHY BOSS DR12 mock  catalogues \citep{Kitaura2016TheRelease}. The amplitudes of the multipoles are overall higher on large scales $s > 120 \hMpc$ with the redshift tails. This may indicate systematics associated with sources at high redshift, but their impact should be negligible as the offset between the two correlation functions is way smaller than the statistical errors. 

From the measured monopole and quadrupole of the correlation function, we constrain galaxy bias $b$ and the structure growth rate, $f \equiv d\ln D/d \ln a$ at a median redshift $z=0.59$ for each case and compare the results. 
In linear theory, the relative amplitudes of multipoles depends only on the combination of $b$ and $f$ as follows:
\bea
\xi_0(s) &=& \left(b^2 + \frac{2}{3}bf + \frac{1}{5}f^2 \right) \xi_m(s)\\
\xi_2(s) &=& \left( \frac{4}{3}bf + \frac{4}{7}f^2 \right) \xi_m(s)~.
\eea
As the linear theory is applicable on the scales we are using, we simply adopt the above equations and expect the potential difference due to the redshift cuts to appear on the constraints of $b$ and $f$. The matter correlation function $\xi_m(s)$ at $z=0.59$ is estimated by \verb|CAMB| \citep{CAMB} with the fiducial cosmology.
%\footnote{http://camb.info/}  

We perform a Markov Chain Monte-Carlo likelihood analysis using \verb|emcee| \citep{EMCEE}. 
The two parameters $b$ and $f$ are varied in the range of $b = [0.5, 3]$ and $f = [0.2, 1.0]$. 
%a maximum likelihood analysis 
The likelihood is taken from $\chi^2$ defined as 
\bea
\chi^2 = \sum_{i,j}^{N_X} [X_{ {\rm obs},i} - X_{{\rm th}, i}]~ \boldsymbol{C}^{-1}_{ij}  ~[ X_{{\rm obs},j} - X_{{\rm th}, j}]
\eea
where $N_X$ is the number of points in the data vector, $X_{\rm th}$ is the vector from the theoretical model, $X_{\rm obs}$ is the vector from the measurement.  
The data points from the multipoles are combined to form a vector $X$ as 
\bea
X = \{ \xi_0(s_0),  \xi_0(s_1), ..., \xi_0(s_N);\xi_2 (s_0), \xi_2 (s_1), ..., \xi_2 (s_N) \}
\eea
where $N$ is the number of bins in each multipole.

\iffalse
we perform Markov Chain Monte-Carlo likelihood analyses %a maximum likelihood analysis 
to constrain $b$ and $f$. 
%to constrain parameters $b$, $\Delta b$ and $\Delta z$. 
The likelihood of the combined cosmological probe is given by the sum of individual log likelihoods given as
\fi

\begin{figure}
\centering
\includegraphics[width=0.43\textwidth]{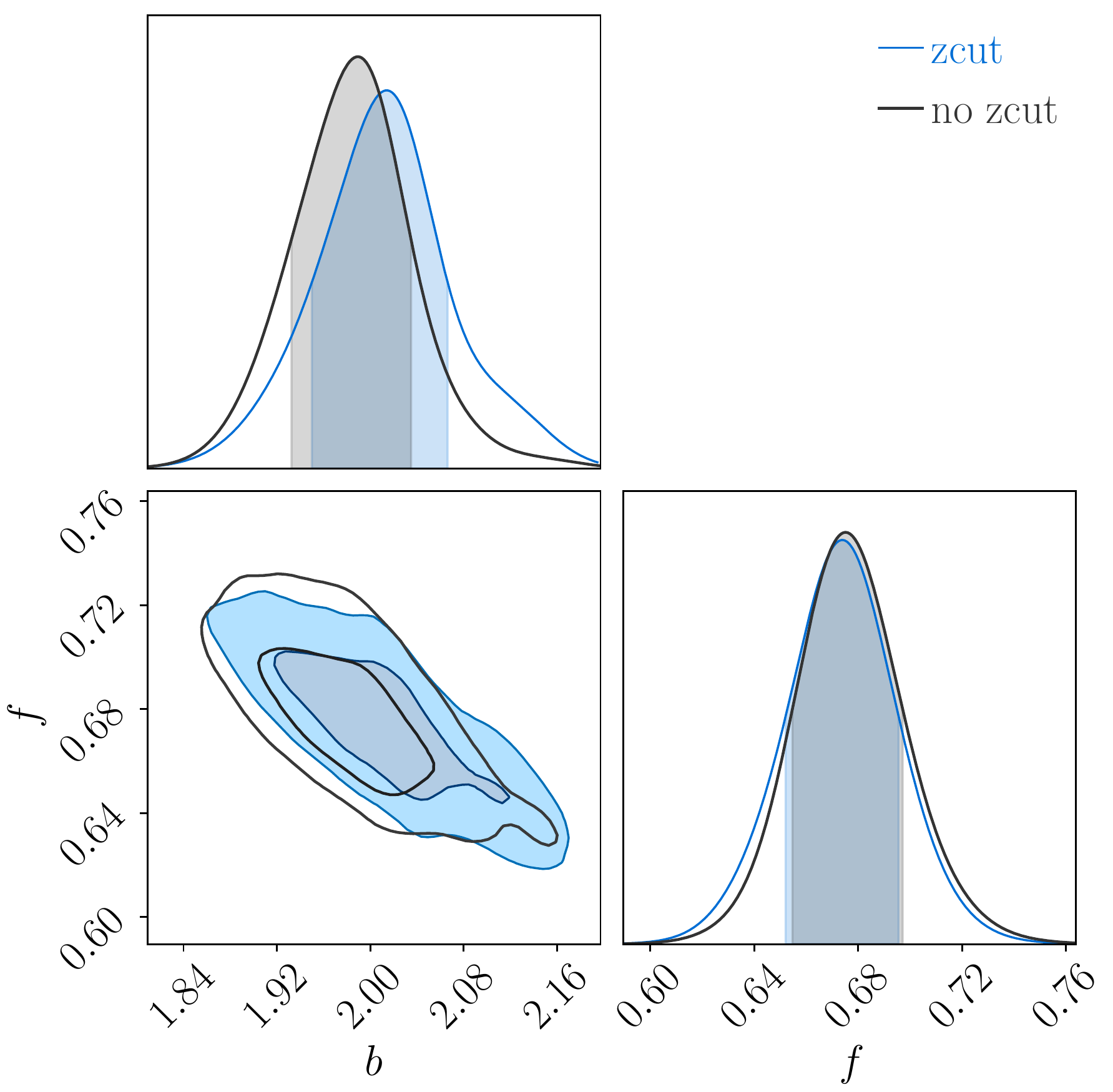}
\caption{Comparison of galaxy bias constraints from the CMASS clustering before (blue) and after (red) applying z-cuts.
}
\label{fig:corr_comparison_monopole}
\end{figure}

Figure \ref{fig:corr_comparison_monopole} shows the constraints of $b$ and $f$ at $z=0.59$ on the two dimensional plane. 
The resulting numbers are 
$b = 2.02^{+0.04}_{-0.07}$ and $f = 0.67\pm0.02$ with the redshift cut, and $b = 2.00^{+0.04}_{-0.06}$ and $f = 0.68\pm0.02$ without the redshift cut.  From these results, we do not find a big discrepancy between the two cases. 
%\notesj{Galaxy bias is not well constrained as shown in Figure A2 (Same as Ibanez et al).Should report $\beta$ instead? Or just go with this plot? } 

The negligible impact of high redshift galaxies has also been studied by the original BOSS analyses. 
\cite{Cuesta2016}, \cite{Gil-Marin2016a} and \cite{Gil-Marin2016b} use CMASS within $0.43 < z < 0.7$ with the effective redshift $z = 0.57$. 
\cite{Alam2015TheSDSS-III} compared the measurements from these analyses with the ones including higher redshift galaxies up to $z < 0.75$, with the effective redshift $z=0.59$ \citep{Chuang2017, Pellejero-Ibanez2016ThePriors}. 
%\cite{Cuesta} and \cite{GilMarin} use CMASS within $0.43 < z < 0.7$ with the effective redshift $z = 0.57$. 
To compare the measurements directly at the same redshift, \cite{Alam2015TheSDSS-III} extrapolated the measurements of \cite{Chuang2017} and \cite{Pellejero-Ibanez2016ThePriors} at $z=0.57$.
A summary of their work can be found in Figures 13 and 15 in \cite{Alam2015TheSDSS-III}. All of the BOSS measurements compared in this work show consistency within the $1\sigma$ level or better.

\section{ Differences between CMASS SGC and NGC }
\label{app:ngc_sgc}

%\reph{[this paragraph will be deleted ]Schlafly et al and Schlafly \& Finkbeiner have estimated the level of color offset between the North and South in the SDSS. The dereddened SGC is redder than dereddened NGC. One can infer that the values of $\dperp$ are offset by 0.0064 magnitudes between the North and South cap. Ashley et al 2011 studied the impact of the offset on the number densities and the angular correlation functions with systematic corrections. By assuming the values in the North are true values, they applied $\dperp > 0.5564$ cuts to the CMASS sample in the South cap. With the inferred offset in $\dperp$, the difference of the number density between north and south decreased from 3.4 \% to below 0.01 \%. The amplitude of the angular correlation function with the new $\dperp$ cut are reduced at large scales and appear to be closer to the correlation function of the full CMASS sample. }

The BOSS CMASS target selection function is applied differently in the South Galactic cap (SGC) and North Galactic cap (NGC) due to the color offsets in the DR8 photometry between two regions. \cite{Schlafly2010THESURVEY} and \cite{Schlafly2011MEASURINGSFD} have estimated the level of color offsets and found that these differences are due to either calibration errors or errors in the galactic extinction corrections (or combination of both). 
%the dereddened SGC is redder than dereddened NGC. One can infer that 
This offset shifts the values of $\dperp$ (the combination of $g-r$ and $r-i$ colors) by 0.0064 magnitudes between the North and South cap, resulting in a few per cent difference in the number density and the amplitude of the angular correlation function. 
%The values of $\dperp$ are offset by 0.0064 magnitudes between the North and South cap. This offsets result in a few per cent difference in number density, therefore, the amplitude of the angular correlation function as well. 
\cite{Ross2011} and \cite{Ross2012}  have shown that the difference in the number density and the angular correlation function can be mitigated 
%from 3.4 \% to below 0.01 \%  
by applying the new cut with the offset, $\dperp > 0.5564$, in the SGC. 
%The amplitude of the angular correlation function with the new $\dperp$ cut are reduced at large scales and appears to be closer to the correlation function of the full CMASS sample. 
However, the final analyses of BOSS-III were completed with the same $\dperp$ cut in both regions. Therefore, we do not consider the color offset either. 
%Therefore, we utilize the same $\dperp$ value to train the extreme deconvolution algorithm. 
%Since the extreme deconvolution model is utilizing the $\dperp$ color in the South galactic cap, 

The resulting DMASS is designed to be closer to CMASS in the SGC than NGC since the extreme deconvolution model is trained with the $\dperp$ color in the SGC. Therefore, we report the measurements of the angular correlation functions and 
galaxy biases of CMASS NGC and SGC here in order to show that discrepancy between DMASS and full CMASS originates from the intrinsic difference within CMASS.

%\caption{$\chi^2/{\rm dof}$ of three probes calculated between two different samples. Values in the parentheses were calculated with systematic weights of DMASS obtained in Section \ref{sec:systematics}. SGC and FULL in bold stand for CMASS in SGC and full CMASS.  }

% \usepackage{booktabs}
% \usepackage{graphicx}
\begin{table}
\centering
\caption{
$\chi^2/{\rm dof}$ of three probes calculated between two different samples.  
$\chi_{\rm sys}^2/{\rm dof}$ in the third column are calculated with systematic weights of DMASS obtained in Section \ref{sec:systematics}. 
Values in the parentheses are corresponding PTE values. 
SGC and FULL in bold stand for CMASS in SGC and full CMASS.  }
\label{tab:chisquare_pte}
\resizebox{0.35\textwidth}{!}{%
\begin{tabular}{@{}lcrrl@{}}
\toprule
 & \textbf{PROBE}                   & \multicolumn{1}{c}{$\chi^2/{\rm dof}$ (PTE)} & \multicolumn{1}{c}{$\chi_{\rm sys}^2/{\rm dof}$ (PTE)} &  \\ \midrule
 & \multicolumn{3}{c}{\textbf{SGC - DMASS}}                                                                                                 &  \\
 & $w^{\delta_g \delta_g}$          & 4.94/10 (90\%)                               & 2.58/10 (99\%)                                         &  \\
 & $w^{\delta_g \delta_{\rm WISE}}$ & 9.04/10 (53\%)                               & 9.70/10 (47\%)                                         &  \\
 & $w^{\delta_g \kappa_{\rm CMB}}$  & 8.92/7   (26\%)                              & 13.25/7 ~   (6\%)                                      &  \\ \midrule
 & \multicolumn{3}{c}{\textbf{FULL - DMASS}}                                                                                                &  \\
 & $w^{\delta_g \delta_g}$          & 10.61/10 (39\%)                              & 8.60/10 (57\%)                                         &  \\
 & $w^{\delta_g \delta_{\rm WISE}}$ & 12.12/10 (28\%)                              & 11.42/10 (33\%)                                        &  \\
 & $w^{\delta_g \kappa_{\rm CMB}}$  & 7.13/7   (42\%)                              & 7.68/7  (36\%)                                         &  \\ \midrule
 & \multicolumn{3}{c}{\textbf{NGC - SGC}}                                                                                                   &  \\
 & $w^{\delta_g \delta_g}$          & 14.53/10 (15\%)                              & \multicolumn{1}{c}{$-$}                                  &  \\
 & $w^{\delta_g \delta_{\rm WISE}}$ & 11.76/10 (30\%)                              & \multicolumn{1}{c}{$-$}                                  &  \\
 & $w^{\delta_g \kappa_{\rm CMB}}$  & 23.95/7 (0.1\%)                              & \multicolumn{1}{c}{$-$}                                  &  \\ \bottomrule
\end{tabular}%
}
\end{table}

\iffalse
% \usepackage{booktabs}
\begin{table}
\centering
\caption{\sjcmt{will be deleted}
$\chi^2/{\rm dof}$ of three probes calculated between two different samples. Values in the parentheses were calculated with systematic weights of DMASS obtained in Section \ref{sec:systematics}. SGC and FULL in bold stand for CMASS in SGC and full CMASS.  }
\label{tab:chisquare}
\resizebox{0.47 \textwidth}{!}{%
\begin{tabular}{@{}clll@{}}
\toprule
\textbf{PROBE}                   & \multicolumn{3}{c}{\textbf{$\chi^2/{\rm dof}$}}                                                                               \\ \midrule
\multicolumn{1}{l}{}             & \multicolumn{1}{c}{\textbf{SGC-DMASS}} & \multicolumn{1}{c}{\textbf{FULL-DMASS}} & \multicolumn{1}{c}{\textbf{NGC - SGC}} \\
$w^{\delta_g \delta_g}$          & ~4.94/10 ~(2.58/10)                        & 10.61/10 ~(8.60/10)                        & ~14.53/10 (-)                           \\
$w^{\delta_g \delta_{g {\rm WISE} }}$ & ~9.04/10 ~(9.70/10)                        & 12.12/10 (11.42/10)                       & ~11.76/10 (-)                           \\
$w^{\delta_g \kappa_{\rm CMB}}$  & 8.92/7 (13.25/7)                      & 5.51/7 (7.74/7)                       & ~23.95/7 (-)                           \\ \bottomrule
\end{tabular}
}
\end{table}
\fi

\begin{figure}
\centering
	\includegraphics[width=0.35\textwidth]{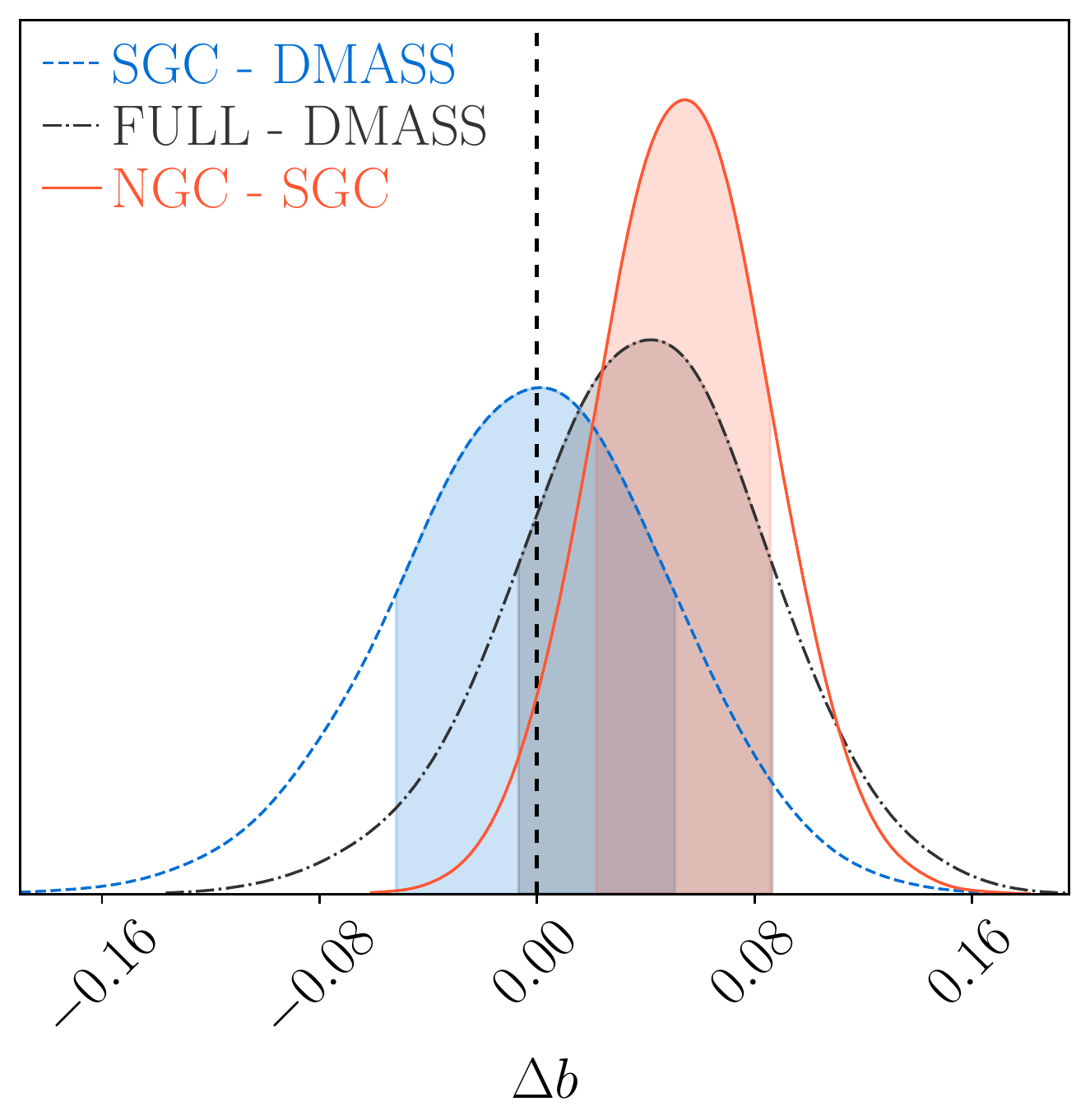}
	\caption{ Difference in galaxy biases constrained by the angular correlation function of CMASS SGC and CMASS NGC (red). 
	$\Delta b_{\rm NGC-SGC} = 0.056^{+0.031}_{-0.033}$.  The blue-dashed and black-dot dashed histograms display $\Delta b_{\rm SGC-DMASS}$ and $\Delta b_{\rm FULL-DMASS}$ obtained in Section \ref{sec:result}, respectively. The redshift bin bias $\Delta z$ of DMASS is marginalized for the latter two cases. 
	%$b_{\rm SGC} = 2.035 \pm 0.02628 $ and $b_{\rm NGC} = 2.088 \pm 0.01685$
	%cmass sgc : bin_bias--b1 = 2.03464 � 0.0262811
	%cmass ngc : bin_bias--b1 = 2.0883 � 0.0168576
	}
\label{fig:appendix_galaxy_bias_difference}
\end{figure}

Table \ref{tab:chisquare_pte} shows the values of $\chi^2/{\rm dof}$ and its corresponding PTE of all three probes computed in Section \ref{sec:result}.  The last column includes $\chi^2/{\rm dof}$ between CMASS in NGC and CMASS in SGC. For all three probes, $\chi^2/{\rm dof}$ of `NGC-SGC' is either larger than any of the other two cases or comparable to the case of `FULL-DMASS'.  

%We constrained the difference in galaxy bias of CMASS SGC and CMASS NGC to show the intrinsic difference within CMASS. 

Galaxy biases are derived from the model of the angular correlation function given as
\bea
w^{\delta_g \delta_g}(\theta ) = \int \ud z~b(z) n(z) \int \ud z'~b(z') n(z') ~\xi_m (R, z, z')
\label{eq:modelw2}
\eea
where $b$ is galaxy bias, $n(z)$ is normalized spectroscopic redshift distribution, $\xi_m$ is matter clustering, and $R$ is the comoving distance defined as $R = (1+z) D_A(z) \theta$.
With the covariance matrices calculated from the QPM mock catalogs in Section \ref{sec:systematics}, we have estimated bestfit values of galaxy bias that minimize $\chi^2$ defined in Equation \eqref{eq:chi2}.
%as follows:
%\bea
%\chi^2 = ({\bf{d}} - {\bf{d}}^T) {\bf{C}}^{-1} ({\bf{d}} - {\bf{d}}^T)~.
%\eea
The data vector $\Delta {\bf{d}}$ in the equation is defined as the residual of the measurement and theoretical prediction given as $\Delta {\bf{d}} = {\bf{d}}_{\rm true} - {\bf{d}}$. 
The vector $\bf{d}$ corresponds to the measurements of the angular correlation function of CMASS SGC and CMASS NGC computed in Section \ref{sec:result}, and ${\bf{d}}_{\rm true}$ is the theoretical data vector from Equation \eqref{eq:modelw2}. 

The final constraints of galaxy biases are $b_{\rm SGC} = 2.035 \pm 0.026$ and $b_{\rm NGC} = 2.088 \pm 0.017$ from CMASS SGC and CMASS NGC, respectively. The derived galaxy bias difference between NGC and SGC is $\Delta b_{\rm NGC-SGC} = 0.056^{+0.031}_{-0.033}$. 
Figure \ref{fig:appendix_galaxy_bias_difference} shows the constraint of $\Delta b_{\rm NGC-SGC}$ (red-solid) plotted with $\Delta b_{\rm SGC-DMASS}$ (blue-dashed) and $\Delta b_{\rm FULL-DMASS}$ (black-dot dashed) obtained in Section \ref{sec:result}. The redshift bin bias $\Delta z$ of DMASS is marginalized for the latter two cases. 
The resulting $\Delta b_{\rm NGC-SGC}$ implies that the color offset between the SGC and NGC naturally yields $\sim 2.6\%$ of the difference in galaxy bias, and the constraints of $\Delta b$ between DMASS and CMASS are safely within this intrinsic difference.

%  Don't change these lines
\bsp   % typesetting comment
\label{lastpage}
\end{document}